\documentclass{article}

\PassOptionsToPackage{numbers, compress}{natbib}

\usepackage[preprint]{neurips_2026}

\usepackage[utf8]{inputenc}
\usepackage[T1]{fontenc}
\usepackage[hidelinks]{hyperref}
\usepackage{url}
\usepackage{booktabs}
\usepackage{amsfonts}
\usepackage{amsmath}
\usepackage{nicefrac}
\usepackage{microtype}
\usepackage{xcolor}
\usepackage{graphicx}
\usepackage{wrapfig}

\usepackage{graphicx}
\usepackage{caption,subcaption}
\usepackage[toc,page]{appendix}
\usepackage{titletoc}
\usepackage{tcolorbox}
\usepackage{fontawesome5}
\usepackage{siunitx}
\usepackage{adjustbox}

\usepackage{multirow}

\usepackage{listings}
\definecolor{codebg}{HTML}{F8F8F8}
\definecolor{codeframe}{HTML}{CCCCCC}
\definecolor{codekeyword}{HTML}{0055AA}
\definecolor{codestring}{HTML}{AA1111}
\definecolor{codecomment}{HTML}{5C8A00}
\definecolor{codelinenumber}{HTML}{AAAAAA}
\definecolor{codenumber}{HTML}{444444}
\definecolor{codedecorator}{HTML}{9B2393}

\newcommand{\standardfootnotesize}{\fontsize{8}{9.5}\selectfont}

\title{Pretraining Strategies and Scaling for ECG Foundation Models: A Systematic Study}

\author{
  M A Al-Masud, Nils Strodthoff\thanks{AI4Health Division --- \url{https://uol.de/en/ai4health}} \\
  AI4Health Division\\
  Carl von Ossietzky Universit\"{a}t Oldenburg \\
  Oldenburg, Lower Saxony, Germany \\
  \texttt{\{m.a.al-masud,nils.strodthoff\}@uol.de} \\
}

\begin{document}

\maketitle

\begin{abstract}
Specialized foundation models are beginning to emerge in various medical subdomains, but pretraining methodologies and parametric scaling with the size of the pretraining dataset are rarely assessed systematically and in a like-for-like manner. This work focuses on foundation models for electrocardiography (ECG) data, one of the most widely captured physiological time series world-wide. We present a comprehensive assessment of pretraining methodologies, covering five different contrastive and non-contrastive self-supervised learning objectives for ECG foundation models, and investigate their scaling behavior with pretraining dataset sizes up to 11M input samples, exclusively from publicly available sources. Pretraining strategy has a meaningful and consistent impact on downstream performance, with contrastive predictive coding (slightly ahead of JEPA) yielding the most transferable representations across diverse clinical tasks. Scaling pretraining data continues to yield meaningful improvements up to 11M samples for most objectives. We also compare model architectures across all pretraining methodologies and find evidence for a clear superiority of structured state space models compared to transformers and CNN models. We hypothesize that the strong inductive biases of structured state space models, rather than pretraining scale alone, are the primary driver of effective ECG representation learning, with important implications for future foundation model development in this and potentially other physiological signal domains.
\end{abstract}

\section{Introduction}
\paragraph{Motivation}
Foundation models (FMs), defined as ``models trained on broad data (generally using self-supervision at scale) that can be adapted to a wide range of downstream tasks" \citep{bommasani2021opportunities} are increasingly gaining traction, also in specialized medical domains \citep{chen2024towards,zhou2023foundation}, offering specific advantages over training from scratch: (i) higher predictive performance (ii) improved label efficiency, and (iii) the ability to serve as frozen feature extractors for downstream tasks. In this submission, we shed light on the training of FMs for electrocardiogram (ECG) data. The ECG represents a widely used non-invasive tool for a first-in-line assessment of the cardiac state and systemic physiology \citep{siontis2021artificial}. For example, the ECG can be used to detect cardiovascular conditions such as myocardial infarction \citep{strodthoff2020deep}, to evaluate cardiovascular risk \citep{bhatia2018screening}, and to guide clinical decisions \citep{rokos2010appropriate}. 

\paragraph{Research gap}
The development of ECG FMs has flourished recently \citep{li2025electrocardiogram, kim2024learning, na2024guiding, liu2024zero, tian2024foundation, coppola2024hubert, mckeen2025ecg, hung2025boosting, lunelli2025benchecg, nolin2026foundation, yu2024ecg, chen2025physiowave}. While comprehensive benchmarking of these models has been a major limitation for a long time, it has been addressed recently by different benchmarking studies \citep{lunelli2025benchecg, al-masud2026benchmarking, wan2025openecgbenchmarkingecgfoundation}. This resolved the question of measuring FM performance in a comprehensive manner, but failed to answer central questions on pretraining methodology and the effect of the scale of pretraining data. This is due to the simple reason that the different FMs used vastly different pretraining methodologies and different training datasets, which entangled pretraining methodology and dataset scale in intricate ways. This submission aims to close this research gap by means of a dedicated study on the pretraining of ECG FMs. We believe this study provides important hints for future FM development in the domain of ECG data and potentially beyond to avoid future FM development with suboptimal setups. Here, we are focusing on FM training purely based on self-supervision. We acknowledge the value of direct/weak supervision, which we envision to be relevant for later training stages of the FM development, but which are not part of this investigation.

\begin{figure}[t]
    \centering
    \begin{minipage}[c][10cm]{\linewidth}
        \includegraphics[width=\linewidth]{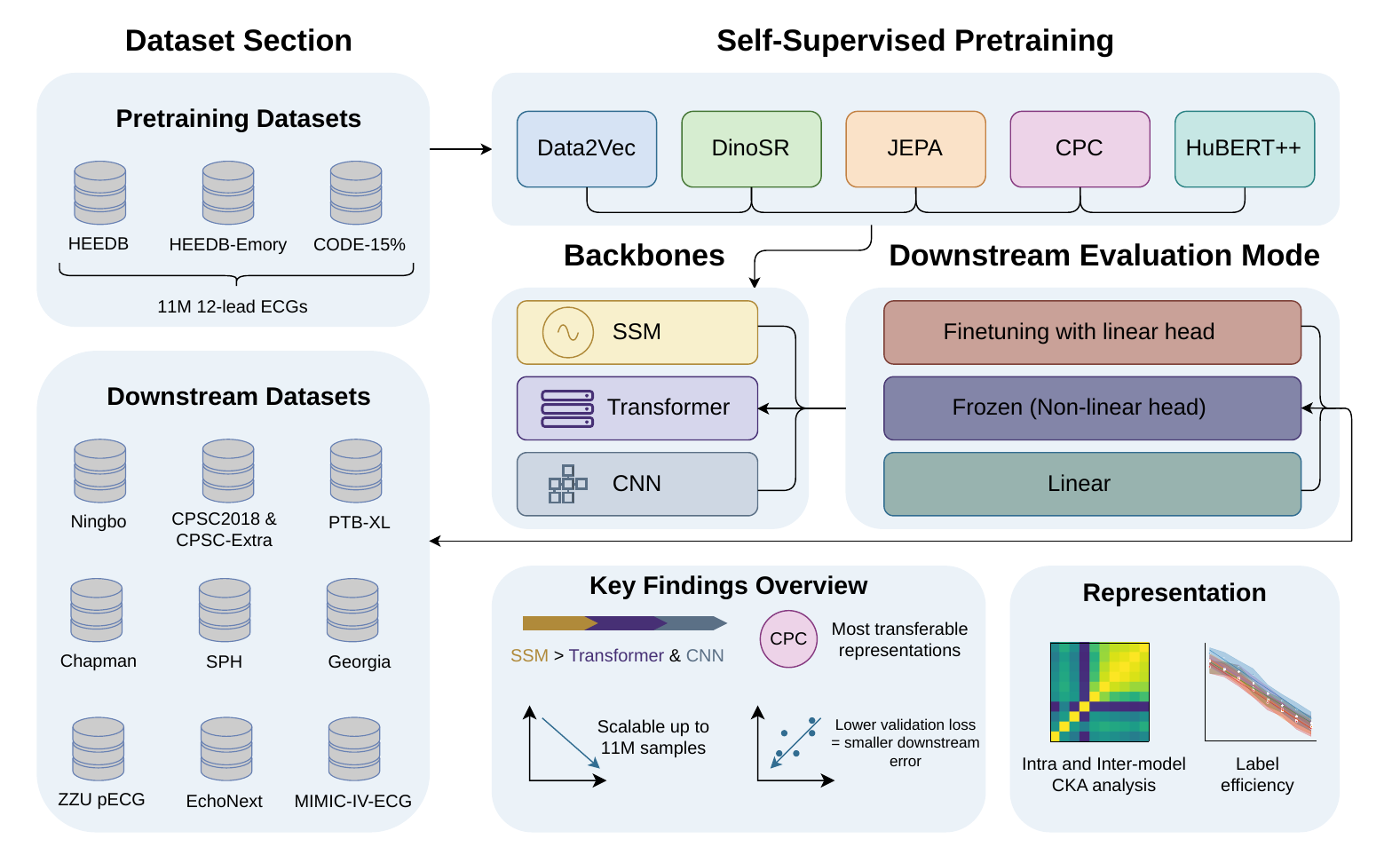}
    \end{minipage}
    \caption{Schematic overview of the design of the study: This work provides a comprehensive assessment of the pretraining of ECG foundation models through self-supervision alone covering pretraining methodologies, model architectures, scaling behavior, label efficiency and representation similarity.}
    \label{fig:abstract}
\end{figure}

\paragraph{Research questions}
More specifically, we try to answer the following research questions: How does the choice of the model architecture impact the performance of ECG FMs? Here we are interested in investigating the broad classes CNN vs.\ transformer vs.\ state space models. Furthermore, what is the impact of the pretraining methodology on the performance and also on the internal structure of ECG FMs? Finally, what is the effect of scale, in terms of the size of the pretraining datasets, on the performance of ECG FMs?

\paragraph{Contributions}
In this submission, we put forward the following technical contributions, which we also summarize in Figure~\ref{fig:abstract}: %\colnst{check again if these are really our final contributions}
\begin{enumerate}
\item We provide the most comprehensive pretraining study on ECG FMs to date, covering five different pretraining methodologies, trained on the largest publicly available pretraining corpus to date comprising over 11M samples, evaluated through a comprehensive evaluation methodology following \citep{al-masud2026benchmarking} in a first like-for-like comparison (see Sec.~\ref{sec:methods}).
\item We provide insights on model architecture comparisons CNN vs.~transformer vs.~state space models, confirming the latter as the superior architecture choice across all pretraining paradigms, in line with comparisons in a supervised setting (see ``Backbone analysis'' in Sec.~\ref{sec:results}).
\item We provide comparative insights on five pretraining methodologies popularized in the general self-supervised learning domain, or for the domain of time series/audio data (see Fig.~\ref{fig:radar_plot}). Pretraining strategy has a meaningful and consistent impact on downstream performance, with CPC showing the strongest and most transferable representations across diverse clinical tasks, while data2vec consistently lags behind across all evaluation modes and scaling regimes.
\item We provide a scaling analysis across all five pretraining methodologies and identify scaling behavior, most clearly for CPC and JEPA. Furthermore, we provide evidence that lower pretraining loss correlates with small residual errors in downstream tasks (see Fig.~\ref{fig:scaling_analysis}).
\item We complement the analysis with a representational similarity analysis, which provides additional hints for the superiority of the CPC approach, label/model efficiency analyses and improved finetuning procedures.
\end{enumerate}

\section{Background}
\paragraph{ECG FMs}
ECG FMs have proliferated rapidly, spanning masked prediction~\citep{na2024guiding, coppola2024hubert}, joint-embedding architectures~\citep{kim2024learning}, and contrastive and multimodal learning~\citep{liu2024zero, yu2024ecg, mckeen2025ecg}, with recent models pretrained on tens of millions of recordings~\citep{li2025electrocardiogram, coppola2024hubert}. The SSL objectives underlying these models are largely inherited from speech (CPC~\citep{oord2018representation}, HuBERT~\citep{hsu2021hubert}) and vision (I-JEPA~\citep{assran2023self}) adapted to physiological time series without systematic evaluation of which objectives or backbone architectures are best suited for ECG data. Backbone choices have similarly followed trends from other domains, with transformers~\citep{vaswani2017attention} dominating despite structured state space models~\citep{gu2022efficiently} showing superior performance on long sequences in supervised ECG settings~\citep{mehari2023towards,strodthoff2024prospects}.

\paragraph{Scaling laws}
Scaling laws relating model performance to pretraining dataset size have been studied extensively in language~\citep{hoffmann2022training} and vision~\citep{nezhurina2026scaling}, typically revealing power-law improvements with increasing data. In the ECG domain, however, whether large pretraining corpora are necessary for strong FM performance or whether carefully designed objectives and architectures can compensate for limited data remains poorly understood, with direct practical implications given the cost of curating large-scale medical datasets.

\paragraph{ECG FM benchmarking}
Benchmarking of ECG FMs has gained increasing attention recently~\citep{lunelli2025benchecg, al-masud2026benchmarking, wan2025openecgbenchmarkingecgfoundation}, establishing standardized evaluation protocols and highlighting the importance of strong supervised baselines. However, existing benchmarks evaluate FMs that differ simultaneously in pretraining objective, backbone architecture, and dataset scale, making it impossible to attribute performance differences to any single factor. This submission directly addresses this gap by conducting the first comprehensive, like-for-like comparison of pretraining objectives and scaling behavior for ECG FMs, holding architecture and training configuration constant across all conditions.

\section{Methods}
\label{sec:methods}
\paragraph{Backbone architectures}
All models share a common encoder comprising a lightweight CNN stem followed by a sequential backbone, with the encoder fixed across all pretraining objectives to enable controlled comparison. The CNN stem consists of four convolutional layers with batch normalization. For the backbone, we evaluate three variants: a S4-based backbone~\citep{gu2022efficiently}, a Transformer~\citep{vaswani2017attention} backbone with RoPE positional encoding~\citep{su2024roformer} and GELU activations~\citep{hendrycks2016gaussian}, and a CNN-based model (Net1D~\citep{li2025electrocardiogram}). We further investigate the effect of S4 backbone depth by comparing 4-layer and 6-layer configurations, and conduct a supervised model dimension ablation across dimensions 512, 768, and 1024 with corresponding state dimensions 8, 12, and 16 to determine the optimal capacity for ECG representation learning. All models operate at 240 Hz on 12-lead ECG inputs. Based on our ablation studies, the S4 backbone with model dimension 512 consistently outperforms larger and alternative configurations, and we therefore adopt the 4-layer S4 with dimension 512 as our default backbone.

\paragraph{Self-supervised pretraining objectives}
We investigate five self-supervised pretraining objectives spanning contrastive, predictive, and clustering-based paradigms within a unified architecture and training framework, enabling controlled comparison of their effectiveness for ECG representation learning. \textbf{data2vec}~\citep{baevski2022data2vec} trains the model to predict the EMA teacher's averaged top-$k$ contextualized layer representations at masked positions, providing continuous rather than discrete supervision, with the teacher averaging the outputs of the top 2 S4 layers. \textbf{DinoSR}~\citep{liu2023dinosr} trains a student encoder to predict EMA teacher-derived discrete cluster assignments at masked positions, combining online $k$-means quantization with a masked prediction objective using two codebooks of sizes 128 and 256. \textbf{JEPA}~\citep{assran2023self} trains a context encoder to predict the EMA teacher's latent representations of masked regions entirely in embedding space via a multi-block masking strategy, avoiding low-level reconstruction. \textbf{CPC}~\citep{oord2018representation} learns representations by predicting future latent states from causal context via a contrastive objective, discriminating true future steps from within-sequence negatives over 5-second input segments. \textbf{HuBERT++} extends the speech HuBERT~\citep{hsu2021hubert} approach by replacing offline hard $k$-means targets with EMA teacher-derived soft cluster assignments computed via Sinkhorn-Knopp optimal transport~\citep{cuturi2013sinkhorn}, encouraging balanced token participation and preventing representation collapse through online prototype updates, see App.~\ref{app:hubertpp} for details. We additionally evaluate \textbf{ECGFounder}~\citep{li2025electrocardiogram} and \textbf{ECG-JEPA}~\citep{kim2024learning}, two of the best-performing external FMs in \citep{al-masud2026benchmarking}, as external FM reference points, and a supervised S4 model trained from scratch as a strong task-specific baseline.
The overall best-performing ECG-CPC model from \citep{al-masud2026benchmarking} is omitted for comparison as the architecture and training setup largely coincides with CPC model studied in this work, see Appendix~\ref{app:cpc_vs_external_cpc} for a detailed comparison.

\paragraph{Pretraining datasets}
We pretrain all models on HEEDB, HEEDB-Emory~\citep{koscova2025bdsp, koscova2026harvard}, and CODE-15\%~\citep{ribeiro_2021_4916206} at 240 Hz, spanning three scales: a small HEEDB subset (106K), a medium HEEDB subset (753K), and the full combined dataset (11M samples). We additionally compare HEEDB against MIMIC-IV-ECG~\citep{PhysioNet-mimic-iv-ecg-1.0} as an alternative pretraining source at matched scale i.e., 753K vs.~759K samples (Appendix~\ref{app:heedb_mimic_240_comparison}) with largely similar downstream performance. 

\paragraph{Downstream datasets and tasks} We follow the evaluation protocol from \citet{al-masud2026benchmarking}, which covers 26 clinically relevant tasks using 10 public datasets comprising 1,622 regression and classification targets. Deviating from \citep{al-masud2026benchmarking}, we removed two evaluation datasets: CODE-15\% as it was used for pretraining in this work and PTB due to its small size. The tasks were assigned to 7 task categories to simplify the interpretation of the results. All datasets and underlying licences used in this study are listed in Appendix~\ref{app:dataset_details}. 

\paragraph{Evaluation}
Pretrained encoders are augmented with a linear prediction head and finetuned on each downstream task using AdamW with layer-dependent learning rates, binary cross-entropy for classification, and MAE for regression with z-normalized targets. For linear evaluation, the encoder is kept frozen and only the linear prediction head is trained. For frozen evaluation, the linear head is replaced with a learnable query-attention head~\citep{bardes2024vjepa}. We report macro-averaged AUROC for classification and MAE for regression, with statistical significance assessed via pairwise empirical bootstrapping on the test set. Rankings account for confidence interval overlaps, with ties indicating no statistically significant difference. (Macro-)AUROC serves as primary classification metric as most widely used metric to characterize the overall discrimative power of a model, even in the presence of label imbalance \citep{mcdermott2024closer}. Full experimental details are provided in Appendix~\ref{app:experimental_setup}.

\section{Results}\label{sec:results}
Comprehensive evaluation results across finetuning, frozen, and linear evaluation modes are compiled in Tables~\ref{tab:finetuning_result}, \ref{tab:frozen_result}, and \ref{tab:linear_result} in the appendix, where boldface entries denote models not performing statistically significantly worse than the best method per task. Statistical rankings aggregated by task category are visualized in Figure~\ref{fig:radar_plot}, with per-task ranked lists provided in Table~\ref{tab:detailed_ranking_table}. Additional results including backbone ablations, label efficiency curves, and pretraining dataset size scaling analyses are provided in Appendices~\ref{app:s4_transformer_predictor_ablation}, \ref{app:echonext_label_efficiency}, and \ref{app:scaling_analysis}.

\paragraph{Backbone analysis: performance}
Across all five pretraining objectives, the 4-layer S4 backbone consistently outperforms both the Transformer and Net1D backbones on the large majority of downstream tasks and performs on par on the remaining tasks (Appendix~\ref{app:s4_transformer_predictor_ablation}). The performance gap over the Transformer is most pronounced for JEPA, CPC, and HuBERT++, and is particularly evident on challenging tasks such as pediatric ECG interpretation and cardiac structure prediction. Among CNN-based backbones, Net1D performs competitively under data2vec but falls consistently below S4 for all other pretraining objectives. This is further corroborated by the underperformance of the CNN-based FMs MERL~\citep{liu2024zero} and ECGFM-KED~\citep{tian2024foundation} relative to our S4-based models (Appendix~\ref{app:mimic_iv_ecg_finetuning}), consistent with recent evidence questioning the suitability of CNNs as backbone architectures for physiological time series in supervised settings \citep{mehari2023towards,strodthoff2024prospects,al-masud2026benchmarking}.

\begin{figure}[htbp]
    \centering
    \begin{subfigure}{0.32\textwidth}
        \centering
        \includegraphics[width=\textwidth]{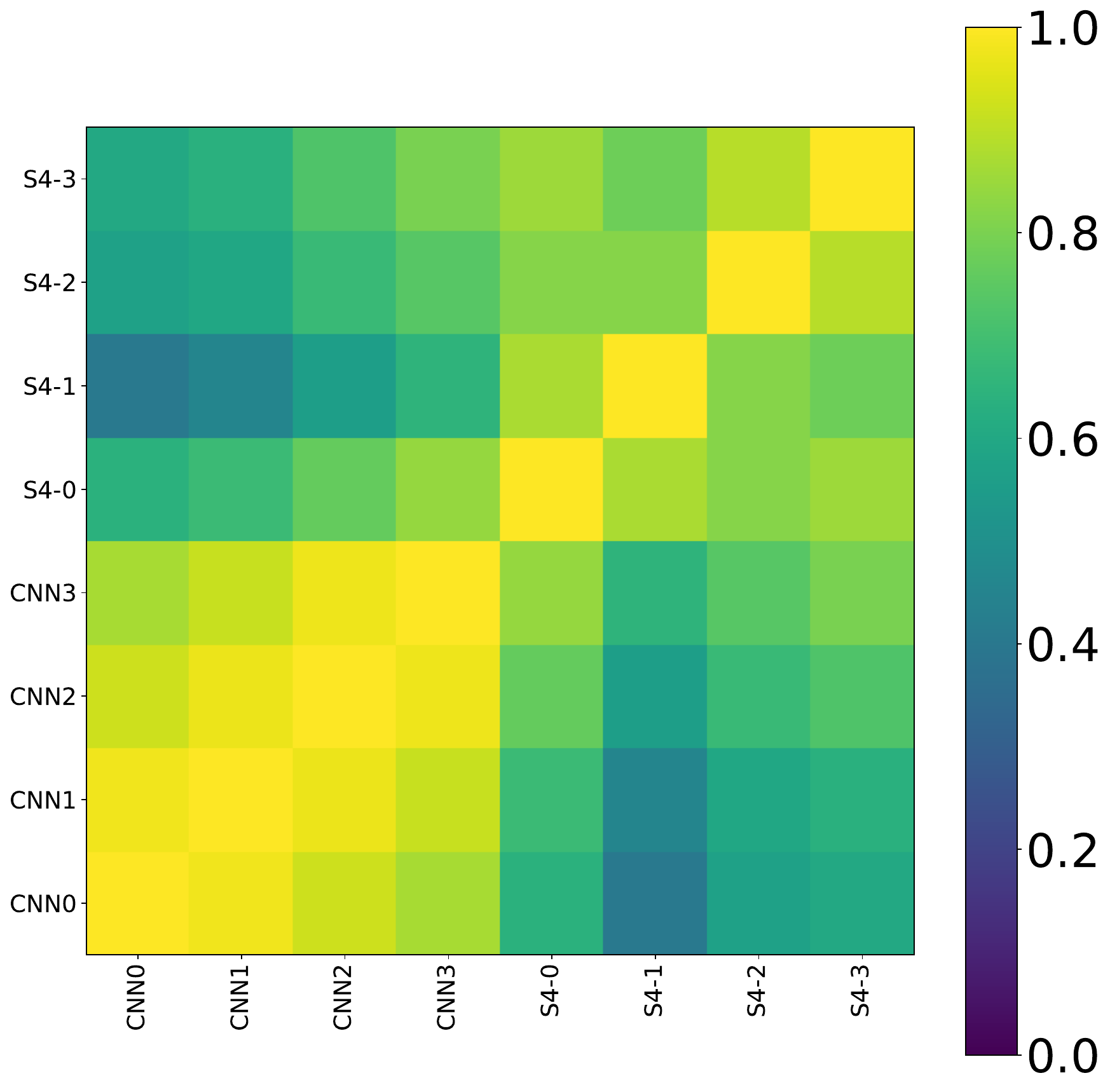}
        \caption{S4}
    \end{subfigure}
    \hfill
    \begin{subfigure}{0.32\textwidth}
        \centering
        \includegraphics[width=\textwidth]{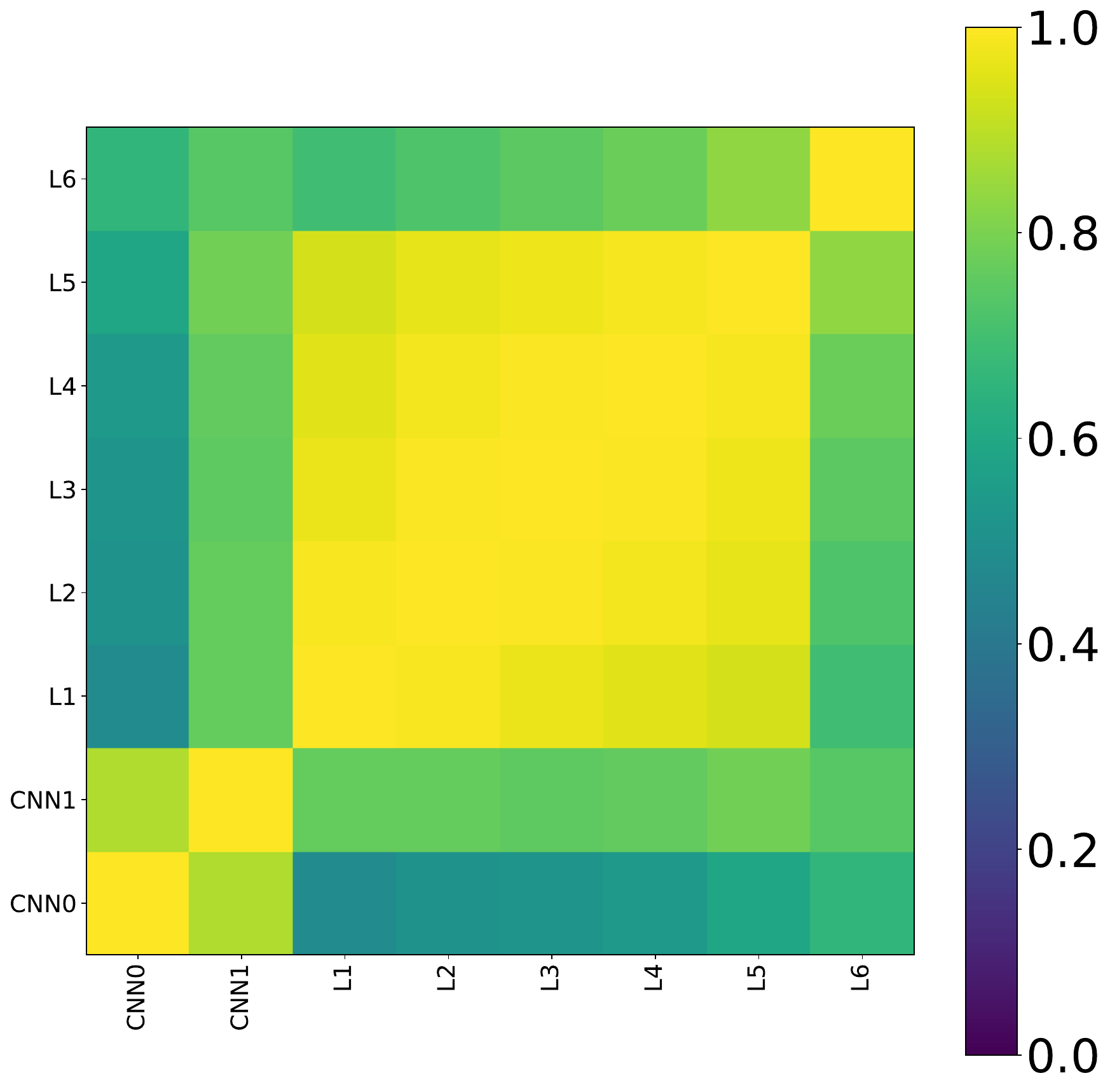}
        \caption{Transformer}
    \end{subfigure}
    \hfill
    \begin{subfigure}{0.32\textwidth}
        \centering
        \includegraphics[width=\textwidth]{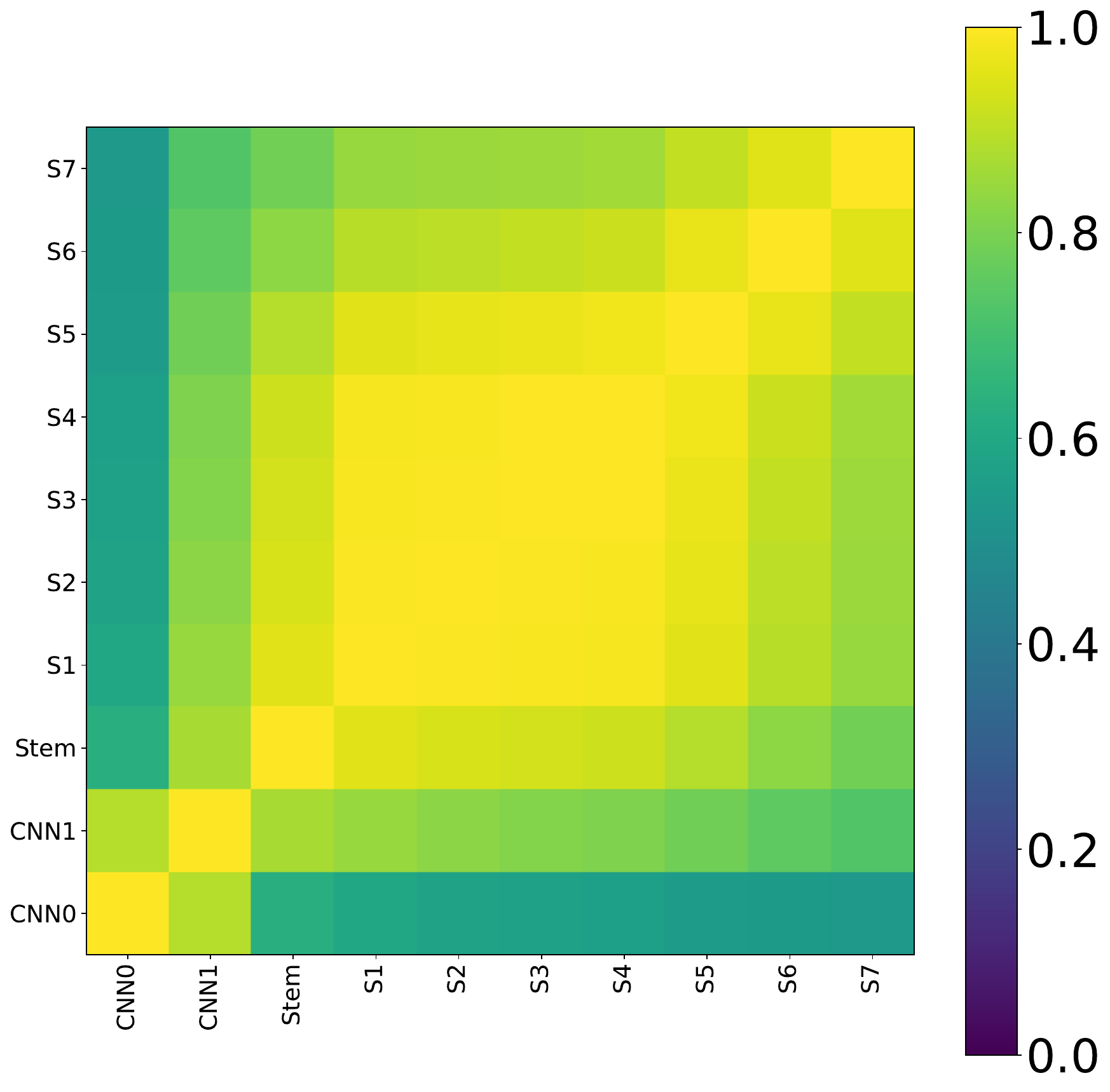}
        \caption{Net1D}
    \end{subfigure}
    \caption{Intra-model layer-wise representational similarity for JEPA}
    \label{fig:main_jepa_cka_s4_transformer_net1d}
\end{figure}

\paragraph{Backbone analysis: representational similarity}
The intra-model CKA analysis (Appendix~\ref{app:backbone_cka_analysis}) provides further insight into the representational structure learned by each backbone. For the S4 backbone, layers develop progressively distinct representations across all pretraining objectives, indicating that each layer captures complementary patterns, a desirable property for hierarchical feature learning. In contrast, Transformer backbones exhibit high inter-layer similarity across intermediate blocks (L1--L5) for most pretraining objectives, with only the final block (L6) showing moderately distinct representations under JEPA and CPC. HuBERT++ with a Transformer backbone shows virtually no layer differentiation across the entire network, suggesting representational collapse. Net1D backbones display uniform similarity in both early CNN layers and later stages across all objectives, pointing to a possible representational bottleneck. The early CNN stem layers show high mutual similarity across all backbone types and pretraining objectives, which is expected given their shared architecture. Overall, the S4 backbone is the only architecture that consistently develops distinct layer-wise representations, with CPC exhibiting the most pronounced differentiation, further supporting S4 as the preferred backbone for ECG FM pretraining.

\begin{figure}[b]
    \centering
    \includegraphics[width=\linewidth]{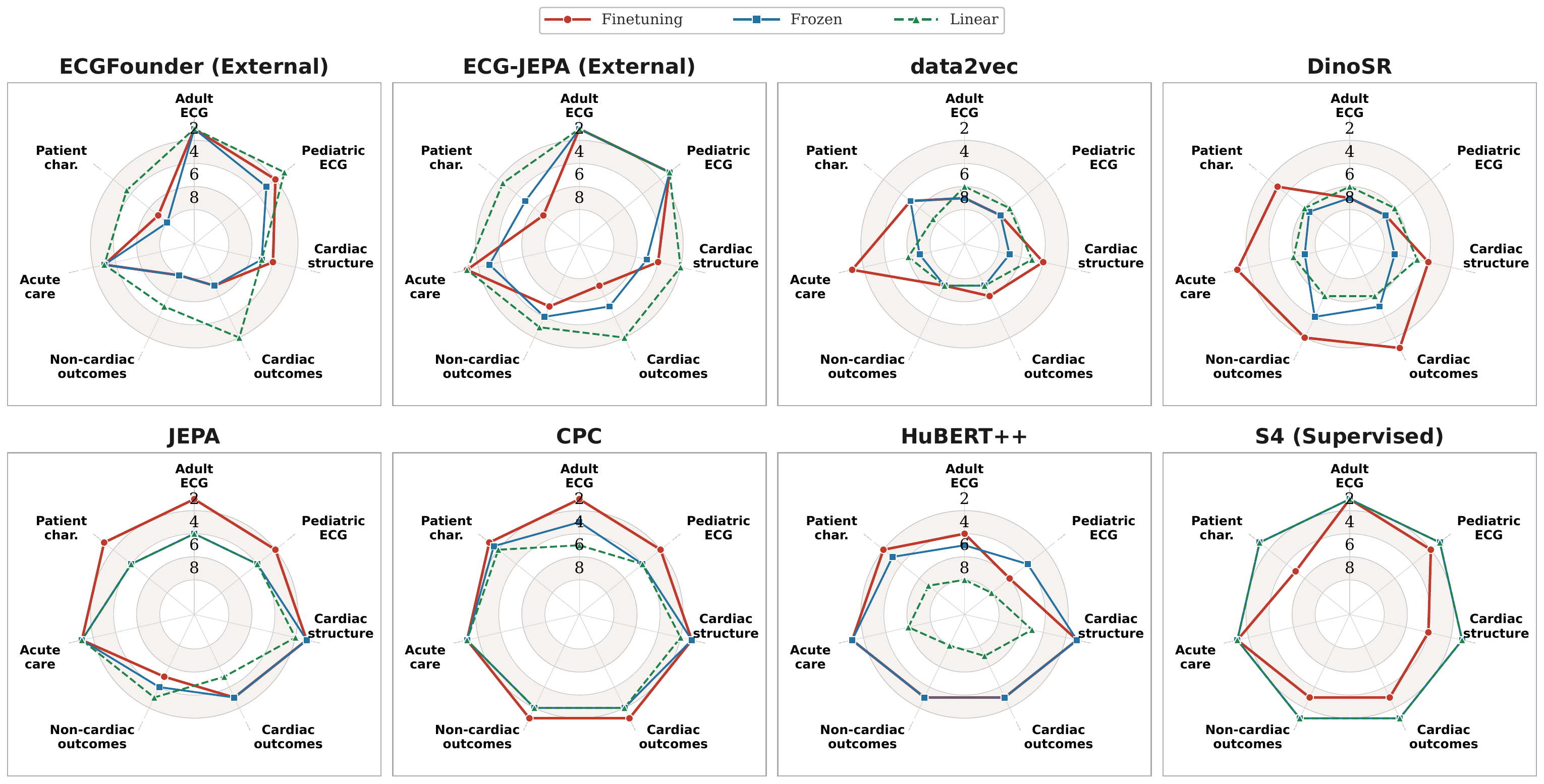}
    \caption{Visualization of performance rankings across seven downstream task categories for the five self-supervised pretraining objectives, where lower rank denotes statistically superior performance.}
    \label{fig:radar_plot}
\end{figure}

\paragraph{Finetuning} Under full finetuning (Appendix~\ref{app:finetuning_full_dataset}), the external FMs ECG-JEPA and ECGFounder consistently rank among the top performers for adult ECG interpretation, frequently matching or surpassing the supervised S4 baseline. ECGFounder was pretrained using a supervised loss on diagnostic statement prediction, which aligns very closely with the adult ECG interpretation task. At this point, we stress that task-specific finetuning or alignment of ECG data with other modalities could still be applied on top of the FMs trained using self-supervision only but is beyond the scope of this submission. ECG-JEPA profits from the fact that its pretraining datasets serve as test datasets in this study. As shown below, FMs profit from continual pretraining on the target dataset before finetuning, see below, which explains performance differences in ECG-JEPAs pretraining datasets in the category of adult ECG interpretation. Among models pretrained for this submission, CPC and JEPA are the strongest competitors, often falling within the top statistical equivalence group across ECG interpretation and outcome prediction tasks. data2vec lag considerably behind across most categories, while HuBERT++ and DinoSR occupy a middle ground. In categories beyond ECG interpretation, covering cardiac structure, clinical outcomes, and patient characteristics, CPC tends to lead among all evaluated approaches, suggesting its pretraining objective yields more transferable representations across diverse clinical tasks.

\paragraph{Frozen and linear evaluation} Rankings under frozen evaluation (Appendix~\ref{app:frozen_full_dataset}) largely mirror finetuning, with ECG-JEPA and ECGFounder retaining strong performance on adult ECG tasks, achieving supervised-level results as frozen feature extractors. CPC maintains its advantage in non-ECG-interpretation categories under frozen evaluation, while data2vec and DinoSR fall further behind relative to finetuning, indicating their representations require task-specific adaptation to be competitive. Under linear evaluation (Appendix~\ref{app:linear_full_dataset}), the supervised S4 baseline proves difficult to surpass across most categories, with CPC showing a notable relative drop compared to its frozen evaluation performance, consistent with its token-level pretraining objective not encouraging globally discriminative pooled representations.

\begin{figure}[tbp]
    \centering
    \begin{subfigure}{0.48\textwidth}
        \centering
        \includegraphics[width=\textwidth]{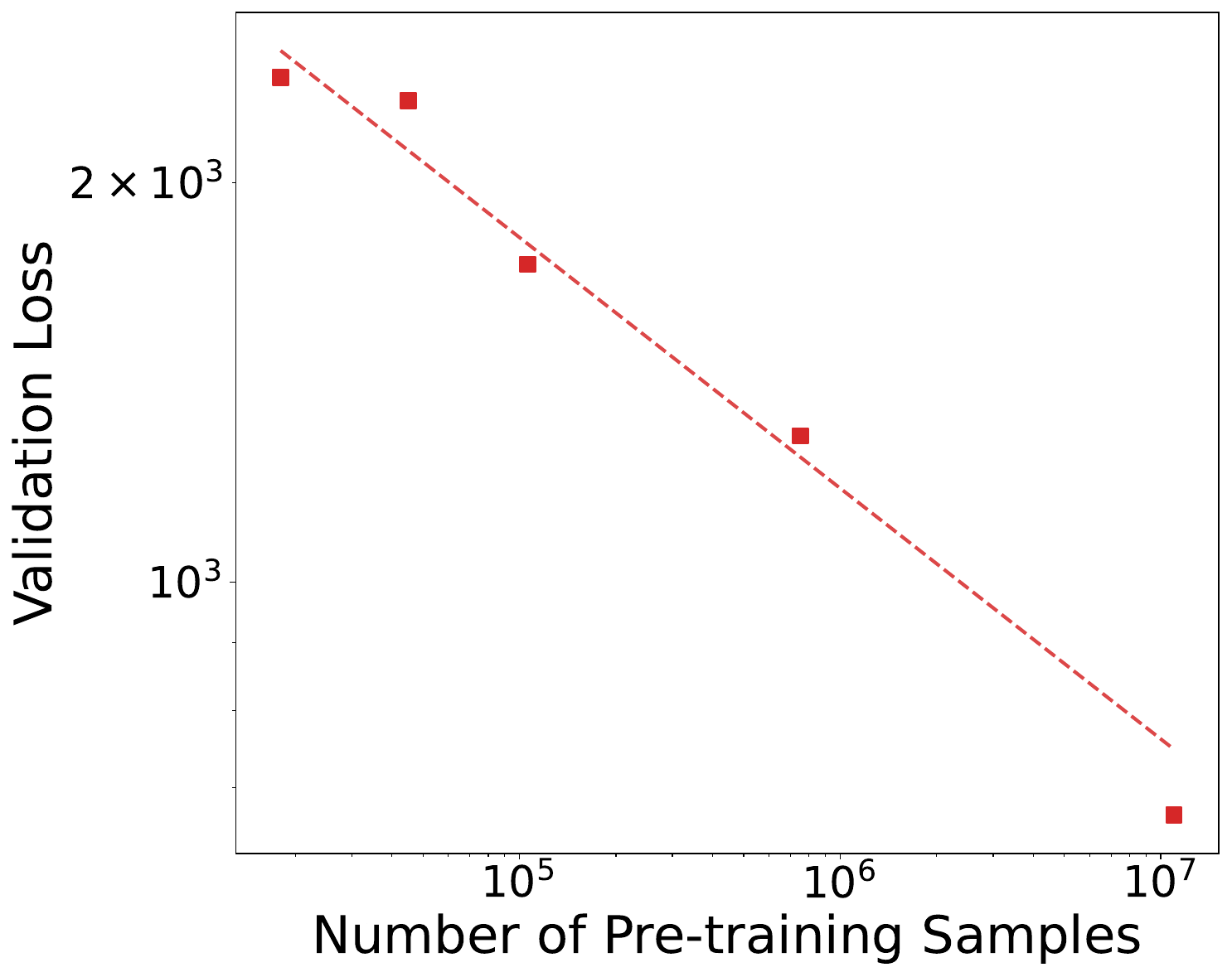}
        \caption{Loss scaling}
        \label{fig:scaling_loss}
    \end{subfigure}
    \quad
    \begin{subfigure}{0.48\textwidth}
        \centering
        \includegraphics[width=\textwidth]{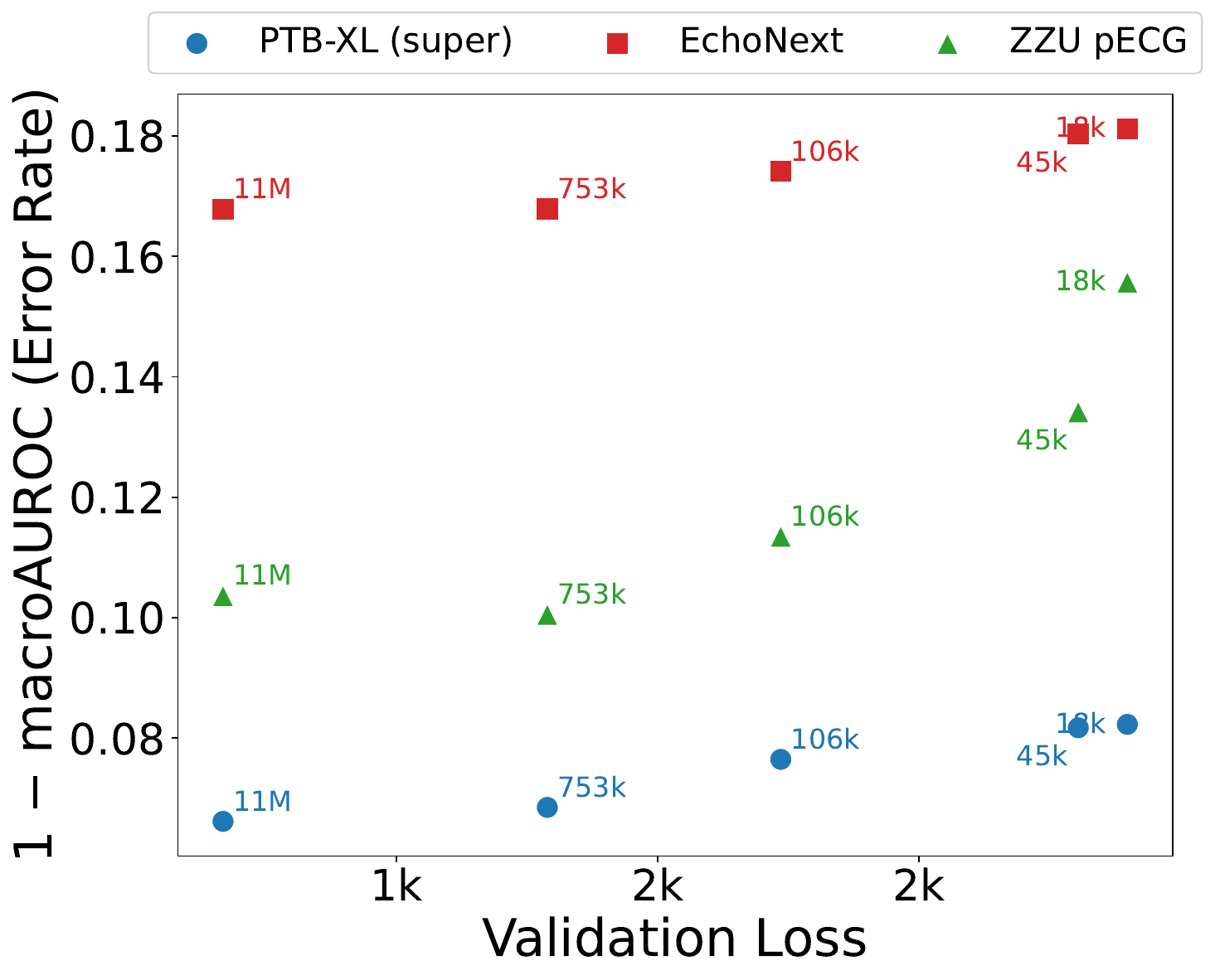}
        \caption{Correlation Loss/error rate}
        \label{fig:scaling_ptbxl_all}
    \end{subfigure}
    \caption{Scaling Analyses for the CPC model investigating the scaling of the pretraining loss with dataset size and the relation between pretraining loss and downstream performance on selected datasets (representative for adult ECG interpretation, pediatric ECG interpretation and cardiac structure evaluation), see Appendix~\ref{app:scaling_analysis} for further results.}
    \label{fig:scaling_analysis}
\end{figure}

\paragraph{Scaling with dataset size}
To investigate the scaling behavior we train models on four subsets of the HEEDB dataset: 18K, 45K, 106K, and 753K, in addition to the full dataset comprising 11M samples, see Appendix~\ref{app:experimental_setup} for the specific dataset composition. 
For each of the pretraining methods, we first consider the conventional loss scaling with dataset size, see Figure~\ref{fig:scaling_loss} and Appendix~\ref{app:val_loss_scaling}. The two best-performing methods, CPC and JEPA, show the most consistent power-law scaling according to fit quality with scaling exponents 0.189 and 0.062, respectively. As a lower pretraining loss not necessarily has to align with a better downstream performance, we also investigate the scaling behavior of the residual error defined as $1-\text{macro AUROC}$ for a given specific downstream dataset/task. We also show scaling curves for the downstream performance in Appendix~\ref{app:scaling_auroc}, the most consistent scaling behavior on PTB-XL (super) and mixed results across different other downstream datasets. Finally, we directly compare the pretraining loss to the residual error on three representative datasets, see Figure~\ref{fig:scaling_ptbxl_all} and Appendix~\ref{app:correlation_plot}. In all cases, but particularly clearly for CPC, JEPA and DinoSR, we see a significant positive correlation of the validation loss with the downstream residual errors across three datasets covering three major task categories.

\begin{figure}[tbp]
    \centering
    \begin{subfigure}[b]{0.23\linewidth}
        \includegraphics[width=\linewidth, height=\linewidth]{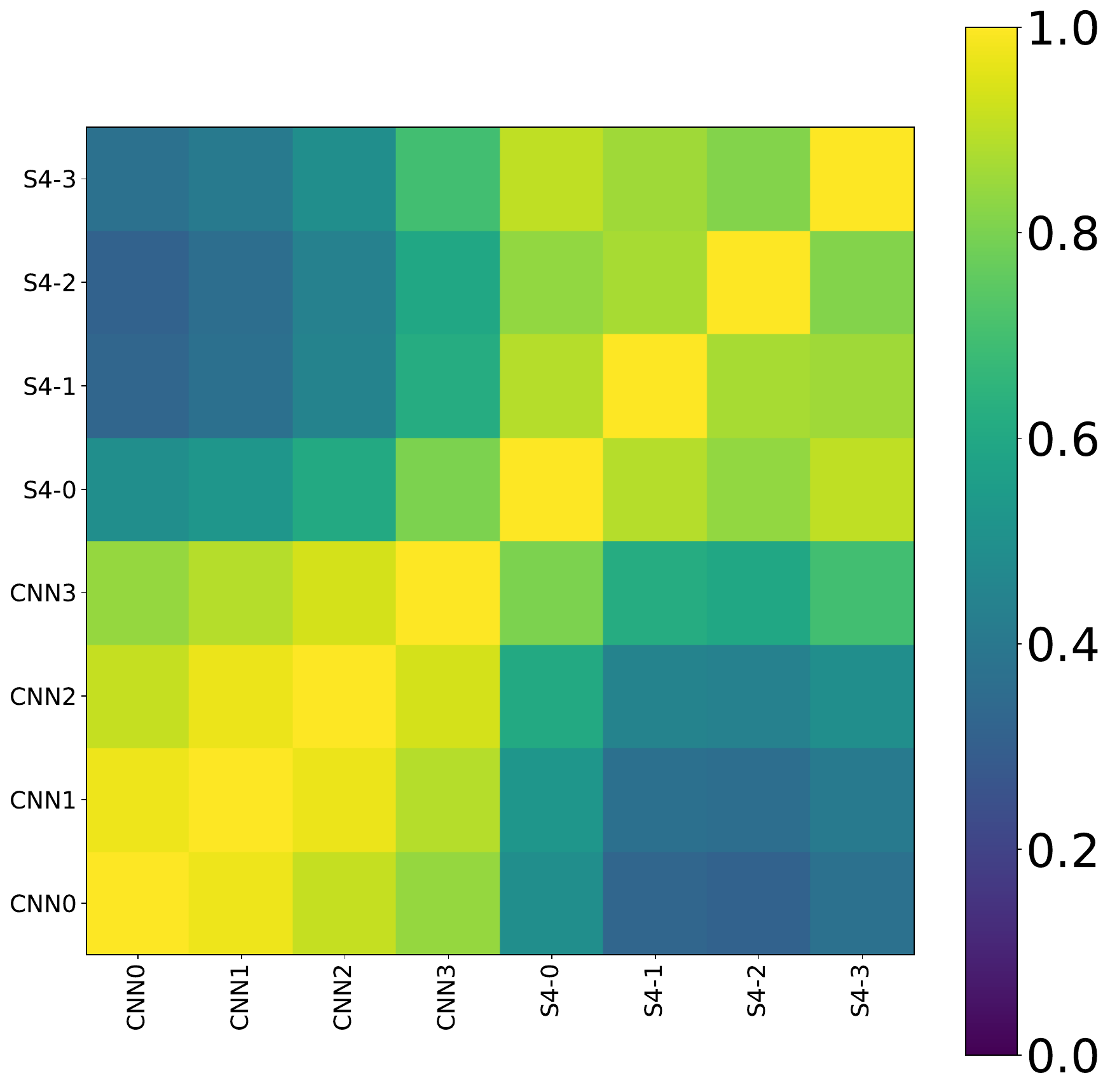}
        \caption{DinoSR}
    \end{subfigure}
    \hfill
    \begin{subfigure}[b]{0.23\linewidth}
        \includegraphics[width=\linewidth, height=\linewidth]{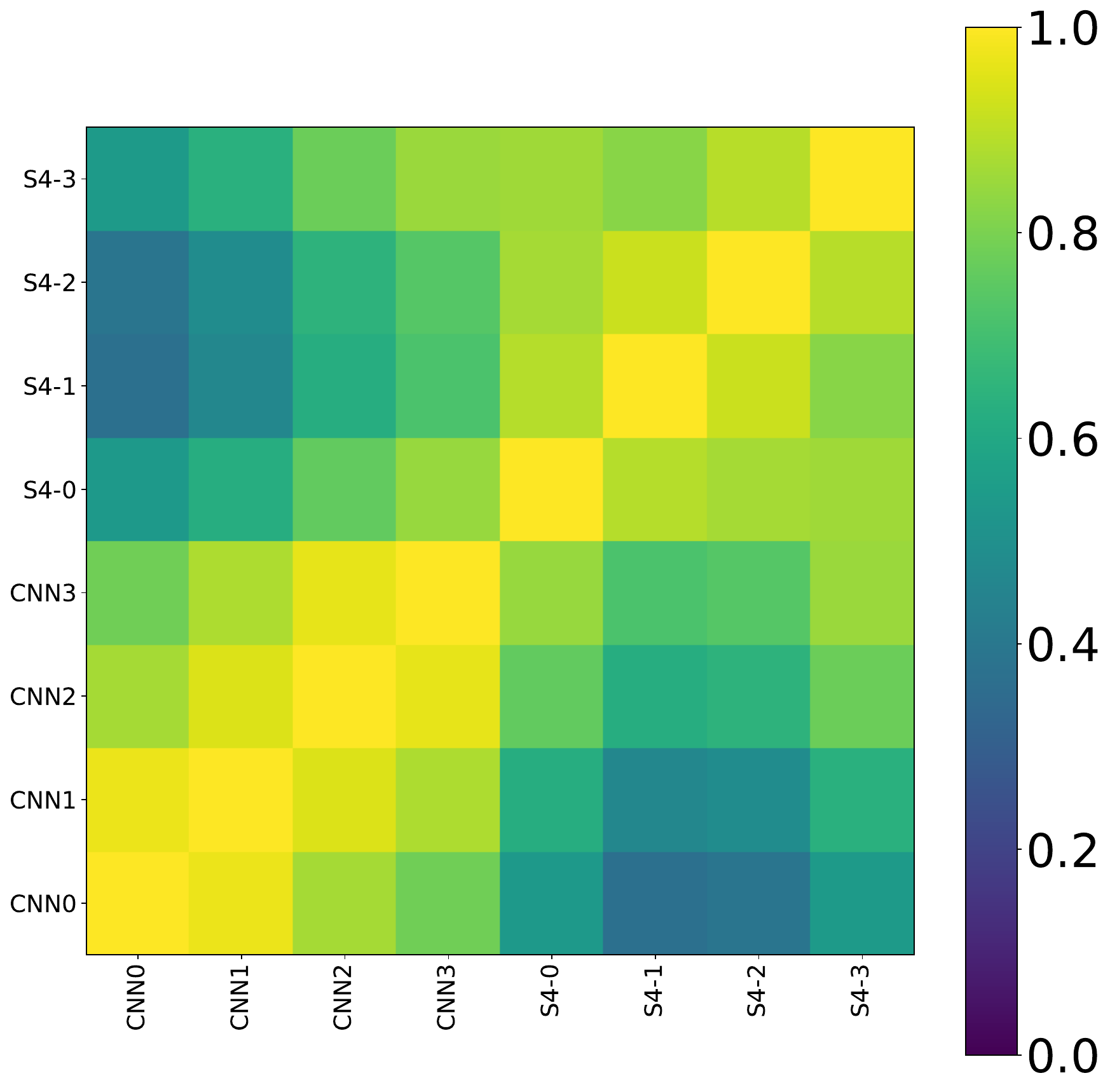}
        \caption{JEPA}
    \end{subfigure}
    \hfill
    \begin{subfigure}[b]{0.23\linewidth}
        \includegraphics[width=\linewidth, height=\linewidth]{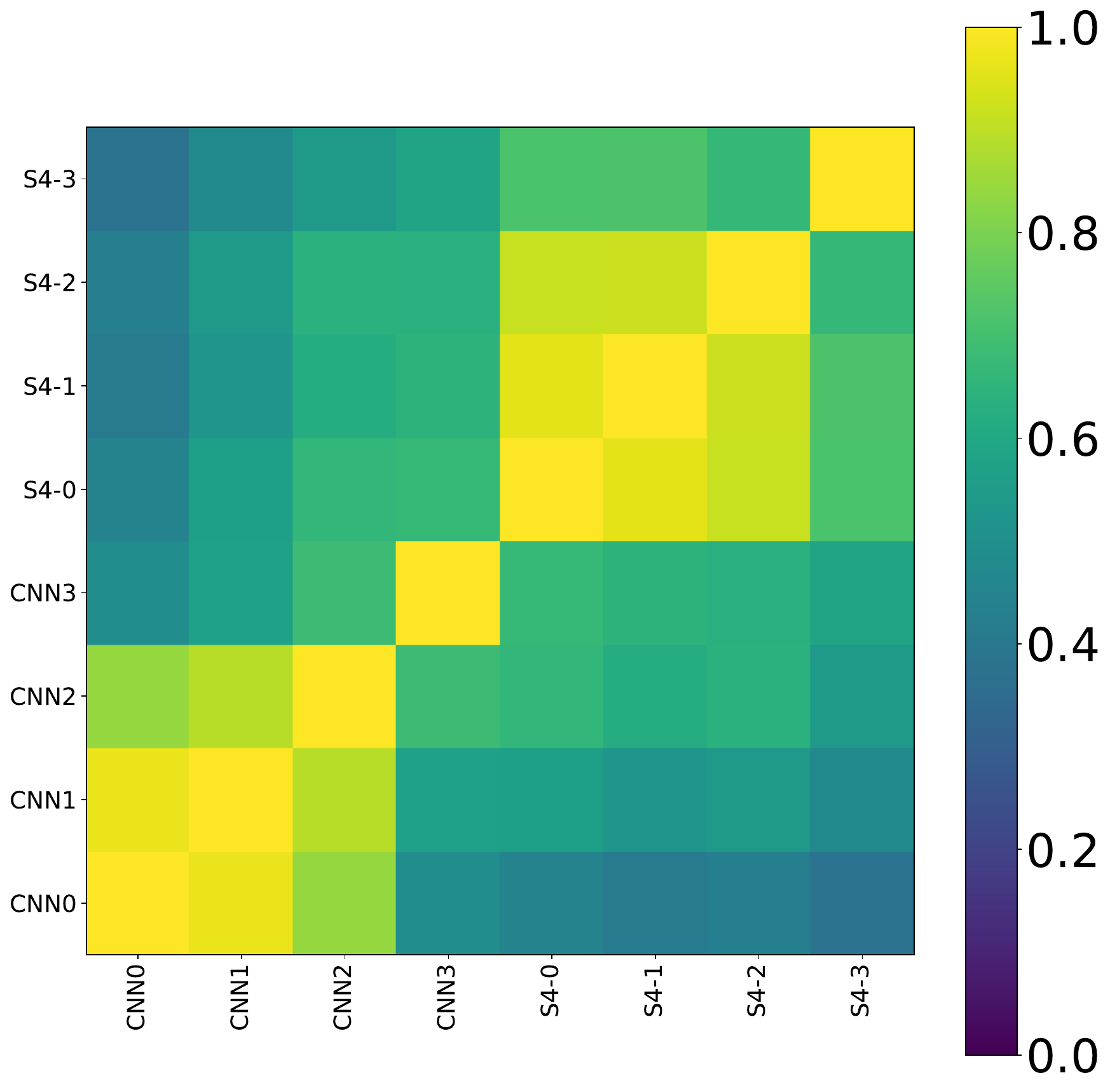}
        \caption{CPC}
    \end{subfigure}
    \hfill
    \begin{subfigure}[b]{0.23\linewidth}
        \includegraphics[width=\linewidth, height=\linewidth]{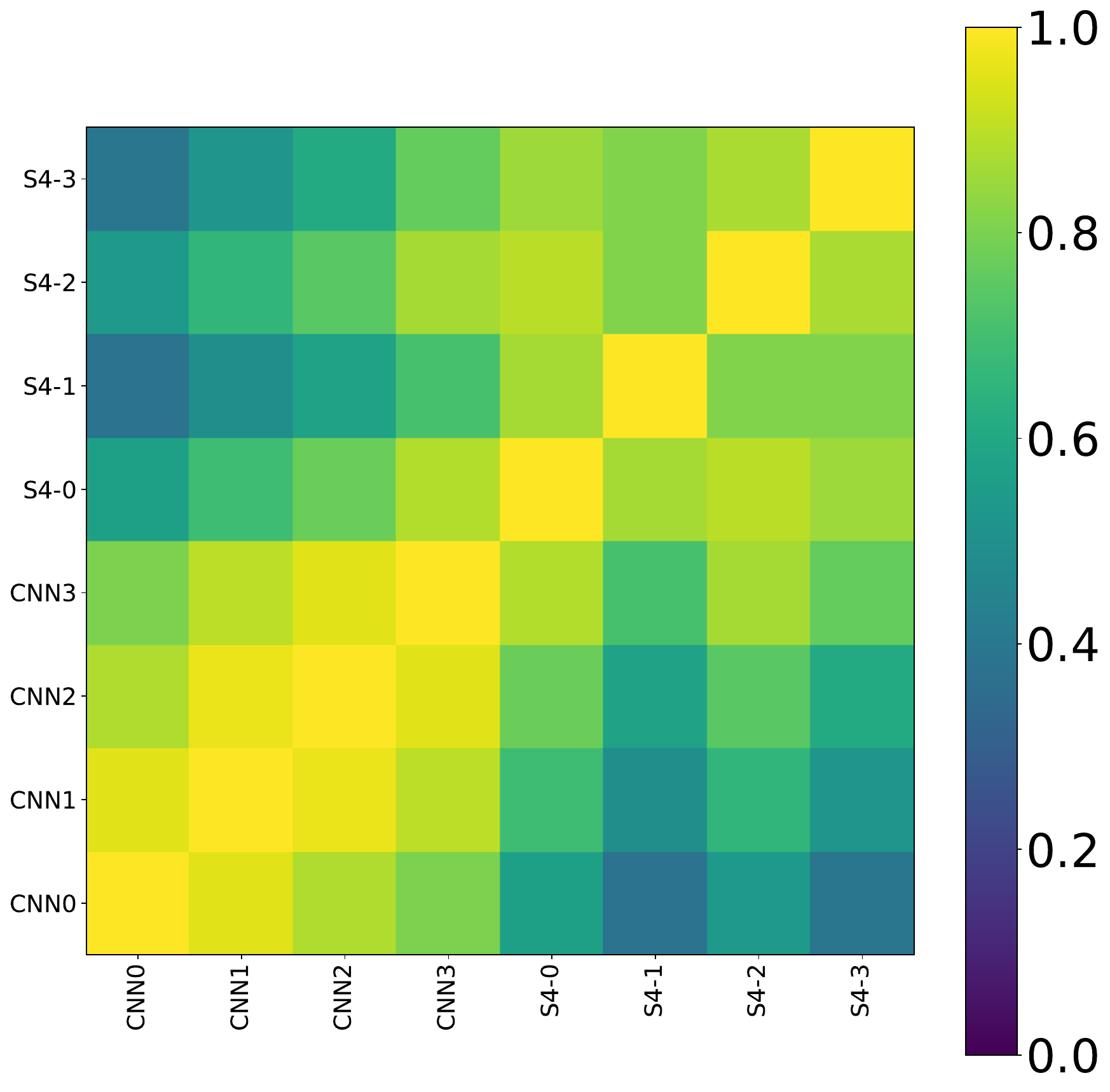}
        \caption{HuBERT++}
    \end{subfigure}
    \caption{Layer-wise representation similarity within each pretraining objective, measured by CKA with a Gaussian RBF kernel ($\sigma=1.0$) on 2,500 PTB-XL samples. Warmer colors indicate greater similarity between layer pairs. The CKA matrix for data2vec was omitted due to place constraints and is shown in Appendix~\ref{app:intra_and_inter_model_rep_sim}.}
    \label{fig:cka}
\end{figure}

\paragraph{Representation similarity analysis}
Intra-model CKA heatmaps for four pretraining objectives are shown in Figure~\ref{fig:cka}. All models exhibit a two-block structure reflecting the hybrid CNN-S4 architecture, with high within-block similarity and lower cross-block similarity between CNN and S4 layers. However, the sharpness of this boundary and the degree of within-block differentiation differ considerably across pretraining strategies. DinoSR shows the sharpest CNN-to-S4 transition with the most distinct separation between the two architectural components, while within each block representations remain relatively homogeneous. JEPA and HuBERT++ exhibit the highest overall representational redundancy, with a notably smoother CNN-to-S4 boundary, suggesting these pretraining objectives do not strongly encourage functional specialization between the two components. CPC stands out with the lowest overall off-diagonal similarity and the most progressive within-block differentiation, where each layer learns increasingly distinct representations forming a clear gradient from local CNN features to higher-level sequential S4 abstractions. This structured representational evolution, largely absent in other pretraining strategies, may underpin CPC's stronger and more consistent downstream performance across diverse clinical tasks.
Inter-model CKA analysis across all five pretraining objectives at early, mid, and late network stages is provided in Appendix~\ref{app:inter_model_cka_analysis}, revealing the degree to which different pretraining strategies converge to or diverge from similar representational spaces across network depth.

\begin{wrapfigure}{r}{0.48\linewidth}
        \centering
        \includegraphics[width=\linewidth]{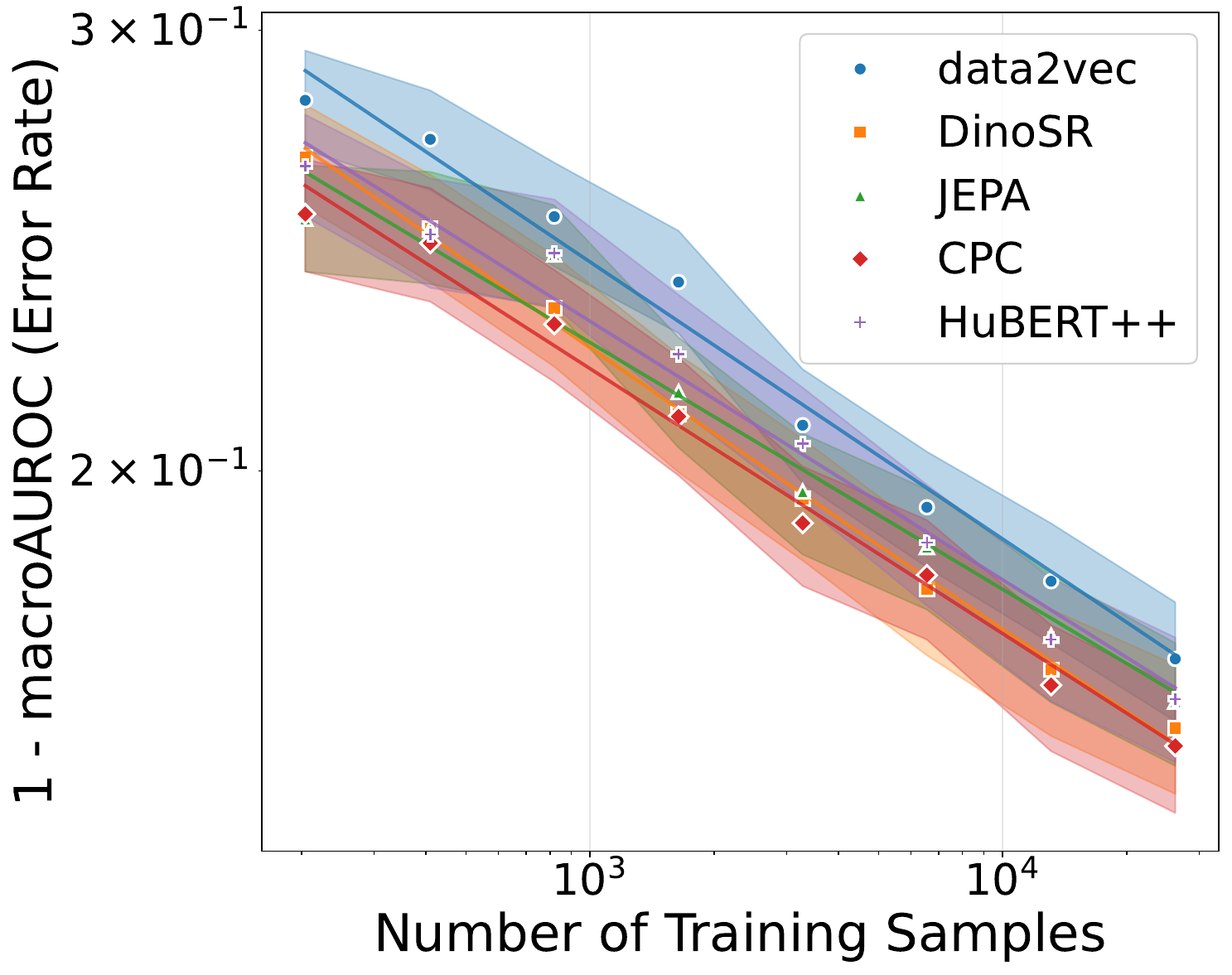}
        \caption{EchoNext label efficiency plot tracing downstream performance in dependence of the number of training samples.}
        \label{fig:echonext_label_efficiency}
\end{wrapfigure}
% \vspace{-10pt}
    
\paragraph{Label efficiency}
In Figure~\ref{fig:echonext_label_efficiency}, we investigate a different scaling behavior, namely the dependence of the downstream performance on the size of the downstream dataset for a given pretrained FM (on the full pretraining dataset), to detect different adaptation behaviors in the small dataset regime. Scaling curves for cardiac structure prediction on EchoNext are shown in Figure~\ref{fig:echonext_label_efficiency}, with fit parameters provided in Appendix~\ref{app:echonext_label_efficiency}. Training subsets are subsampled in powers of 2 and we fit scaling curves as above. CPC achieves the lowest error rate consistently across the entire range of training set sizes, establishing it as the most label-efficient model for this task. DinoSR starts with a relatively high error rate in the low-data regime but improves steeply with additional data, eventually surpassing JEPA and approaching CPC at the large-data end, consistent with its highest scaling exponent among all evaluated models. JEPA, by contrast, starts favorably in the low-data regime but scales more slowly, eventually converging toward HuBERT++ at larger dataset sizes. Data2vec lags behind all other models throughout the entire scaling range, indicating that its pretraining objective transfers poorly to cardiac structure prediction regardless of the amount of available labeled data. Overall, the results highlight that the choice of pretraining strategy has a pronounced effect on label efficiency, with differences being most consequential in the low-data regime.

\paragraph{Model efficiency}
Detailed computational efficiency metrics are provided in Appendix~\ref{app:computation_efficiency_analysis}. All five pretraining strategies share the same lightweight SSM backbone with only 3M parameters, operating on 600 timesteps, making them substantially more compact than the reference FMs ECGFounder (33.8M) and ECG-JEPA (87.2M). Consequently, all proposed models are considerably more efficient in terms of FLOPS, GPU memory, and inference latency than ECG-JEPA, and remain competitive with ECGFounder despite the latter's CNN-based architectural advantage in raw throughput. Across the five pretraining strategies, computational cost is virtually identical since the backbone is shared, meaning that performance differences observed in downstream tasks are attributable entirely to the pretraining objective rather than model capacity or computational budget. This makes our benchmark a controlled and fair comparison of pretraining strategies, and highlights that strong downstream performance can be achieved at minimal computational cost.

\paragraph{Continual pretraining for improved finetuning}
In line with positive results from other fields \citep{howard2018universal}, we propose to use continual pretraining on the target dataset as first finetuning phase before the actual classifier training. Domain-adaptive pretraining followed by a linear head consistently outperforms finetuning both with a standard linear and a non-linear query-attention head~\citep{bardes2024vjepa} across nearly all tasks (Appendix~\ref{app:finetuning_strategy_comparison}), in line with results of \citet{do2026domainadaptedfinetuningecgfoundation}, suggesting that closing the distribution gap between pretraining and target domain is more beneficial than increasing prediction head capacity. These findings advocate for domain-adaptive pretraining as a practical strategy when downstream domain data is available. For reasons of comparability, the finetuning results underlying Figure~\ref{fig:radar_plot} refer to standard finetuning, but we strongly advise to use continual pretraining for any downstream tasks. 

\paragraph{HuBERT-ECG vs. HuBERT++}
HuBERT++ consistently outperforms HuBERT-ECG~\citep{coppola2024hubert} across all evaluated tasks (Appendix~\ref{app:hubert_ecg_vs_hubert++}), with particularly pronounced improvements on adult ECG interpretation and cardiac structure prediction. We attribute this to the combination of Sinkhorn-Knopp cluster assignment preventing prototype collapse, EMA-based target generation providing more stable training signals, and the choice of S4 over Transformer as the backbone, demonstrating that careful refinement of existing objectives can yield meaningful gains without wholesale architectural innovation.

\section{Discussion}
\paragraph{SSMs as preferred backbone}
The consistent superiority of the S4 backbone over Transformer and CNN alternatives across all five pretraining objectives provides strong evidence that architectural inductive biases are the dominant factor in ECG representation learning. The performance gap is most pronounced for JEPA, CPC, and HuBERT++, and is particularly evident on challenging tasks such as pediatric ECG interpretation and cardiac structure prediction, suggesting that the stable long-range memory and spectral filtering properties of SSMs align naturally with the temporal structure of ECG signals. This is further corroborated by the underperformance of CNN-based external FMs relative to our lightweight S4-based models despite their substantially larger parameter counts, and is mechanistically supported by the CKA analysis showing that S4 is the only backbone that consistently develops progressive and distinct layer-wise representations across all pretraining objectives. These results suggest that backbone architecture should be treated as a first-order design decision in ECG FM development.

\paragraph{Pretraining strategies}
Pretraining strategies show meaningful and consistent differences in downstream performance. CPC consistently achieves the strongest results across most task categories and evaluation modes, with particular advantages in patient characterization and outcome prediction. JEPA occupies a strong second position, frequently matching CPC under finetuning but falling behind under frozen and linear evaluation. We hypothesize that the sequential nature of CPC's prediction task aligns better with the time series modality than arbitrary block masks used by other approaches. DinoSR and HuBERT++ occupy a middle ground in the performance, with selected strengths but also significant weakness in the core domain of adult and pediatric ECG interpretation. %DinoSR and HuBERT++ perform competitively on adult ECG interpretation but trail on more diverse clinical tasks, while 
Data2vec lags most consistently across all evaluation modes. Notably, performance gaps widen under frozen and linear evaluation, indicating that representation quality differs substantially across pretraining strategies beyond mere finetuning adaptability. Nevertheless, the consistent superiority of the S4 backbone across all five objectives suggests that architectural choice remains the dominant factor, with pretraining strategy playing an important but secondary role.

\paragraph{Pretraining configuration}
Careful hyperparameter configuration proves critical regardless of the chosen pretraining objective. A backbone model dimension of 512 consistently outperforms both smaller and larger alternatives, and a 4-layer S4 backbone outperforms its 6-layer counterpart across most tasks, indicating that neither capacity nor depth should be maximized indiscriminately. The masking configuration is particularly consequential for JEPA, where the original hyperparameters transfer well to ECG data and yield strong downstream performance.

\paragraph{Scaling}
The consistent scaling of the pretraining loss, most notably in the cases of the two best-performing methods CPC and JEPA, is a positive sign for the further development of ECG foundation models, and the first systematic investigation of this matter in this specific domain. Downstream residual errors correlates positively with pretraining loss across all models but again most clearly for CPC and JEPA, which represents the second hopeful signal for foundation model development in the field.

\paragraph{Limitations} This work is subject to a number of limitations. First, the submission focuses deliberately on foundation model pretraining through pure self-supervision, without leveraging (weak) supervision through text or other signals. Second, this work does not provided deeper insights into the learned representations apart from linear probing. Third, the poor performance of data2vec could potentially be improved through further hyperparameter optimization, which was not explored as thoroughly as for some of the competitors due to computational limitations.

\paragraph{Broader impact statement} This work provides guidance on essential design parameters of the pretraining of ECG FMs, which can inform future FM development and avoid suboptimal design decisions. The development of medical FMs is clearly beneficial to the society due to the advantages laid out in the introduction. We do not see any immediate negative consequences in terms of broader impact. However, we stress that the trained FMs are purely intended for research use and have not been validated for clinical application.

\section{Conclusion}
We present the first comprehensive, like-for-like study on pretraining strategies and scaling behavior for ECG FMs, covering five self-supervised objectives trained on a unified pretraining corpus of up to 11M samples and evaluated across a diverse set of clinical downstream tasks. Our results establish three key findings. First, backbone architecture is a dominant factor in ECG representation learning, with the S4-based SSM consistently outperforming Transformer and CNN alternatives across all pretraining objectives, a finding mechanistically supported by CKA analysis revealing superior layer-wise representational differentiation in S4 models. Second, pretraining strategy matters: CPC yields the most transferable representations across diverse clinical tasks, with performance gaps that widen under frozen and linear evaluation, while data2vec consistently underperforms regardless of scale. Third, we present the first analysis of scaling behavior in the domain of ECG data, most clearly seen for the two best-performing approaches CPC and JEPA, and provide evindence for correlation of pretraining and downstream performance.

Taken together, these findings provide actionable guidance for future ECG FM development, advocating for SSM-based backbones, carefully configured pretraining objectives, and domain-adaptive finetuning as the most impactful levers for improving downstream performance. LLMs were used solely for language refinement. Code and model weights are available at \url{https://anonymous.4open.science/r/ecg-pretraining-strategies-4DE3}.

\begin{ack}
This work was supported by the German Research Foundation (DFG) through the grant SELPHY-TS (project ID: 553038473).
\end{ack}

% \newpage
{
\small
    \bibliographystyle{unsrtnat}
    \bibliography{references}
}

\newpage
\appendix
{\LARGE\bfseries Appendices}
\vspace{1em}

\startcontents[appendices]
\printcontents[appendices]{}{1}{\setcounter{tocdepth}{2}}

\newpage
\section{S4 Model Dimension Ablation}
\label{app:s4_model_dimension_ablation}

\begin{table}[htbp]
    \centering
    \caption{Comparison of aggregated macro-AUROC for S4 trained from scratch in a supervised fashion with different model dimensions (512, 768, 1024) and corresponding state dimensions (8, 12, 16). Models were trained for 100 epochs using 6 S4 layers and learning rate 1e-3. The best-performing result is highlighted in boldface and underlined, while models that do not perform statistically significantly worse are also highlighted in boldface.\\}
    \label{tab:s4_model_dimension_supervised_result}
    \begin{tabular}{lccc}
        \toprule
        \multicolumn{4}{c}{\textbf{S4 (Supervised)}} \\
        & \textbf{512} & \textbf{768} & \textbf{1024} \\
        \midrule
        
        \multicolumn{4}{c}{\textbf{Adult ECG interpretation }} \\
        Ningbo & \underline{\textbf{0.973}} & \textbf{0.973} & \textbf{0.972} \\
        CPSC2018 & \underline{\textbf{0.967}} & \textbf{0.964} & \textbf{0.967} \\
        CPSC-Extra & \textbf{0.888} & 0.861 & \underline{\textbf{0.889}} \\
        Georgia & \underline{\textbf{0.917}} & 0.898 & 0.902 \\
        Chapman & \textbf{0.957} & \textbf{0.956} & \underline{\textbf{0.960}} \\
        SPH & \textbf{0.978} & \textbf{0.979} & \underline{\textbf{0.980}} \\
        % CODE-15\% & \textbf{0.991} & \underline{\textbf{0.991}} & \textbf{0.991} \\
        PTB-XL (all) & \underline{\textbf{0.940}} & 0.929 & \textbf{0.938} \\
        PTB-XL (sub) & \underline{\textbf{0.935}} & \textbf{0.933} & \textbf{0.932} \\
        PTB-XL (super) & \textbf{0.931} & \textbf{0.931} & \underline{\textbf{0.932}} \\
        \midrule
        
        \multicolumn{4}{c}{\textbf{Pediatric ECG interpretation }} \\
        ZZU pECG & 0.879 & \underline{\textbf{0.893}} & 0.880 \\
        \midrule
        
        \multicolumn{4}{c}{\textbf{Cardiac structure \& function }} \\
        EchoNext (Echo) & \underline{\textbf{0.823}} & \textbf{0.823} & \textbf{0.822} \\
        
        \bottomrule
    \end{tabular}
    \bigskip
    
    \textit{Note:} Model dimension 512 (state dimension 8) achieves optimal performance across most tasks.
\end{table}

\newpage
\section{Learning Rate Ablation for SSL Pretraining}
\label{app:learning_rate_ablation}

\begin{table}[htbp]
    \centering
    \caption{Comparison of aggregated macro-AUROC for data2vec pretrained with different learning rates (3e-3, 1e-3, 3e-4) and finetuned on downstream tasks. Models were pretrained on HEEDB subset (753K samples) for 20 epochs with S4 backbone (6 layers). The best-performing result is highlighted in boldface and underlined, while models that do not perform statistically significantly worse are also highlighted in boldface.\\}
    \label{tab:data2vec_learning_rate_finetuning_result}
    \begin{tabular}{lccc}
        \toprule
        \multicolumn{4}{c}{\textbf{data2vec (Finetuned)}} \\
        & \textbf{3e-3} & \textbf{1e-3} & \textbf{3e-4} \\
        \midrule
        
        \multicolumn{4}{c}{\textbf{Adult ECG interpretation }} \\
        Ningbo & \textbf{0.960} & \textbf{0.960} & \underline{\textbf{0.963}} \\
        CPSC2018 & \textbf{0.958} & \underline{\textbf{0.960}} & 0.954 \\
        CPSC-Extra & \underline{\textbf{0.860}} & \textbf{0.853} & 0.833 \\
        Georgia & \textbf{0.879} & \underline{\textbf{0.883}} & \textbf{0.877} \\
        Chapman & \textbf{0.927} & \textbf{0.929} & \underline{\textbf{0.931}} \\
        SPH & \textbf{0.965} & \textbf{0.965} & \underline{\textbf{0.969}} \\
        % CODE-15\% & \textbf{0.990} & \underline{\textbf{0.991}} & \textbf{0.990} \\
        PTB-XL (all) & \textbf{0.908} & \textbf{0.908} & \underline{\textbf{0.914}} \\
        PTB-XL (sub) & \underline{\textbf{0.907}} & \textbf{0.897} & 0.892 \\
        PTB-XL (super) & \textbf{0.911} & \underline{\textbf{0.912}} & 0.909 \\
        \midrule
        
        \multicolumn{4}{c}{\textbf{Pediatric ECG interpretation }} \\
        ZZU pECG & \textbf{0.853} & \underline{\textbf{0.857}} & 0.849 \\
        \midrule
        
        \multicolumn{4}{c}{\textbf{Cardiac structure \& function }} \\
        EchoNext (Echo) & \textbf{0.818} & \underline{\textbf{0.820}} & \textbf{0.818} \\
        
        \bottomrule
    \end{tabular}

    \bigskip
    
    \textit{Note:} Learning rates 3e-3 and 1e-3 achieve optimal performance across most tasks.
\end{table}

\begin{table}[htbp]
    \centering
    \caption{Comparison of aggregated macro-AUROC for DinoSR pretrained with different learning rates (3e-3, 1e-3, 3e-4) and finetuned on downstream tasks. Models were pretrained on HEEDB subset (753K samples) for 20 epochs with S4 backbone (6 layers). The best-performing result is highlighted in boldface and underlined, while models that do not perform statistically significantly worse are also highlighted in boldface.\\}
    \label{tab:dinosr_learning_rate_finetuning_result}
    \begin{tabular}{lccc}
        \toprule
        \multicolumn{4}{c}{\textbf{DinoSR (Finetuned)}} \\
        & \textbf{3e-3} & \textbf{1e-3} & \textbf{3e-4} \\
        \midrule
        
        \multicolumn{4}{c}{\textbf{Adult ECG interpretation }} \\
        Ningbo & \textbf{0.965} & \underline{\textbf{0.967}} & 0.962 \\
        CPSC2018 & \textbf{0.965} & \underline{\textbf{0.966}} & \textbf{0.963} \\
        CPSC-Extra & \textbf{0.865} & \underline{\textbf{0.867}} & \textbf{0.861} \\
        Georgia & \underline{\textbf{0.909}} & 0.903 & 0.886 \\
        Chapman & \underline{\textbf{0.948}} & \textbf{0.946} & 0.934 \\
        SPH & \textbf{0.971} & \underline{\textbf{0.977}} & \textbf{0.975} \\
        % CODE-15\% & \textbf{0.989} & \underline{\textbf{0.989}} & \textbf{0.988} \\
        PTB-XL (all) & \textbf{0.927} & \underline{\textbf{0.927}} & 0.921 \\
        PTB-XL (sub) & \underline{\textbf{0.930}} & \textbf{0.925} & 0.915 \\
        PTB-XL (super) & \underline{\textbf{0.920}} & 0.917 & 0.915 \\
        \midrule
        
        \multicolumn{4}{c}{\textbf{Pediatric ECG interpretation }} \\
        ZZU pECG & \underline{\textbf{0.878}} & \textbf{0.870} & \textbf{0.870} \\
        \midrule
        
        \multicolumn{4}{c}{\textbf{Cardiac structure \& function }} \\
        EchoNext (Echo) & \textbf{0.826} & \underline{\textbf{0.827}} & 0.815 \\
        
        \bottomrule            
    \end{tabular}

    \bigskip
    
    \textit{Note:} Learning rate 3e-3 achieves optimal performance across most tasks.
\end{table}

\begin{table}[htbp]
    \centering
    \caption{Comparison of aggregated macro-AUROC for JEPA pretrained with different learning rates (3e-3, 1e-3, 3e-4) and finetuned on downstream tasks. Models were pretrained on HEEDB subset (753K samples) for 20 epochs with S4 backbone (6 layers). The best-performing result is highlighted in boldface and underlined, while models that do not perform statistically significantly worse are also highlighted in boldface.\\}
    \label{tab:jepa_learning_rate_finetuning_result}
    \begin{tabular}{lccc}
        \toprule
        \multicolumn{4}{c}{\textbf{JEPA (Finetuned)}} \\
        & \textbf{3e-3} & \textbf{1e-3} & \textbf{3e-4} \\
        \midrule
        
        \multicolumn{4}{c}{\textbf{Adult ECG interpretation }} \\
        Ningbo & \underline{\textbf{0.962}} & 0.954 & 0.947 \\
        CPSC2018 & \underline{\textbf{0.961}} & \textbf{0.958} & 0.953 \\
        CPSC-Extra & \textbf{0.846} & \textbf{0.837} & \underline{\textbf{0.849}} \\
        Georgia & \textbf{0.873} & \textbf{0.877} & \underline{\textbf{0.878}} \\
        Chapman & \textbf{0.934} & \underline{\textbf{0.937}} & 0.924 \\
        SPH & \underline{\textbf{0.965}} & \textbf{0.960} & 0.958 \\
        % CODE-15\% & \underline{\textbf{0.991}} & \textbf{0.990} & \textbf{0.990} \\
        PTB-XL (all) & \underline{\textbf{0.925}} & 0.913 & 0.905 \\
        PTB-XL (sub) & \underline{\textbf{0.905}} & 0.894 & 0.884 \\
        PTB-XL (super) & \textbf{0.912} & \underline{\textbf{0.912}} & 0.909 \\
        \midrule
        
        \multicolumn{4}{c}{\textbf{Pediatric ECG interpretation }} \\
        ZZU pECG & \underline{\textbf{0.845}} & \textbf{0.835} & \textbf{0.841} \\
        \midrule
        
        \multicolumn{4}{c}{\textbf{Cardiac structure \& function }} \\
        EchoNext (Echo) & \textbf{0.817} & \underline{\textbf{0.821}} & \textbf{0.817} \\
        \bottomrule
    \end{tabular}

    \bigskip
    
    \textit{Note:} Learning rate 3e-3 achieves optimal performance across most tasks.
\end{table}

\begin{table}[htbp]
    \centering
    \caption{Comparison of aggregated macro-AUROC for CPC pretrained with different learning rates (3e-3, 1e-3, 3e-4) and finetuned on downstream tasks. Models were pretrained on HEEDB subset (753K samples) for 20 epochs with S4 backbone (6 layers). The best-performing result is highlighted in boldface and underlined, while models that do not perform statistically significantly worse are also highlighted in boldface.\\}
    \label{tab:cpc_learning_rate_finetuning_result}
    \begin{tabular}{lccc}
        \toprule
        \multicolumn{4}{c}{\textbf{CPC (Finetuned)}} \\
        & \textbf{3e-3} & \textbf{1e-3} & \textbf{3e-4} \\
        \midrule
        
        \multicolumn{4}{c}{\textbf{Adult ECG interpretation }} \\
        Ningbo & \underline{\textbf{0.971}} & \textbf{0.968} & 0.965 \\
        CPSC2018 & \textbf{0.969} & \underline{\textbf{0.973}} & \textbf{0.970} \\
        CPSC-Extra & \textbf{0.882} & \underline{\textbf{0.893}} & \textbf{0.893} \\
        Georgia & \underline{\textbf{0.904}} & \textbf{0.901} & 0.896 \\
        Chapman & \underline{\textbf{0.958}} & 0.952 & 0.947 \\
        SPH & \underline{\textbf{0.979}} & \textbf{0.978} & \textbf{0.976} \\
        % CODE-15\% & \textbf{0.991} & \underline{\textbf{0.992}} & \textbf{0.991} \\
        PTB-XL (all) & \textbf{0.936} & \underline{\textbf{0.938}} & \textbf{0.933} \\
        PTB-XL (sub) & \underline{\textbf{0.933}} & \textbf{0.931} & 0.915 \\
        PTB-XL (super) & \underline{\textbf{0.921}} & \textbf{0.920} & 0.915 \\
        \midrule
        
        \multicolumn{4}{c}{\textbf{Pediatric ECG interpretation }} \\
        ZZU pECG & \textbf{0.876} & \underline{\textbf{0.889}} & \textbf{0.882} \\
        \midrule
        
        \multicolumn{4}{c}{\textbf{Cardiac structure \& function }} \\
        EchoNext (Echo) & \underline{\textbf{0.824}} & \textbf{0.822} & \textbf{0.821} \\
        
        \bottomrule            
    \end{tabular}
    
    \bigskip
    
    \textit{Note:} Learning rate 3e-3 achieves optimal performance across most tasks.
\end{table}

\begin{table}[htbp]
    \centering
    \caption{Comparison of aggregated macro-AUROC for HuBERT++ pretrained with different learning rates (3e-3, 1e-3, 3e-4) and finetuned on downstream tasks. Models were pretrained on HEEDB subset (753K samples) for 20 epochs with S4 backbone (6 layers). The best-performing result is highlighted in boldface and underlined, while models that do not perform statistically significantly worse are also highlighted in boldface.\\}
    \label{tab:hubert_learning_rate_finetuning_result}
    \begin{tabular}{lccc}
        \toprule
        \multicolumn{4}{c}{\textbf{HuBERT++ (Finetuned)}} \\
        & \textbf{3e-3} & \textbf{1e-3} & \textbf{3e-4} \\
        \midrule
        
        \multicolumn{4}{c}{\textbf{Adult ECG interpretation }} \\
        Ningbo & \underline{\textbf{0.970}} & \textbf{0.970} & \textbf{0.968} \\
        CPSC2018 & \underline{\textbf{0.972}} & \textbf{0.970} & 0.966 \\
        CPSC-Extra & \textbf{0.884} & \underline{\textbf{0.892}} & 0.875 \\
        Georgia & \underline{\textbf{0.906}} & 0.896 & 0.889 \\
        Chapman & \underline{\textbf{0.955}} & \textbf{0.954} & 0.942 \\
        SPH & 0.979 & \underline{\textbf{0.981}} & 0.974 \\
        % CODE-15\% & \underline{\textbf{0.991}} & \textbf{0.991} & \textbf{0.988} \\
        PTB-XL (all) & \textbf{0.932} & \underline{\textbf{0.935}} & 0.926 \\
        PTB-XL (sub) & \textbf{0.931} & \underline{\textbf{0.931}} & 0.923 \\
        PTB-XL (super) & \underline{\textbf{0.919}} & 0.916 & 0.912 \\
        \midrule
        
        \multicolumn{4}{c}{\textbf{Pediatric ECG interpretation }} \\
        ZZU pECG & \underline{\textbf{0.898}} & 0.889 & 0.880 \\
        \midrule
        
        \multicolumn{4}{c}{\textbf{Cardiac structure \& function }} \\
        EchoNext (Echo) & \underline{\textbf{0.824}} & \textbf{0.823} & \textbf{0.821} \\
        
        \bottomrule            
    \end{tabular}
    
    \bigskip
    
    \textit{Note:} Learning rate 3e-3 achieves optimal performance across most tasks.
\end{table}

\newpage
\section{HuBERT++ SSL Head and Codebook Size Ablation}
\label{app:hubert_ssl_head_codebook_ablation}

\begin{table}[htbp]
    \centering
    \caption{Comparison of aggregated macro-AUROC for HuBERT++ pretrained with different configurations (\textbf{Base:} MLP SSL head, codebook size [128, 256]; \textbf{S4:} S4 SSL head, codebook size [128, 256]; \textbf{[256, 512]:} MLP SSL head, codebook size [256, 512]; \textbf{S4 + [256, 512]:} S4 SSL head, codebook size [256, 512]) and finetuned on downstream tasks. Models were pretrained on HEEDB subset (753K samples) for 20 epochs with a learning rate of 3e-3 using an S4 backbone (6 layers). The best-performing result is highlighted in boldface and underlined, while models that do not perform statistically significantly worse are also highlighted in boldface.\\}
    \label{tab:hubert_configurations_finetuning_result}
    \begin{tabular}{lcccc}
        \toprule
        \multicolumn{5}{c}{\textbf{HuBERT++ (Finetuned)}} \\
        & \textbf{Base} & \textbf{S4} & \textbf{[256, 512]} & \textbf{S4 + [256, 512]} \\
        \midrule
        
        \multicolumn{5}{c}{\textbf{Adult ECG interpretation }} \\
        Ningbo & \underline{\textbf{0.970}} & \textbf{0.967} & \textbf{0.970} & \textbf{0.967} \\
        CPSC2018 & \underline{\textbf{0.972}} & 0.968 & \textbf{0.971} & 0.968 \\
        CPSC-Extra & \textbf{0.884} & \textbf{0.883} & \textbf{0.882} & \underline{\textbf{0.887}} \\
        Georgia & \textbf{0.906} & \textbf{0.903} & \underline{\textbf{0.910}} & 0.901 \\
        Chapman & \underline{\textbf{0.955}} & \textbf{0.954} & \textbf{0.953} & 0.948 \\
        SPH & \textbf{0.979} & 0.976 & \underline{\textbf{0.981}} & \textbf{0.978} \\
        % CODE-15\% & \underline{\textbf{0.991}} & \textbf{0.990} & 0.989 & \textbf{0.990} \\
        PTB-XL (all) & \textbf{0.932} & 0.932 & \underline{\textbf{0.936}} & \textbf{0.933} \\
        PTB-XL (sub) & \textbf{0.931} & \underline{\textbf{0.934}} & 0.927 & \textbf{0.931} \\
        PTB-XL (super) & \underline{\textbf{0.919}} & \textbf{0.918} & \textbf{0.918} & \textbf{0.918} \\
        \midrule
        
        \multicolumn{5}{c}{\textbf{Pediatric ECG interpretation }} \\
        ZZU pECG & \underline{\textbf{0.898}} & 0.887 & \textbf{0.891} & 0.888 \\
        \midrule
        
        \multicolumn{5}{c}{\textbf{Cardiac structure \& function }} \\
        EchoNext (Echo) & \textbf{0.824} & \underline{\textbf{0.826}} & \textbf{0.824} & \textbf{0.825} \\
        
        \bottomrule            
    \end{tabular}

    \bigskip
    
    \textit{Note:} Optimal configuration uses MLP SSL head with codebook size [128, 256].
\end{table}

\newpage
\section{Backbone Architecture Comparison}
\label{app:s4_transformer_predictor_ablation}

\subsection{Finetuning Result}

\begin{table}[htbp]
    \centering
    \caption{Comparison of aggregated macro-AUROC for data2vec pretrained with different configurations (\textbf{S4:} S4 backbone with 4 layers; \textbf{Transformer:} Transformer backbone with 6 blocks; \textbf{Net1D:} Net1D backbone with 7 stages) and finetuned on downstream tasks. Models were pretrained on HEEDB subset (753K samples) for 20 epochs with a learning rate of 3e-3. The best-performing result is highlighted in boldface and underlined, while models that do not perform statistically significantly worse are also highlighted in boldface.\\}
    \label{tab:data2vec_s4_transformer_finetuning_result}
    \begin{tabular}{lccc}
        \toprule
        \multicolumn{4}{c}{\textbf{data2vec (Finetuned)}} \\
        & \textbf{S4} & \textbf{Transformer} & \textbf{Net1D} \\
        \midrule
        
        \multicolumn{4}{c}{\textbf{Adult ECG interpretation }} \\
        Ningbo & \underline{\textbf{0.965}} & \textbf{0.962} & \textbf{0.964} \\
        CPSC2018 & \underline{\textbf{0.960}} & 0.950 & \textbf{0.959} \\
        CPSC-Extra & \textbf{0.839} & \underline{\textbf{0.858}} & \textbf{0.858} \\
        Georgia & \underline{\textbf{0.894}} & 0.867 & \textbf{0.892} \\
        Chapman & \underline{\textbf{0.946}} & 0.935 & \textbf{0.937} \\
        SPH & \underline{\textbf{0.976}} & 0.964 & \textbf{0.975} \\
        PTB-XL (all) & \textbf{0.924} & 0.905 & \underline{\textbf{0.925}} \\
        PTB-XL (sub) & \underline{\textbf{0.934}} & 0.915 & \textbf{0.930} \\
        PTB-XL (super) & \underline{\textbf{0.926}} & 0.911 & 0.917 \\
        \midrule
        
        \multicolumn{4}{c}{\textbf{Pediatric ECG interpretation }} \\
        ZZU pECG & \textbf{0.868} & 0.850 & \underline{\textbf{0.873}} \\
        \midrule
        
        \multicolumn{4}{c}{\textbf{Cardiac structure \& function }} \\
        EchoNext & \textbf{0.816} & \textbf{0.821} & \underline{\textbf{0.823}} \\
        
        \bottomrule            
    \end{tabular}
    \bigskip
    
    \textit{Note:} S4 backbone outperforms Transformer and Net1D across most tasks.
\end{table}

\begin{table}[htbp]
    \centering
    \caption{Comparison of aggregated macro-AUROC for DinoSR pretrained with different configurations (\textbf{S4:} S4 backbone with 4 layers; \textbf{Transformer:} Transformer backbone with 6 blocks; \textbf{Net1D:} Net1D backbone with 7 stages) and finetuned on downstream tasks. Models were pretrained on HEEDB subset (753K samples) for 20 epochs with a learning rate of 3e-3. The best-performing result is highlighted in boldface and underlined, while models that do not perform statistically significantly worse are also highlighted in boldface.\\}
    \label{tab:dinosr_s4_transformer_finetuning_result}
    \begin{tabular}{lccc}
        \toprule
        \multicolumn{4}{c}{\textbf{DinoSR (Finetuned)}} \\
        & \textbf{S4} & \textbf{Transformer} & \textbf{Net1D} \\
        \midrule
        
        \multicolumn{4}{c}{\textbf{Adult ECG interpretation }} \\
        Ningbo & \textbf{0.964} & 0.956 & \underline{\textbf{0.967}} \\
        CPSC2018 & \underline{\textbf{0.968}} & 0.942 & 0.962 \\
        CPSC-Extra & \textbf{0.883} & 0.843 & \underline{\textbf{0.884}} \\
        Georgia & \underline{\textbf{0.906}} & 0.849 & 0.889 \\
        Chapman & \underline{\textbf{0.960}} & 0.925 & 0.952 \\
        SPH & \underline{\textbf{0.975}} & 0.956 & 0.968 \\
        PTB-XL (all) & \underline{\textbf{0.935}} & 0.907 & 0.920 \\
        PTB-XL (sub) & \underline{\textbf{0.938}} & 0.921 & 0.925 \\
        PTB-XL (super) & \underline{\textbf{0.932}} & 0.905 & 0.913 \\
        \midrule
        
        \multicolumn{4}{c}{\textbf{Pediatric ECG interpretation }} \\
        ZZU pECG & \underline{\textbf{0.888}} & 0.861 & \textbf{0.882} \\
        \midrule
        
        \multicolumn{4}{c}{\textbf{Cardiac structure \& function }} \\
        EchoNext & \underline{\textbf{0.829}} & 0.813 & 0.817 \\
        
        \bottomrule            
    \end{tabular}
    \bigskip
    
    \textit{Note:} S4 backbone outperforms Transformer and Net1D across all tasks (except for one, where it performs on par).
\end{table}

\begin{table}[htbp]
    \centering
    \caption{Comparison of aggregated macro-AUROC for JEPA pretrained with different configurations (\textbf{S4:} S4 backbone with 4 layers; \textbf{Transformer:} Transformer backbone with 6 blocks; \textbf{Net1D:} Net1D backbone with 7 stages) and finetuned on downstream tasks. Models were pretrained on HEEDB subset (753K samples) for 20 epochs with a learning rate of 3e-3. The best-performing result is highlighted in boldface and underlined, while models that do not perform statistically significantly worse are also highlighted in boldface.\\}
    \label{tab:jepa_s4_transformer_finetuning_result}
    \begin{tabular}{lccc}
        \toprule
        \multicolumn{4}{c}{\textbf{JEPA (Finetuned)}} \\
        & \textbf{S4} & \textbf{Transformer} & \textbf{Net1D} \\
        \midrule
        
        \multicolumn{4}{c}{\textbf{Adult ECG interpretation }} \\
        Ningbo & \underline{\textbf{0.971}} & 0.952 & 0.936 \\
        CPSC2018 & \underline{\textbf{0.970}} & 0.949 & 0.945 \\
        CPSC-Extra & \underline{\textbf{0.888}} & 0.853 & 0.829 \\
        Georgia & \underline{\textbf{0.913}} & 0.868 & 0.844 \\
        Chapman & \underline{\textbf{0.957}} & 0.934 & 0.927 \\
        SPH & \underline{\textbf{0.976}} & 0.961 & 0.921 \\
        PTB-XL (all) & \underline{\textbf{0.943}} & 0.907 & 0.896 \\
        PTB-XL (sub) & \underline{\textbf{0.924}} & 0.901 & 0.875 \\
        PTB-XL (super) & \underline{\textbf{0.932}} & 0.896 & 0.898 \\
        \midrule
        
        \multicolumn{4}{c}{\textbf{Pediatric ECG interpretation }} \\
        ZZU pECG & \underline{\textbf{0.891}} & 0.847 & 0.841 \\
        \midrule
        
        \multicolumn{4}{c}{\textbf{Cardiac structure \& function }} \\
        EchoNext & \underline{\textbf{0.826}} & 0.802 & 0.801 \\
        
        \bottomrule            
    \end{tabular}
    \bigskip
    
    \textit{Note:} S4 backbone outperforms Transformer and Net1D for all tasks.
\end{table}

\begin{table}[htbp]
    \centering
    \caption{Comparison of aggregated macro-AUROC for CPC pretrained with different configurations (\textbf{S4:} S4 backbone with 4 layers; \textbf{Transformer:} Transformer backbone with 6 blocks; \textbf{Net1D:} Net1D backbone with 7 stages) and finetuned on downstream tasks. Models were pretrained on HEEDB subset (753K samples) for 20 epochs with a learning rate of 3e-3. The best-performing result is highlighted in boldface and underlined, while models that do not perform statistically significantly worse are also highlighted in boldface.\\}
    \label{tab:cpc_s4_transformer_finetuning_result}
    \begin{tabular}{lccc}
        \toprule
        \multicolumn{4}{c}{\textbf{CPC (Finetuned)}} \\
        & \textbf{S4} & \textbf{Transformer} & \textbf{Net1D} \\
        \midrule
        
        \multicolumn{4}{c}{\textbf{Adult ECG interpretation }} \\
        Ningbo & \underline{\textbf{0.974}} & 0.962 & 0.945 \\
        CPSC2018 & \underline{\textbf{0.972}} & 0.950 & 0.942 \\
        CPSC-Extra & \underline{\textbf{0.906}} & 0.814 & 0.840 \\
        Georgia & \underline{\textbf{0.912}} & 0.875 & 0.878 \\
        Chapman & \underline{\textbf{0.959}} & 0.932 & 0.929 \\
        SPH & \underline{\textbf{0.982}} & 0.965 & 0.943 \\
        PTB-XL (all) & \underline{\textbf{0.943}} & 0.907 & 0.902 \\
        PTB-XL (sub) & \underline{\textbf{0.937}} & 0.911 & 0.921 \\
        PTB-XL (super) & \underline{\textbf{0.932}} & 0.914 & 0.907 \\
        \midrule
        
        \multicolumn{4}{c}{\textbf{Pediatric ECG interpretation }} \\
        ZZU pECG & \underline{\textbf{0.899}} & 0.864 & 0.852 \\
        \midrule
        
        \multicolumn{4}{c}{\textbf{Cardiac structure \& function }} \\
        EchoNext & \underline{\textbf{0.832}} & 0.812 & 0.816 \\
        
        \bottomrule            
    \end{tabular}
    \bigskip
    
    \textit{Note:} S4 backbone outperforms Transformer and Net1D for all tasks.
\end{table}

\begin{table}[htbp]
    \centering
    \caption{Comparison of aggregated macro-AUROC for HuBERT++ pretrained with different configurations (\textbf{S4:} S4 backbone with 4 layers; \textbf{Transformer:} Transformer backbone with 6 blocks; \textbf{Net1D:} Net1D backbone with 7 stages) and finetuned on downstream tasks. Models were pretrained on HEEDB subset (753K samples) for 20 epochs with a learning rate of 3e-3. The best-performing result is highlighted in boldface and underlined, while models that do not perform statistically significantly worse are also highlighted in boldface.\\}
    \label{tab:hubert_s4_transformer_finetuning_result}
    \begin{tabular}{lccc}
        \toprule
        \multicolumn{4}{c}{\textbf{HuBERT++ (Finetuned)}} \\
        & \textbf{S4} & \textbf{Transformer} & \textbf{Net1D} \\
        \midrule
        
        \multicolumn{4}{c}{\textbf{Adult ECG interpretation }} \\
        Ningbo & \underline{\textbf{0.970}} & 0.940 & 0.952 \\
        CPSC2018 & \underline{\textbf{0.970}} & 0.926 & 0.941 \\
        CPSC-Extra & \underline{\textbf{0.891}} & 0.774 & 0.847 \\
        Georgia & \underline{\textbf{0.915}} & 0.833 & 0.861 \\
        Chapman & \underline{\textbf{0.958}} & 0.906 & 0.916 \\
        SPH & \underline{\textbf{0.982}} & 0.948 & 0.951 \\
        PTB-XL (all) & \underline{\textbf{0.939}} & 0.904 & 0.905 \\
        PTB-XL (sub) & \underline{\textbf{0.940}} & 0.888 & 0.906 \\
        PTB-XL (super) & \underline{\textbf{0.930}} & 0.888 & 0.903 \\
        \midrule
        
        \multicolumn{4}{c}{\textbf{Pediatric ECG interpretation }} \\
        ZZU pECG & \underline{\textbf{0.896}} & 0.839 & 0.834 \\
        \midrule
        
        \multicolumn{4}{c}{\textbf{Cardiac structure \& function }} \\
        EchoNext & \underline{\textbf{0.830}} & 0.805 & 0.808 \\
        
        \bottomrule            
    \end{tabular}
    \bigskip
    
    \textit{Note:} S4 backbone outperforms Transformer and Net1D for all tasks.
\end{table}

\newpage

\subsection{CKA Analysis}
\label{app:backbone_cka_analysis}
In this section, we present intra-model layer-wise representational similarity analyses for all five pretraining objectives (data2vec, DinoSR, JEPA, CPC, and HuBERT++), each evaluated across S4, Transformer, and Net1D backbones. For each configuration, CKA heatmaps are computed with a Gaussian RBF kernel ($\sigma=1.0$) on 2,500 PTB-XL samples, where warmer colors indicate greater similarity between layer pairs.

\subsubsection{data2vec}
\begin{figure}[htbp]
    \centering
    \begin{subfigure}{0.32\textwidth}
        \centering
        \includegraphics[width=\textwidth]{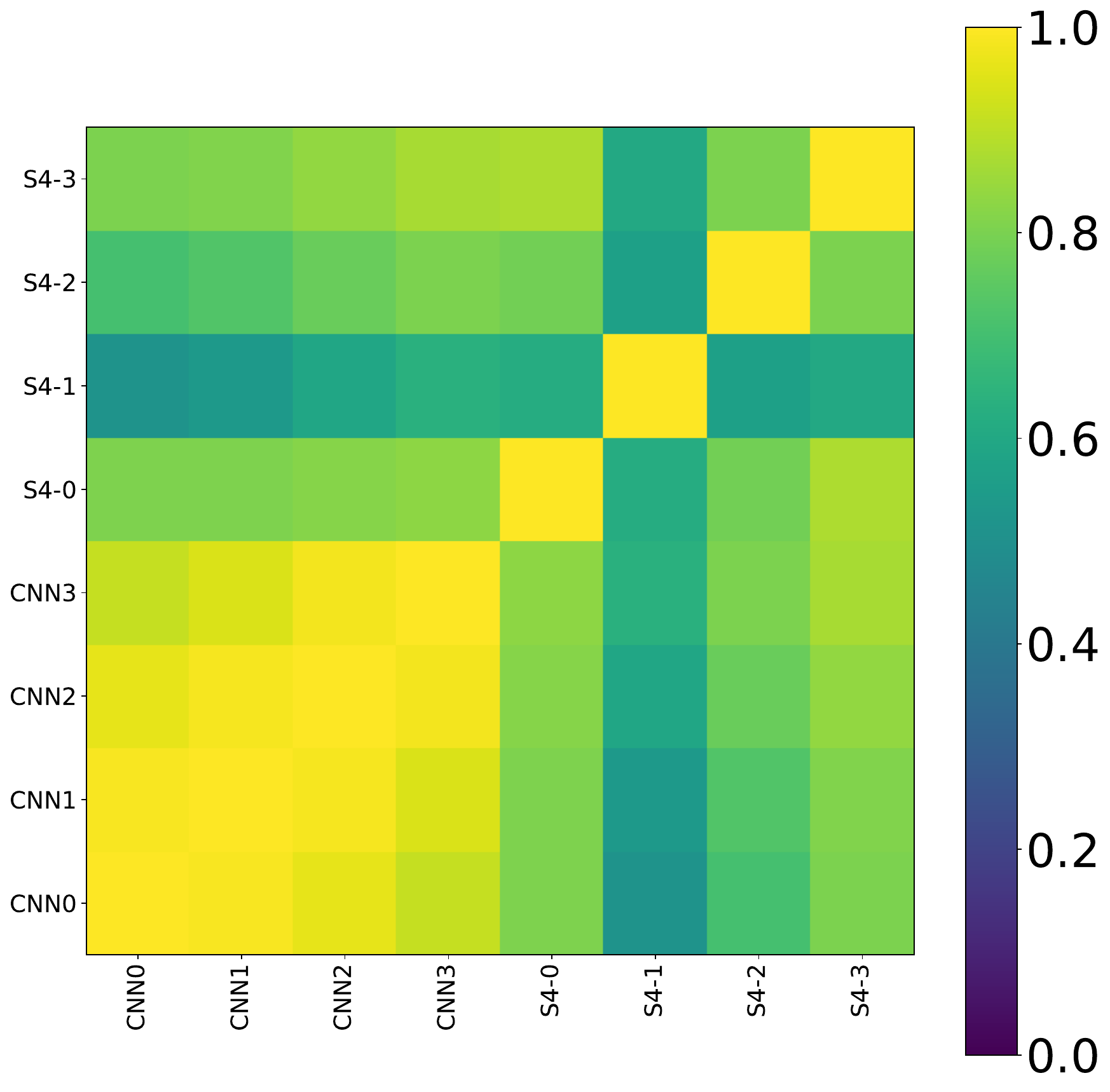}
        \caption{S4}
        \label{fig:d2v_753k_s4_4}
    \end{subfigure}
    \hfill
    \begin{subfigure}{0.32\textwidth}
        \centering
        \includegraphics[width=\textwidth]{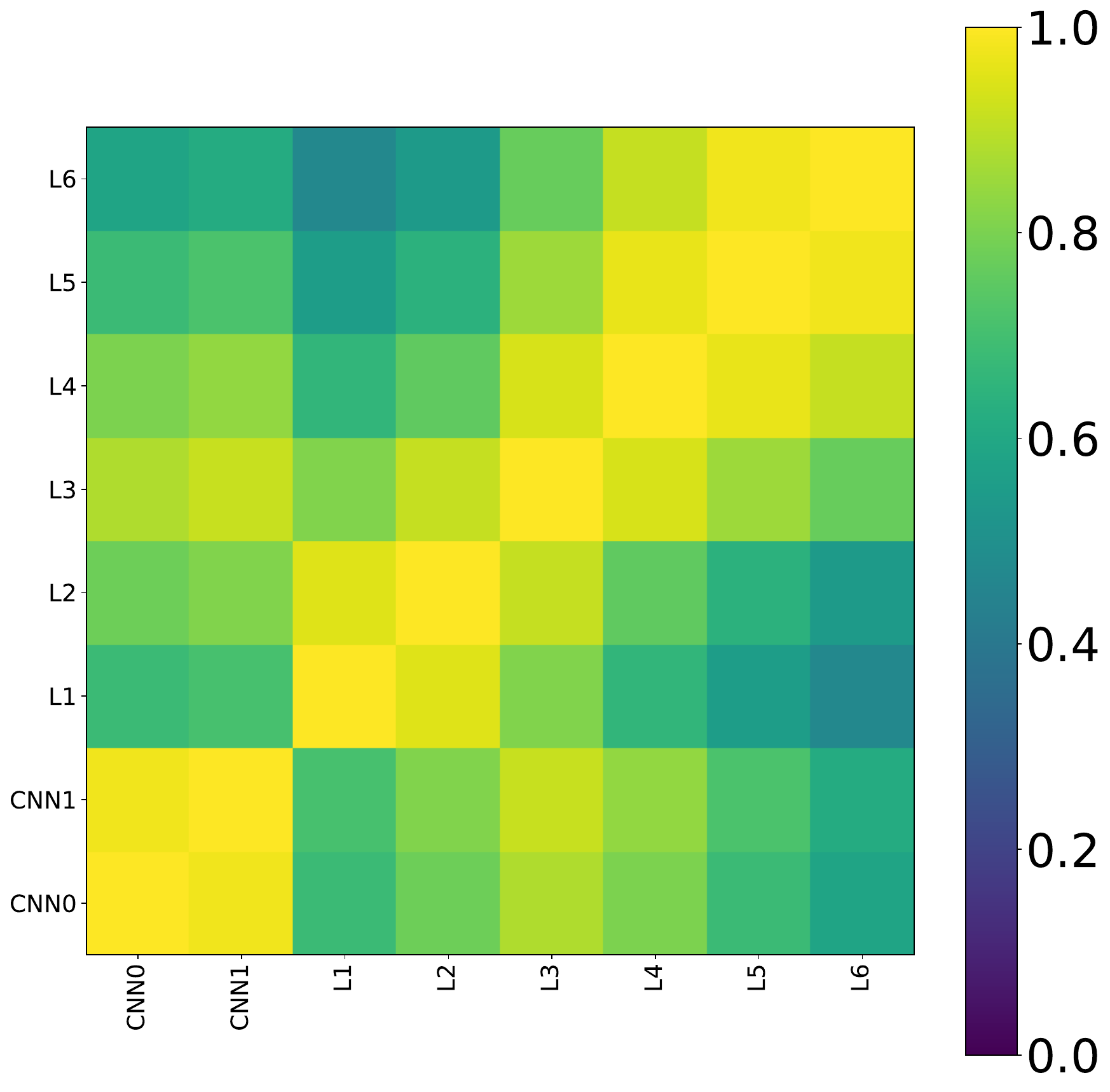}
        \caption{Transformer}
        \label{fig:d2v_753k_transformer}
    \end{subfigure}
    \hfill
    \begin{subfigure}{0.32\textwidth}
        \centering
        \includegraphics[width=\textwidth]{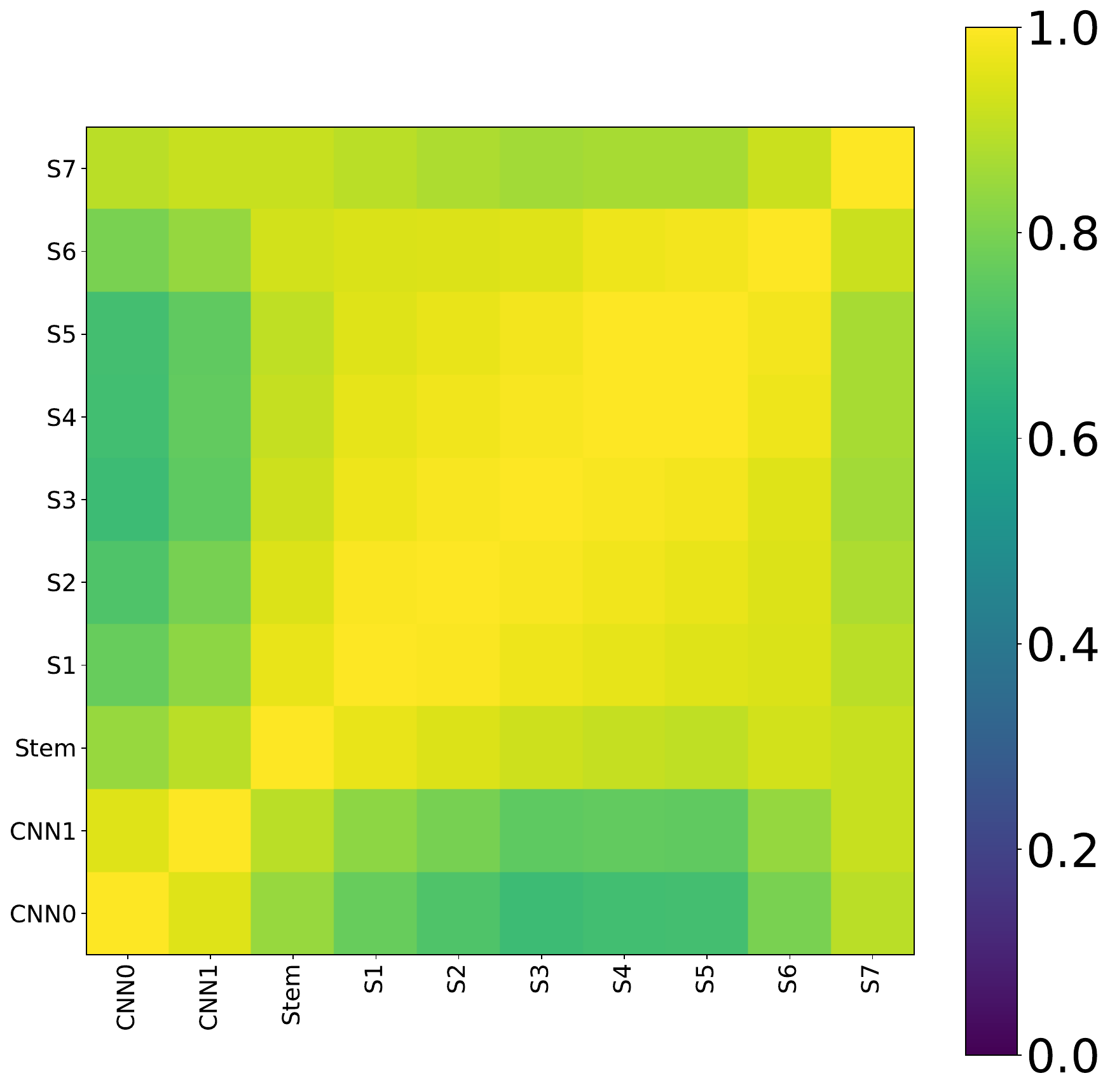}
        \caption{Net1D}
        \label{fig:d2v_753k_net1d}
    \end{subfigure}
    \caption{Intra-model layer-wise representational similarity for data2vec}
    \label{fig:data2vec_cka_s4_transformer_net1d}
\end{figure}

\subsubsection{DinoSR}
\begin{figure}[htbp]
    \centering
    \begin{subfigure}{0.32\textwidth}
        \centering
        \includegraphics[width=\textwidth]{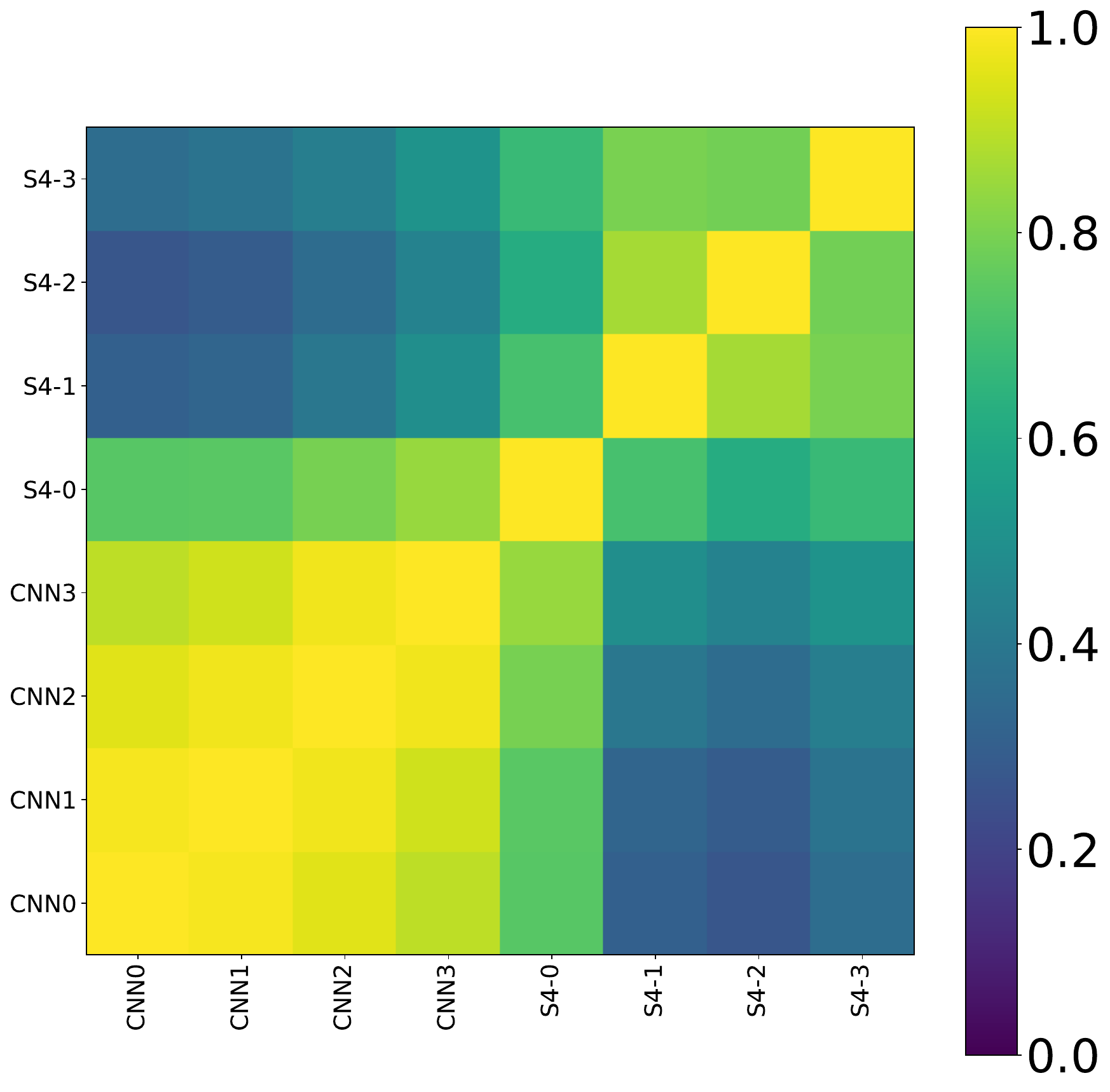}
        \caption{S4}
        \label{fig:dsr_753k_s4_4}
    \end{subfigure}
    \hfill
    \begin{subfigure}{0.32\textwidth}
        \centering
        \includegraphics[width=\textwidth]{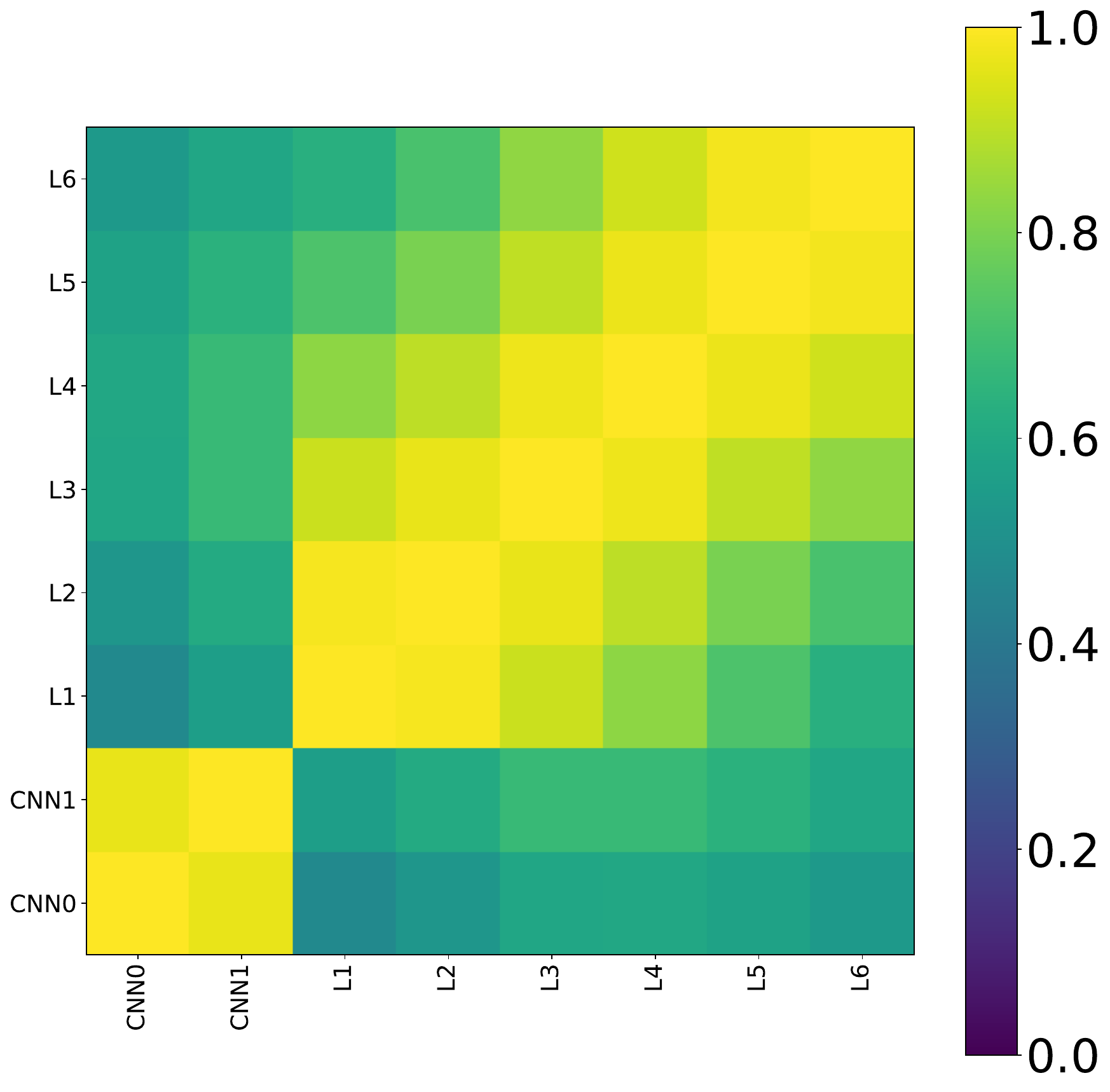}
        \caption{Transformer}
        \label{fig:dsr_753k_transformer}
    \end{subfigure}
    \hfill
    \begin{subfigure}{0.32\textwidth}
        \centering
        \includegraphics[width=\textwidth]{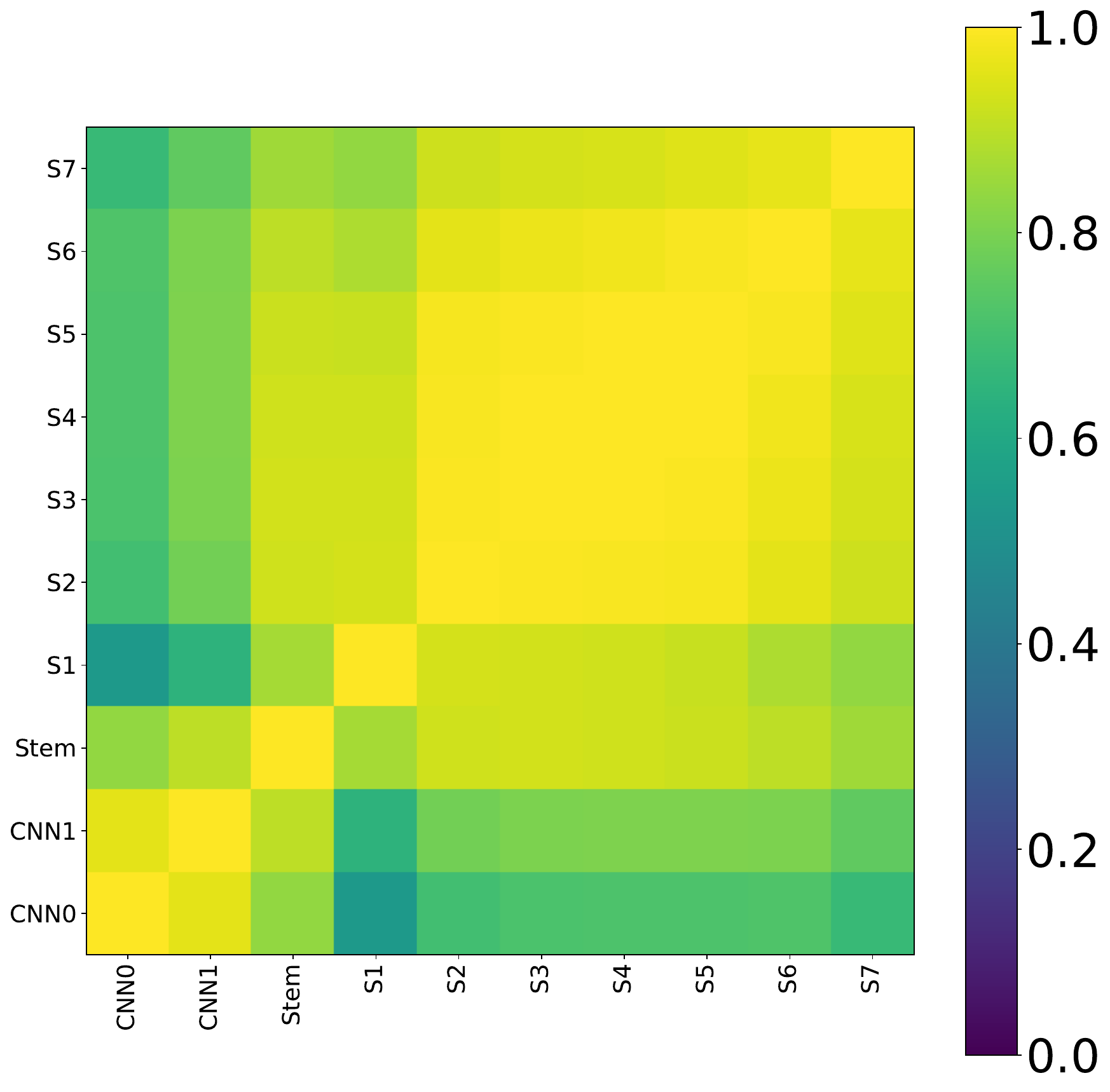}
        \caption{Net1D}
        \label{fig:dsr_753k_net1d}
    \end{subfigure}
    \caption{Intra-model layer-wise representational similarity for DinoSR}
    \label{fig:dinosr_cka_s4_transformer_net1d}
\end{figure}

\newpage
\subsubsection{JEPA}
\begin{figure}[htbp]
    \centering
    \begin{subfigure}{0.32\textwidth}
        \centering
        \includegraphics[width=\textwidth]{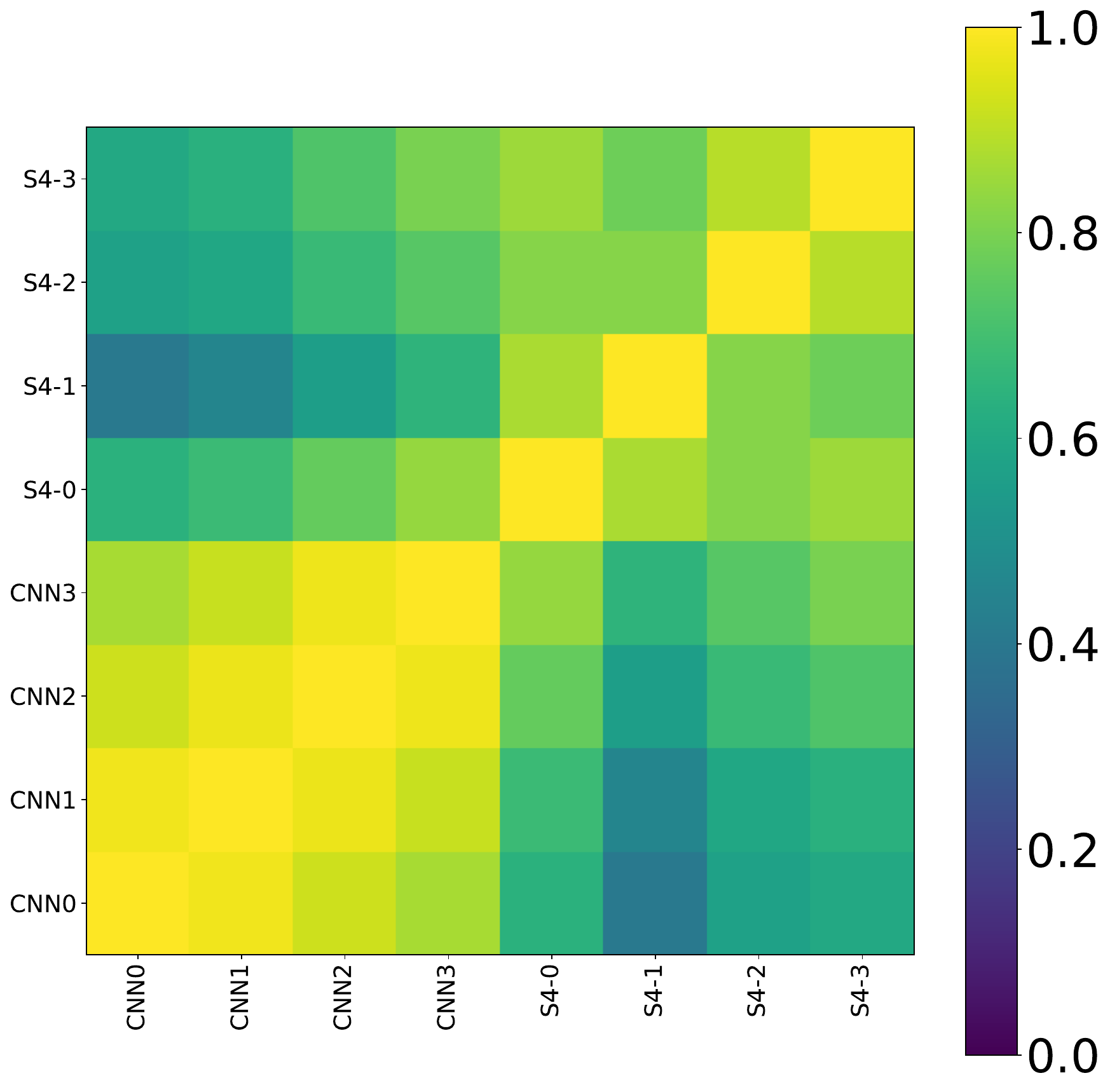}
        \caption{S4}
        \label{fig:jep_753k_s4_4}
    \end{subfigure}
    \hfill
    \begin{subfigure}{0.32\textwidth}
        \centering
        \includegraphics[width=\textwidth]{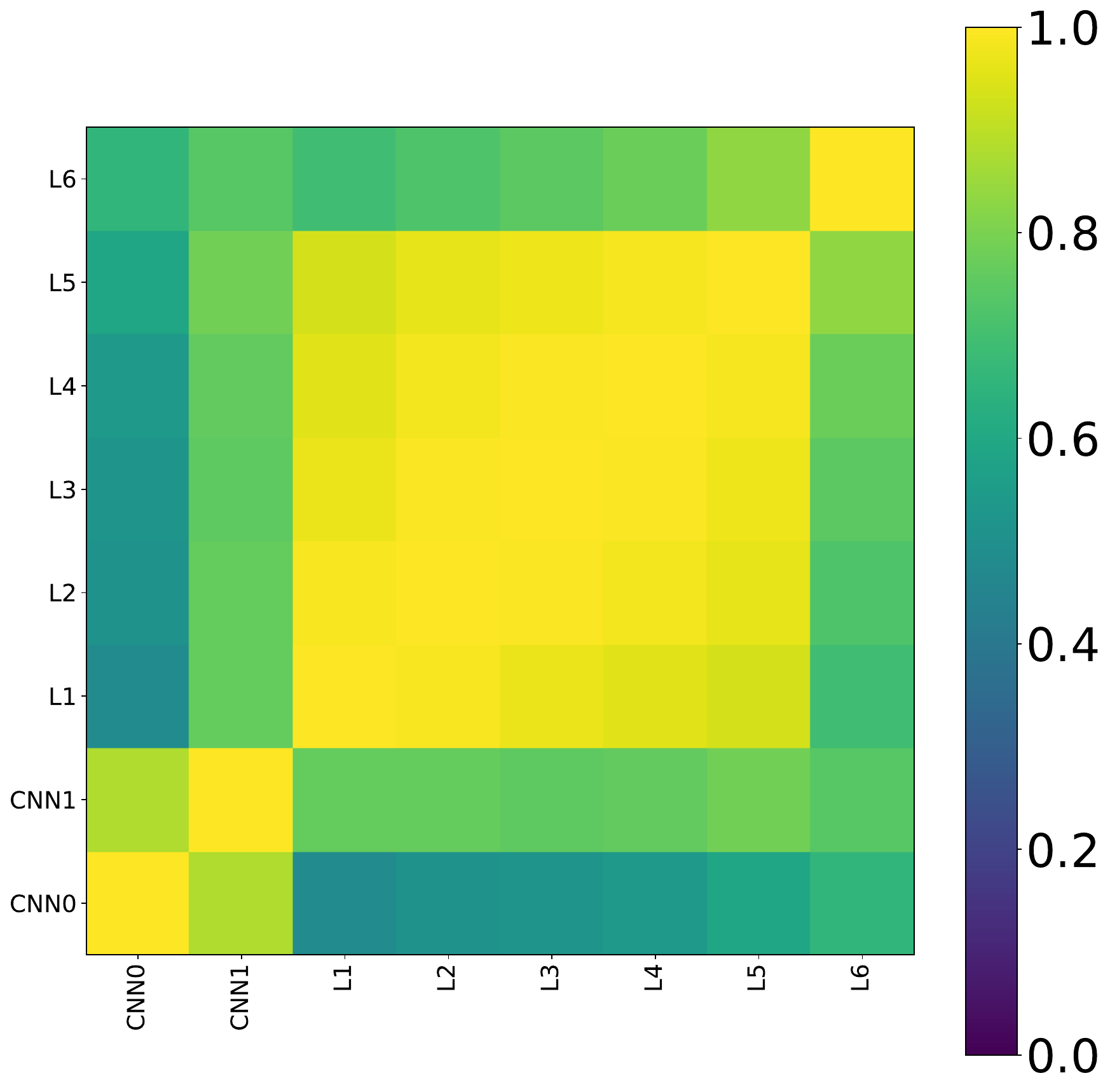}
        \caption{Transformer}
        \label{fig:jep_753k_transformer}
    \end{subfigure}
    \hfill
    \begin{subfigure}{0.32\textwidth}
        \centering
        \includegraphics[width=\textwidth]{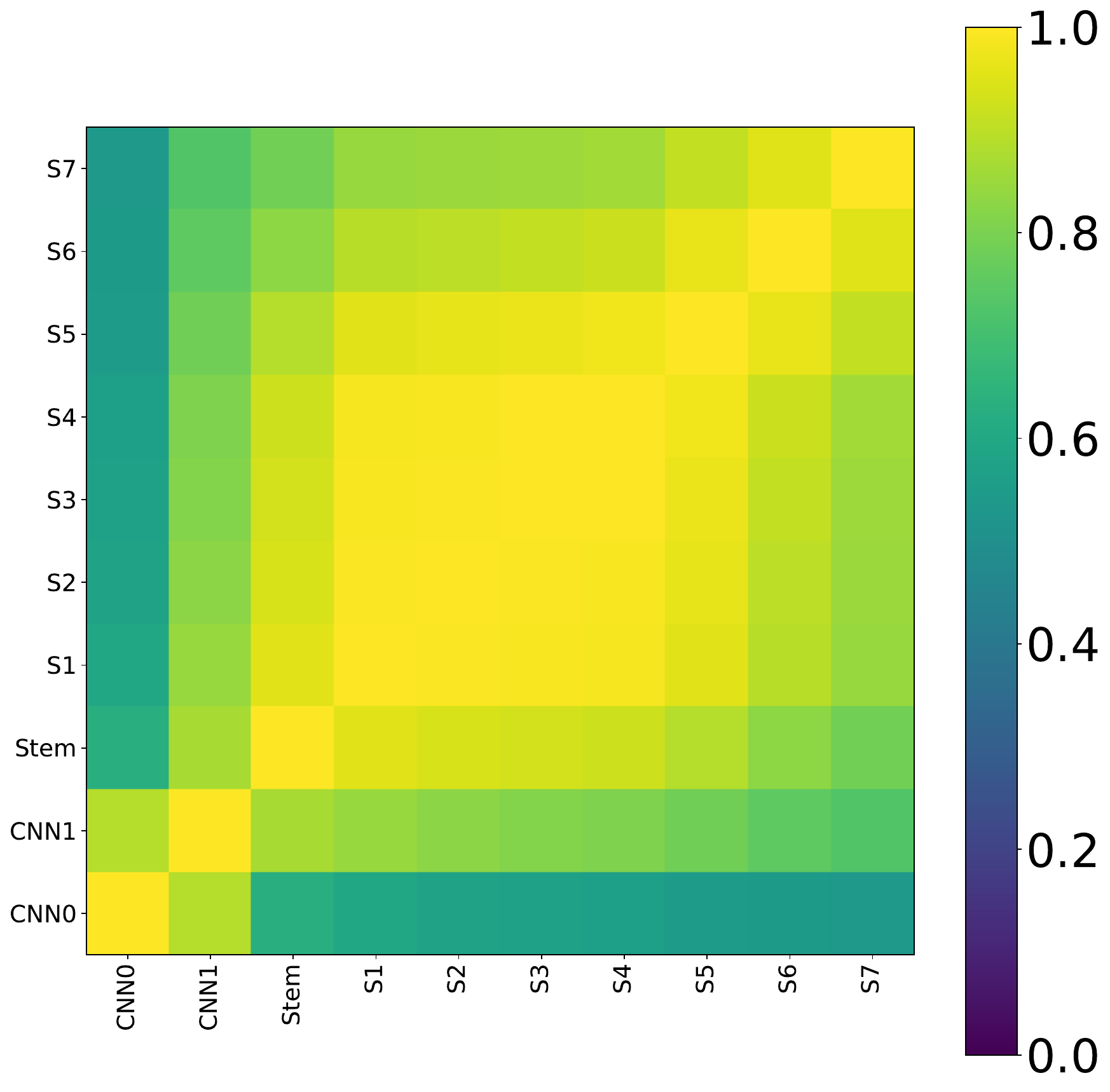}
        \caption{Net1D}
        \label{fig:jep_753k_net1d}
    \end{subfigure}
    \caption{Intra-model layer-wise representational similarity for JEPA}
    \label{fig:jepa_cka_s4_transformer_net1d}
\end{figure}

\subsubsection{CPC}
\begin{figure}[htbp]
    \centering
    \begin{subfigure}{0.32\textwidth}
        \centering
        \includegraphics[width=\textwidth]{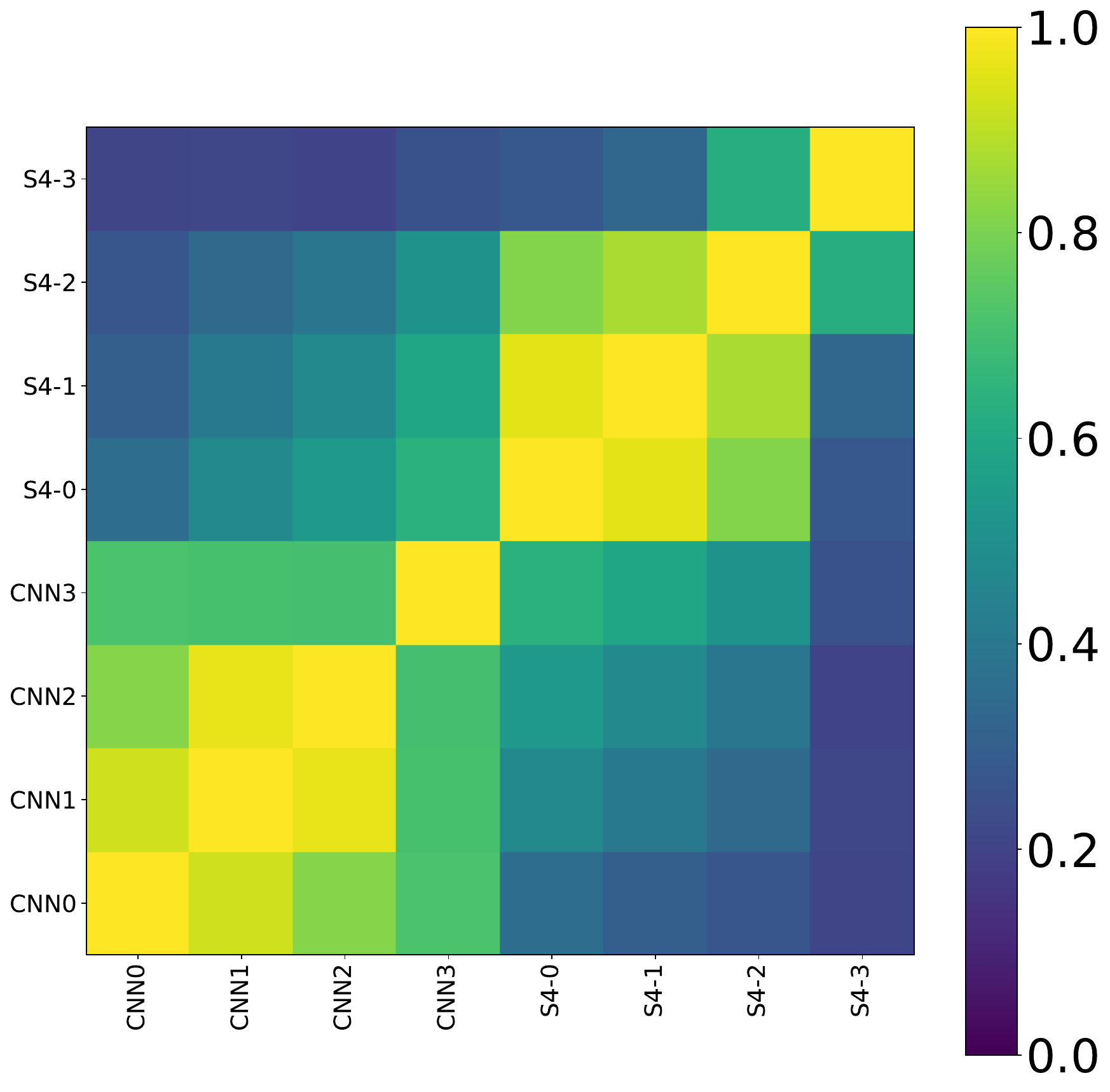}
        \caption{S4}
        \label{fig:cpc_753k_s4_4}
    \end{subfigure}
    \hfill
    \begin{subfigure}{0.32\textwidth}
        \centering
        \includegraphics[width=\textwidth]{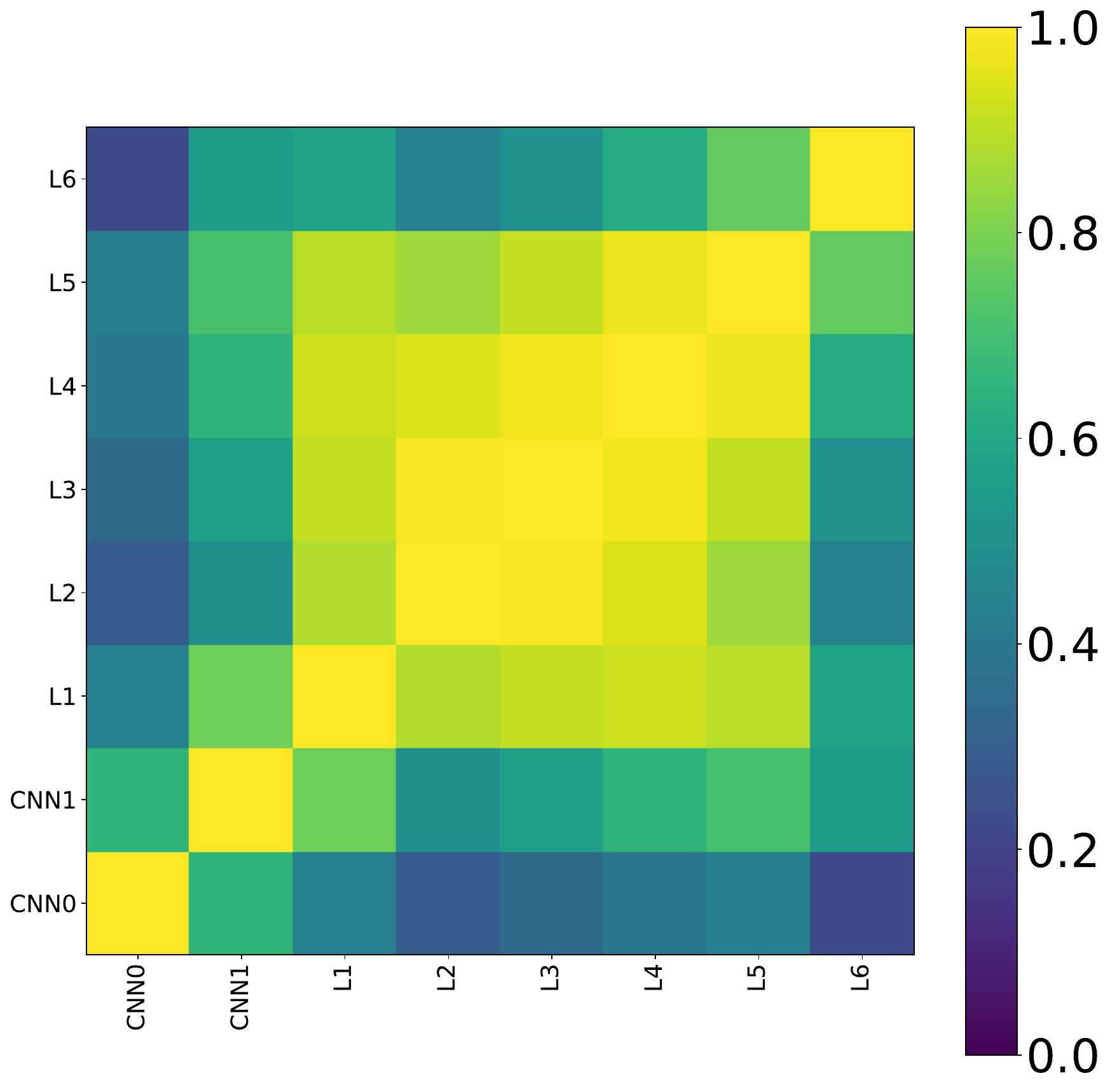}
        \caption{Transformer}
        \label{fig:cpc_753k_transformer}
    \end{subfigure}
    \hfill
    \begin{subfigure}{0.32\textwidth}
        \centering
        \includegraphics[width=\textwidth]{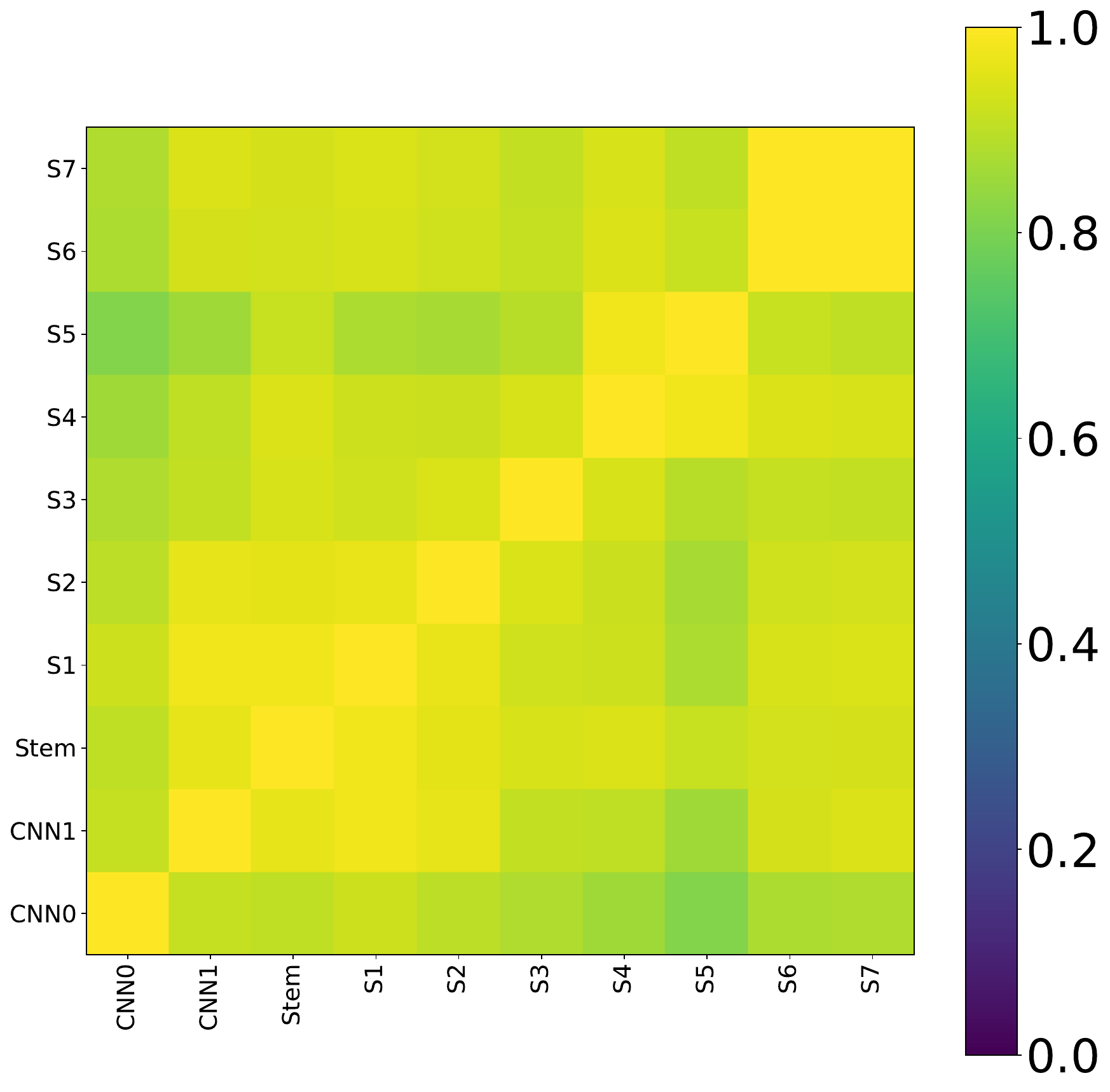}
        \caption{Net1D}
        \label{fig:cpc_753k_net1d}
    \end{subfigure}
    \caption{Intra-model layer-wise representational similarity for CPC}
    \label{fig:cpc_cka_s4_transformer_net1d}
\end{figure}

\subsubsection{HuBERT++}
\begin{figure}[htbp]
    \centering
    \begin{subfigure}{0.32\textwidth}
        \centering
        \includegraphics[width=\textwidth]{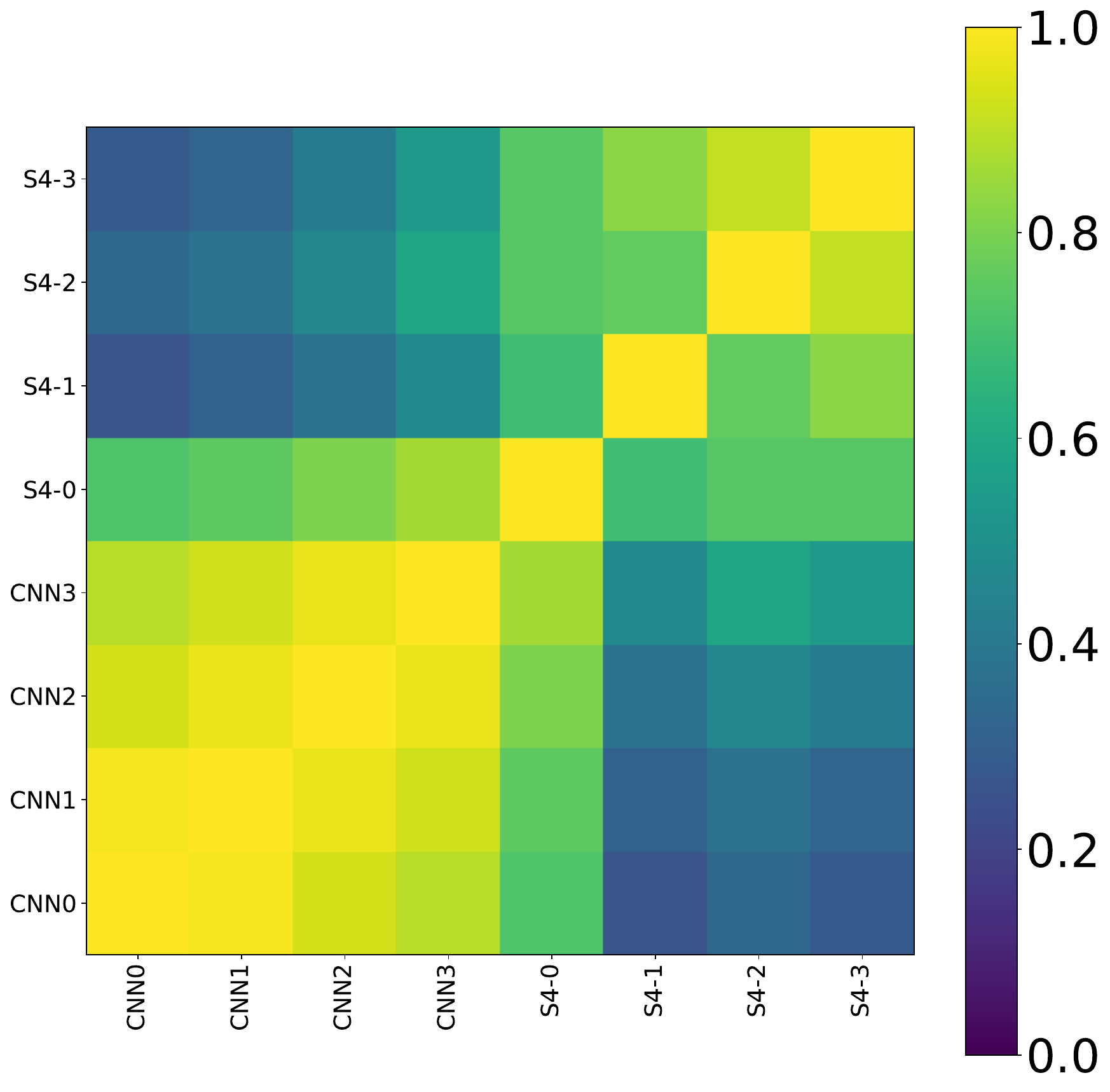}
        \caption{S4}
        \label{fig:hub_753k_s4_4}
    \end{subfigure}
    \hfill
    \begin{subfigure}{0.32\textwidth}
        \centering
        \includegraphics[width=\textwidth]{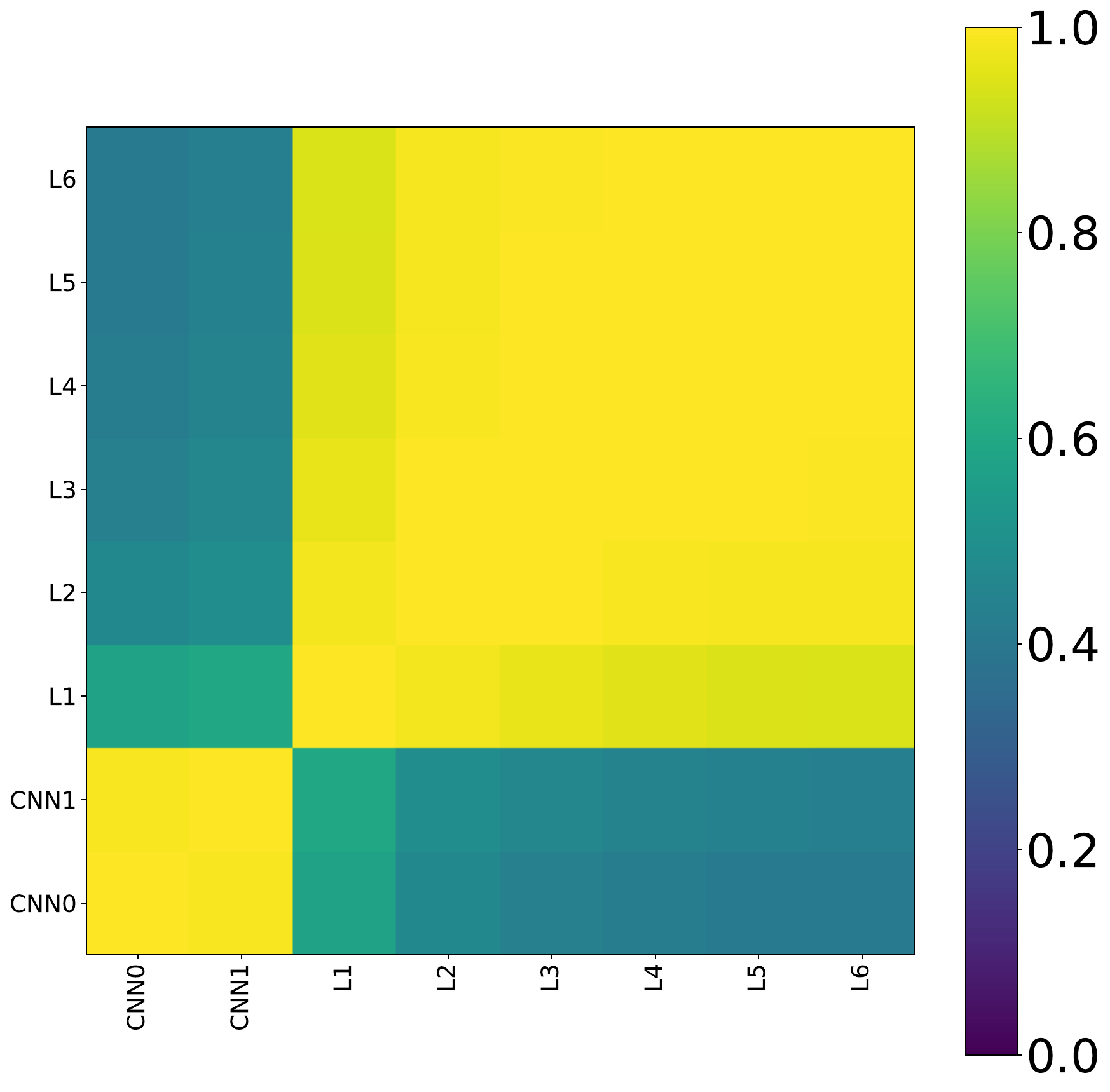}
        \caption{Transformer}
        \label{fig:hub_753k_transformer}
    \end{subfigure}
    \hfill
    \begin{subfigure}{0.32\textwidth}
        \centering
        \includegraphics[width=\textwidth]{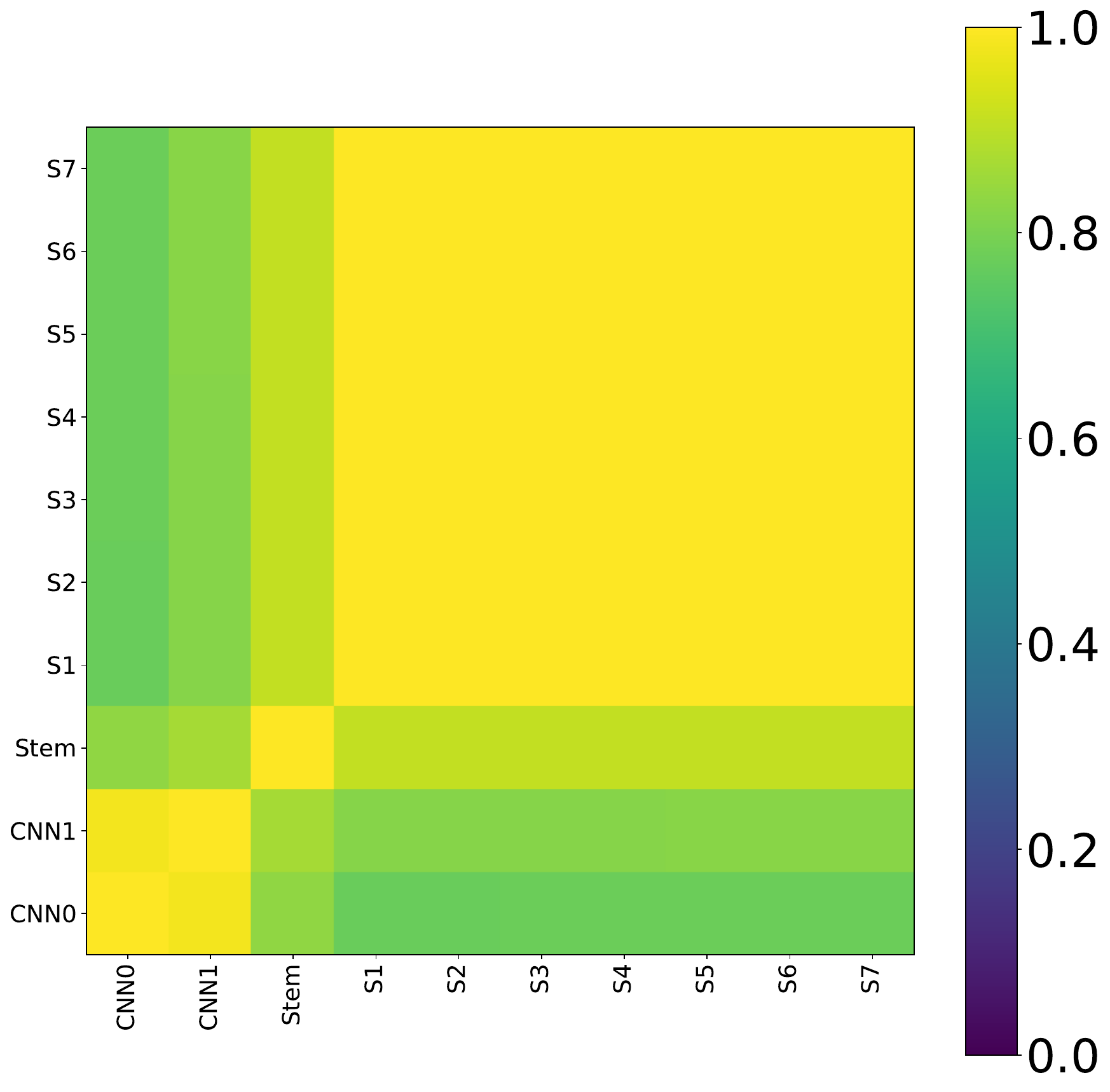}
        \caption{Net1D}
        \label{fig:hub_753k_net1d}
    \end{subfigure}
    \caption{Intra-model layer-wise representational similarity for HuBERT++}
    \label{fig:hubert_cka_s4_transformer_net1d}
\end{figure}

\newpage
\subsection{Computational Efficiency Analysis}
\label{app:backbone_comaprison_computation_efficiency_analysis}

\begin{table}[htbp]
    \centering
    \scriptsize
    \caption{Unified comparison of computational cost, memory usage, and inference efficiency for all ECG representation learning models. Reported metrics include: (i) GFLOPs for forward (F) and backward (B) passes (lower is better), measured with batch size 1 on an NVIDIA L40; (ii) peak GPU memory during inference (lower is better), measured on PTB-XL (all) with batch size 64; and (iii) throughput (samples/s; higher is better) and latency (ms/sample; lower is better) under the same hardware and batch size. Parameter counts include all trainable weights. The best model is highlighted in bold face and underlined.\\}
    \begin{tabular}{llcccc}
        \toprule
        \textbf{Model} & \textbf{Backbone} & \textbf{Parameters} $\downarrow$ & \textbf{GFLOP (F/B)} $\downarrow$ & \textbf{GPU Mem (MB)} $\downarrow$ & \textbf{Thr}$\uparrow$/ \textbf{Lat}$\downarrow$ \\
        \midrule
        \multirow{3}{*}{data2vec}
        & S4          & \underline{\textbf{3M}} & \underline{\textbf{1.741}} / \underline{\textbf{5.213}} & \underline{\textbf{482.59}} & \underline{\textbf{441.10}} / \underline{\textbf{2.267}} \\
        & Transformer & 19.2M & 27.410 / 82.207 & 1952.84 & 239.12 / 4.182 \\
        & Net1D       & 10M & 8.845 / 73.817 & 582.76 & 103.92 / 9.623 \\
        \midrule
        \multirow{3}{*}{DinoSR}
        & S4          & \underline{\textbf{3M}} & \underline{\textbf{1.741}} / \underline{\textbf{5.213}} & \underline{\textbf{482.59}} & \underline{\textbf{436.51}} / \underline{\textbf{2.291}} \\
        & Transformer & 19.2M & 27.410 / 82.207 & 1952.84 & 215.89 / 4.632 \\
        & Net1D       & 10M & 8.845 / 73.817 & 582.76 & 192.12 / 5.205 \\
        \midrule
        \multirow{3}{*}{JEPA}
        & S4          & \underline{\textbf{3M}} & \underline{\textbf{1.741}} / \underline{\textbf{5.213}} & \underline{\textbf{482.59}} & \underline{\textbf{409.39}} / \underline{\textbf{2.443}} \\
        & Transformer & 19.2M & 27.410 / 82.207 & 1952.84 & 223.23 / 4.480 \\
        & Net1D       & 10M & 8.845 / 73.817 & 582.76 & 171.55 / 5.829 \\
        \midrule
        \multirow{3}{*}{CPC}
        & S4          & \underline{\textbf{3M}} & \underline{\textbf{1.741}} / \underline{\textbf{5.213}} & \underline{\textbf{482.59}} & \underline{\textbf{401.14}} / \underline{\textbf{2.493}} \\
        & Transformer & 19.2M & 27.410 / 82.207 & 1953.18 & 229.64 / 4.355 \\
        & Net1D       & 10M & 8.845 / 73.817 & 582.76 & 186.24 / 5.369 \\
        \midrule
        \multirow{3}{*}{HuBERT++}
        & S4          & \underline{\textbf{3M}} & \underline{\textbf{1.741}} / \underline{\textbf{5.213}} & \underline{\textbf{482.59}} & \underline{\textbf{418.21}} / \underline{\textbf{2.391}} \\
        & Transformer & 19.2M & 27.410 / 82.207 & 1952.84 & 226.81 / 4.409 \\
        & Net1D       & 10M & 8.845 / 73.817 & 582.76 & 164.67 / 6.073 \\
        \bottomrule
    \end{tabular}
    \label{tab:backbone_efficiency}
\end{table}

\newpage
\section{S4 Backbone Model Dimension Ablation for SSL Pretraining}
\label{app:model_dimension_ablation}

\begin{table}[htbp]
    \centering
    \caption{Comparison of aggregated macro-AUROC for data2vec pretrained with different model dimensions (256, 512) and finetuned on downstream tasks. Models were pretrained on HEEDB subset (753K samples) for 20 epochs with an S4 backbone (6 layers). The best-performing result is highlighted in boldface and underlined, while models that do not perform statistically significantly worse are also highlighted in boldface.\\}
    \label{tab:data2vec_model_dimension_finetuning_result}
    \begin{tabular}{lcc}
        \toprule
        \multicolumn{3}{c}{\textbf{data2vec (Finetuned)}} \\
        & \textbf{256} & \textbf{512} \\
        \midrule
        
        \multicolumn{3}{c}{\textbf{Adult ECG interpretation }} \\
        Ningbo & 0.949 & \underline{\textbf{0.960}} \\
        CPSC2018 & 0.938 & \underline{\textbf{0.960}} \\
        CPSC-Extra & 0.843 & \underline{\textbf{0.853}} \\
        Georgia & 0.866 & \underline{\textbf{0.883}} \\
        Chapman & \underline{\textbf{0.931}} & \textbf{0.929} \\
        SPH & \textbf{0.959} & \underline{\textbf{0.965}} \\
        % CODE-15\% & 0.987 & \underline{\textbf{0.991}} \\
        PTB-XL (all) & \textbf{0.906} & \underline{\textbf{0.908}} \\
        PTB-XL (sub) & \textbf{0.887} & \underline{\textbf{0.897}} \\
        PTB-XL (super) & 0.906 & \underline{\textbf{0.912}} \\
        \midrule
        
        \multicolumn{3}{c}{\textbf{Pediatric ECG interpretation }} \\
        ZZU pECG & 0.837 & \underline{\textbf{0.857}} \\
        \midrule
        
        \multicolumn{3}{c}{\textbf{Cardiac structure \& function }} \\
        EchoNext (Echo) & \textbf{0.817} & \underline{\textbf{0.820}} \\
        
        \bottomrule
    \end{tabular}
    \bigskip
    
    \textit{Note:} The 512-dimensional predictor outperforms its 256-dimensional counterpart across most tasks.
\end{table}

\begin{table}[htbp]
    \centering
    \caption{Comparison of aggregated macro-AUROC for DinoSR pretrained with different model dimensions (256, 512) and finetuned on downstream tasks. Models were pretrained on HEEDB subset (753K samples) for 20 epochs with an S4 backbone (6 layers). The best-performing result is highlighted in boldface and underlined, while models that do not perform statistically significantly worse are also highlighted in boldface.\\}
    \label{tab:dinosr_model_dimension_finetuning_result}
    \begin{tabular}{lcc}
        \toprule
        \multicolumn{3}{c}{\textbf{DinoSR (Finetuned)}} \\
        & \textbf{256} & \textbf{512} \\
        \midrule
        
        \multicolumn{3}{c}{\textbf{Adult ECG interpretation }} \\
        Ningbo & 0.959 & \underline{\textbf{0.965}} \\
        CPSC2018 & \textbf{0.956} & \underline{\textbf{0.965}} \\
        CPSC-Extra & \underline{\textbf{0.866}} & \textbf{0.865} \\
        Georgia & 0.874 & \underline{\textbf{0.909}} \\
        Chapman & 0.937 & \underline{\textbf{0.948}} \\
        SPH & \underline{\textbf{0.971}} & \textbf{0.971} \\
        % CODE-15\% & \textbf{0.988} & \underline{\textbf{0.989}} \\
        PTB-XL (all) & \textbf{0.922} & \underline{\textbf{0.927}} \\
        PTB-XL (sub) & \textbf{0.922} & \underline{\textbf{0.930}} \\
        PTB-XL (super) & 0.916 & \underline{\textbf{0.920}} \\
        \midrule
        
        \multicolumn{3}{c}{\textbf{Pediatric ECG interpretation }} \\
        ZZU pECG & 0.864 & \underline{\textbf{0.878}} \\
        \midrule
        
        \multicolumn{3}{c}{\textbf{Cardiac structure \& function }} \\
        EchoNext (Echo) & 0.819 & \underline{\textbf{0.826}} \\
        
        \bottomrule
    \end{tabular}
    \bigskip
    
    \textit{Note:} The 512-dimensional predictor outperforms its 256-dimensional counterpart across most tasks.
\end{table}

\begin{table}[htbp]
    \centering
    \caption{Comparison of aggregated macro-AUROC for JEPA pretrained with different model dimensions (256, 512) and finetuned on downstream tasks. Models were pretrained on HEEDB subset (753K samples) for 20 epochs with an S4 backbone (6 layers). The best-performing result is highlighted in boldface and underlined, while models that do not perform statistically significantly worse are also highlighted in boldface.\\}
    \label{tab:jepa_model_dimension_finetuning_result}
    \begin{tabular}{lcc}
        \toprule
        \multicolumn{3}{c}{\textbf{JEPA (Finetuned)}} \\
        & \textbf{256} & \textbf{512} \\
        \midrule
        
        \multicolumn{3}{c}{\textbf{Adult ECG interpretation }} \\
        Ningbo & 0.942 & \underline{\textbf{0.962}} \\
        CPSC2018 & 0.945 & \underline{\textbf{0.961}} \\
        CPSC-Extra & \underline{\textbf{0.851}} & \textbf{0.846} \\
        Georgia & \underline{\textbf{0.882}} & \textbf{0.873} \\
        Chapman & \underline{\textbf{0.935}} & \textbf{0.934} \\
        SPH & 0.949 & \underline{\textbf{0.965}} \\
        % CODE-15\% & \textbf{0.989} & \underline{\textbf{0.991}} \\
        PTB-XL (all) & 0.901 & \underline{\textbf{0.925}} \\
        PTB-XL (sub) & \underline{\textbf{0.907}} & \textbf{0.905} \\
        PTB-XL (super) & \underline{\textbf{0.913}} & \textbf{0.912} \\
        \midrule
        
        \multicolumn{3}{c}{\textbf{Pediatric ECG interpretation }} \\
        ZZU pECG & \textbf{0.839} & \underline{\textbf{0.845}} \\
        \midrule
        
        \multicolumn{3}{c}{\textbf{Cardiac structure \& function }} \\
        EchoNext & \textbf{0.814} & \underline{\textbf{0.817}} \\
        
        \bottomrule
    \end{tabular}
    \bigskip
    
    \textit{Note:} The 512-dimensional predictor outperforms its 256-dimensional counterpart across most tasks.
\end{table}

\begin{table}[htbp]
    \centering
    \caption{Comparison of aggregated macro-AUROC for CPC pretrained with different model dimensions (256, 512) and finetuned on downstream tasks. Models were pretrained on HEEDB subset (753K samples) for 20 epochs with an S4 backbone (6 layers). The best-performing result is highlighted in boldface and underlined, while models that do not perform statistically significantly worse are also highlighted in boldface.\\}
    \label{tab:cpc_model_dimension_finetuning_result}
    \begin{tabular}{lcc}
        \toprule
        \multicolumn{3}{c}{\textbf{CPC (Finetuned)}} \\
        & \textbf{256} & \textbf{512} \\
        \midrule
        
        \multicolumn{3}{c}{\textbf{Adult ECG interpretation }} \\
        Ningbo & 0.963 & \underline{\textbf{0.971}} \\
        CPSC2018 & \underline{\textbf{0.969}} & \textbf{0.969} \\
        CPSC-Extra & \underline{\textbf{0.901}} & \textbf{0.882} \\
        Georgia & 0.892 & \underline{\textbf{0.904}} \\
        Chapman & \textbf{0.955} & \underline{\textbf{0.958}} \\
        SPH & 0.976 & \underline{\textbf{0.979}} \\
        % CODE-15\% & 0.990 & \underline{\textbf{0.991}} \\
        PTB-XL (all) & \underline{\textbf{0.936}} & \textbf{0.936} \\
        PTB-XL (sub) & \textbf{0.929} & \underline{\textbf{0.933}} \\
        PTB-XL (super) & 0.918 & \underline{\textbf{0.921}} \\
        \midrule
        
        \multicolumn{3}{c}{\textbf{Pediatric ECG interpretation }} \\
        ZZU pECG & \underline{\textbf{0.883}} & \textbf{0.876} \\
        \midrule
        
        \multicolumn{3}{c}{\textbf{Cardiac structure \& function }} \\
        EchoNext & \underline{\textbf{0.825}} & \textbf{0.824} \\
        
        \bottomrule
    \end{tabular}
    \bigskip
    
    \textit{Note:} The 512-dimensional predictor outperforms its 256-dimensional counterpart across most tasks.
\end{table}

\begin{table}[htbp]
    \centering
    \caption{Comparison of aggregated macro-AUROC for HuBERT++ pretrained with different model dimensions (256, 512) and finetuned on downstream tasks. Models were pretrained on HEEDB subset (753K samples) for 20 epochs with an S4 backbone (6 layers). The best-performing result is highlighted in boldface and underlined, while models that do not perform statistically significantly worse are also highlighted in boldface.\\}
    \label{tab:hubert_model_dimension_finetuning_result}
    \begin{tabular}{lcc}
        \toprule
        \multicolumn{3}{c}{\textbf{HuBERT++ (Finetuned)}} \\
        & \textbf{256} & \textbf{512} \\
        \midrule
        
        \multicolumn{3}{c}{\textbf{Adult ECG interpretation }} \\
        Ningbo & 0.961 & \underline{\textbf{0.970}} \\
        CPSC2018 & 0.948 & \underline{\textbf{0.972}} \\
        CPSC-Extra & 0.859 & \underline{\textbf{0.884}} \\
        Georgia & 0.875 & \underline{\textbf{0.906}} \\
        Chapman & 0.933 & \underline{\textbf{0.955}} \\
        SPH & \textbf{0.962} & \underline{\textbf{0.979}} \\
        % CODE-15\% & \textbf{0.988} & \underline{\textbf{0.991}} \\
        PTB-XL (all) & 0.918 & \underline{\textbf{0.932}} \\
        PTB-XL (sub) & 0.902 & \underline{\textbf{0.931}} \\
        PTB-XL (super) & 0.904 & \underline{\textbf{0.919}} \\
        \midrule
        
        \multicolumn{3}{c}{\textbf{Pediatric ECG interpretation }} \\
        ZZU pECG & 0.850 & \underline{\textbf{0.898}} \\
        \midrule
        
        \multicolumn{3}{c}{\textbf{Cardiac structure \& function }} \\
        EchoNext & 0.812 & \underline{\textbf{0.824}} \\
        
        \bottomrule
    \end{tabular}
    \bigskip
    
    \textit{Note:} The 512-dimensional predictor outperforms its 256-dimensional counterpart across most tasks.
\end{table}

\newpage
\section{Pretraining Dataset Comparison: HEEDB vs.\ MIMIC-IV-ECG}
\label{app:heedb_mimic_240_comparison}

\begin{table}[htbp]
    \centering
    \caption{Comparison of aggregated macro-AUROC for data2vec pretrained with different datasets (\textbf{HEEDB:} 753K-sample subset at 240 Hz; \textbf{MIMIC:} MIMIC-IV-ECG with 759K samples, downsampled to 240 Hz) and finetuned on downstream tasks. Models were pretrained for 20 epochs with an S4 backbone (6 layers). The best-performing result is highlighted in boldface and underlined, while models that do not perform statistically significantly worse are also highlighted in boldface.\\}
    \label{tab:data2vec_heedb_mimic_240_comparison}
    \begin{tabular}{lcc}
        \toprule
        \multicolumn{3}{c}{\textbf{data2vec (Finetuned)}} \\
        & \textbf{HEEDB} & \textbf{MIMIC} \\
        \midrule
        
        \multicolumn{3}{c}{\textbf{Adult ECG interpretation }} \\
        Ningbo & \textbf{0.960} & \underline{\textbf{0.962}} \\
        CPSC2018 & \textbf{0.960} & \underline{\textbf{0.961}} \\
        CPSC-Extra & \textbf{0.853} & \underline{\textbf{0.855}} \\
        Georgia & \textbf{0.883} & \underline{\textbf{0.886}} \\
        Chapman & \textbf{0.929} & \underline{\textbf{0.932}} \\
        SPH & \underline{\textbf{0.965}} & \textbf{0.963} \\
        % CODE-15\% & \underline{\textbf{0.991}} & \textbf{0.990} \\
        PTB-XL (all) & \textbf{0.908} & \underline{\textbf{0.910}} \\
        PTB-XL (sub) & \textbf{0.897} & \underline{\textbf{0.909}} \\
        PTB-XL (super) & \textbf{0.912} & \underline{\textbf{0.913}} \\
        \midrule
        
        \multicolumn{3}{c}{\textbf{Pediatric ECG interpretation }} \\
        ZZU pECG & \underline{\textbf{0.857}} & 0.848 \\
        \midrule
        
        \multicolumn{3}{c}{\textbf{Cardiac structure \& function }} \\
        EchoNext (Echo) & \textbf{0.820} & \underline{\textbf{0.821}} \\
        
        \bottomrule            
    \end{tabular}
    \bigskip
    
    \textit{Note:} MIMIC-pretrained models slightly outperform HEEDB-pretrained models across most datasets.
\end{table}

\begin{table}[htbp]
    \centering
    \caption{Comparison of aggregated macro-AUROC for DinoSR pretrained with different datasets (\textbf{HEEDB:} 753K-sample subset at 240 Hz; \textbf{MIMIC:} MIMIC-IV-ECG with 759K samples, downsampled to 240 Hz) and finetuned on downstream tasks. Models were pretrained for 20 epochs with an S4 backbone (6 layers). The best-performing result is highlighted in boldface and underlined, while models that do not perform statistically significantly worse are also highlighted in boldface.\\}
    \label{tab:dinosr_heedb_mimic_240_comparison}
    \begin{tabular}{lcc}
        \toprule
        \multicolumn{3}{c}{\textbf{DinoSR (Finetuned)}} \\
        & \textbf{HEEDB} & \textbf{MIMIC} \\
        \midrule
        
        \multicolumn{3}{c}{\textbf{Adult ECG interpretation }} \\
        Ningbo & 0.965 & \underline{\textbf{0.968}} \\
        CPSC2018 & \textbf{0.965} & \underline{\textbf{0.967}} \\
        CPSC-Extra & \underline{\textbf{0.865}} & \textbf{0.864} \\
        Georgia & \textbf{0.909} & \underline{\textbf{0.913}} \\
        Chapman & \textbf{0.948} & \underline{\textbf{0.951}} \\
        SPH & \textbf{0.971} & \underline{\textbf{0.978}} \\
        % CODE-15\% & \underline{\textbf{0.989}} & \textbf{0.989} \\
        PTB-XL (all) & 0.927 & \underline{\textbf{0.939}} \\
        PTB-XL (sub) & \textbf{0.930} & \underline{\textbf{0.933}} \\
        PTB-XL (super) & \textbf{0.920} & \underline{\textbf{0.921}} \\
        \midrule
        
        \multicolumn{3}{c}{\textbf{Pediatric ECG interpretation }} \\
        ZZU pECG & \textbf{0.878} & \underline{\textbf{0.879}} \\
        \midrule
        
        \multicolumn{3}{c}{\textbf{Cardiac structure \& function }} \\
        EchoNext (Echo) & \textbf{0.826} & \underline{\textbf{0.830}} \\
        
        \bottomrule            
    \end{tabular}
    \bigskip
    
    \textit{Note:} MIMIC-pretrained models slightly outperform HEEDB-pretrained models across most datasets.
\end{table}

\begin{table}[htbp]
    \centering
    \caption{Comparison of aggregated macro-AUROC for JEPA pretrained with different datasets (\textbf{HEEDB:} 753K-sample subset at 240 Hz; \textbf{MIMIC:} MIMIC-IV-ECG with 759K samples, downsampled to 240 Hz) and finetuned on downstream tasks. Models were pretrained for 20 epochs with an S4 backbone (6 layers). The best-performing result is highlighted in boldface and underlined, while models that do not perform statistically significantly worse are also highlighted in boldface.\\}
    \label{tab:jepa_heedb_mimic_240_comparison}
    \begin{tabular}{lcc}
        \toprule
        \multicolumn{3}{c}{\textbf{JEPA (Finetuned)}} \\
        & \textbf{HEEDB} & \textbf{MIMIC} \\
        \midrule
        
        \multicolumn{3}{c}{\textbf{Adult ECG interpretation }} \\
        Ningbo & \underline{\textbf{0.962}} & 0.955 \\
        CPSC2018 & \textbf{0.961} & \underline{\textbf{0.965}} \\
        CPSC-Extra & \textbf{0.846} & \underline{\textbf{0.868}} \\
        Georgia & \textbf{0.873} & \underline{\textbf{0.881}} \\
        Chapman & \underline{\textbf{0.934}} & \textbf{0.934} \\
        SPH & \textbf{0.965} & \underline{\textbf{0.974}} \\
        % CODE-15\% & \textbf{0.991} & \underline{\textbf{0.991}} \\
        PTB-XL (all) & \underline{\textbf{0.925}} & 0.920 \\
        PTB-XL (sub) & \textbf{0.905} & \underline{\textbf{0.910}} \\
        PTB-XL (super) & \textbf{0.912} & \underline{\textbf{0.913}} \\
        \midrule
        
        \multicolumn{3}{c}{\textbf{Pediatric ECG interpretation }} \\
        ZZU pECG & 0.845 & \underline{\textbf{0.875}} \\
        \midrule
        
        \multicolumn{3}{c}{\textbf{Cardiac structure \& function }} \\
        EchoNext (Echo) & 0.817 & \underline{\textbf{0.822}} \\
        
        \bottomrule            
    \end{tabular}
    \bigskip
    
    \textit{Note:} MIMIC-pretrained models slightly outperform HEEDB-pretrained models across most datasets.
\end{table}

\begin{table}[htbp]
    \centering
    \caption{Comparison of aggregated macro-AUROC for CPC pretrained with different datasets (\textbf{HEEDB:} 753K-sample subset at 240 Hz; \textbf{MIMIC:} MIMIC-IV-ECG with 759K samples, downsampled to 240 Hz) and finetuned on downstream tasks. Models were pretrained for 20 epochs with an S4 backbone (6 layers). The best-performing result is highlighted in boldface and underlined, while models that do not perform statistically significantly worse are also highlighted in boldface.\\}
    \label{tab:cpc_heedb_mimic_240_comparison}
    \begin{tabular}{lcc}
        \toprule
        \multicolumn{3}{c}{\textbf{CPC (Finetuned)}} \\
        & \textbf{HEEDB} & \textbf{MIMIC} \\
        \midrule
        
        \multicolumn{3}{c}{\textbf{Adult ECG interpretation }} \\
        Ningbo & \textbf{0.971} & \underline{\textbf{0.973}} \\
        CPSC2018 & \textbf{0.969} & \underline{\textbf{0.973}} \\
        CPSC-Extra & 0.882 & \underline{\textbf{0.900}} \\
        Georgia & \textbf{0.904} & \underline{\textbf{0.904}} \\
        Chapman & \textbf{0.958} & \underline{\textbf{0.959}} \\
        SPH & \textbf{0.979} & \underline{\textbf{0.979}} \\
        % CODE-15\% & \underline{\textbf{0.991}} & \textbf{0.989} \\
        PTB-XL (all) & 0.936 & \underline{\textbf{0.943}} \\
        PTB-XL (sub) & \textbf{0.933} & \underline{\textbf{0.933}} \\
        PTB-XL (super) & \underline{\textbf{0.921}} & \textbf{0.920} \\
        \midrule
        
        \multicolumn{3}{c}{\textbf{Pediatric ECG interpretation }} \\
        ZZU pECG & 0.876 & \underline{\textbf{0.892}} \\
        \midrule
        
        \multicolumn{3}{c}{\textbf{Cardiac structure \& function }} \\
        EchoNext (Echo) & \textbf{0.824} & \underline{\textbf{0.829}} \\
        
        \bottomrule            
    \end{tabular}
    \bigskip
    
    \textit{Note:} MIMIC-pretrained models slightly outperform HEEDB-pretrained models across most datasets.
\end{table}

\begin{table}[htbp]
    \centering
    \caption{Comparison of aggregated macro-AUROC for HuBERT++ pretrained with different datasets (\textbf{HEEDB:} 753K-sample subset at 240 Hz; \textbf{MIMIC:} MIMIC-IV-ECG with 759K samples, downsampled to 240 Hz) and finetuned on downstream tasks. Models were pretrained for 20 epochs with an S4 backbone (6 layers). The best-performing result is highlighted in boldface and underlined, while models that do not perform statistically significantly worse are also highlighted in boldface.\\}
    \label{tab:hubert_heedb_mimic_240_comparison}
    \begin{tabular}{lcc}
        \toprule
        \multicolumn{3}{c}{\textbf{HuBERT++ (Finetuned)}} \\
        & \textbf{HEEDB} & \textbf{MIMIC} \\
        \midrule
        
        \multicolumn{3}{c}{\textbf{Adult ECG interpretation }} \\
        Ningbo & 0.970 & \underline{\textbf{0.973}} \\
        CPSC2018 & \textbf{0.972} & \underline{\textbf{0.973}} \\
        CPSC-Extra & \textbf{0.884} & \underline{\textbf{0.890}} \\
        Georgia & \underline{\textbf{0.906}} & \textbf{0.900} \\
        Chapman & \textbf{0.955} & \underline{\textbf{0.960}} \\
        SPH & \underline{\textbf{0.979}} & \textbf{0.977} \\
        % CODE-15\% & \underline{\textbf{0.991}} & \textbf{0.991} \\
        PTB-XL (all) & \textbf{0.932} & \underline{\textbf{0.937}} \\
        PTB-XL (sub) & \underline{\textbf{0.931}} & \textbf{0.929} \\
        PTB-XL (super) & \underline{\textbf{0.919}} & \textbf{0.917} \\
        \midrule
        
        \multicolumn{3}{c}{\textbf{Pediatric ECG interpretation }} \\
        ZZU pECG & \textbf{0.898} & \underline{\textbf{0.901}} \\
        \midrule
        
        \multicolumn{3}{c}{\textbf{Cardiac structure \& function }} \\
        EchoNext (Echo) & \textbf{0.824} & \underline{\textbf{0.826}} \\
        
        \bottomrule            
    \end{tabular}
    \bigskip
    
    \textit{Note:} MIMIC-pretrained models slightly outperform HEEDB-pretrained models across most datasets.
\end{table}

\newpage
\section{JEPA SSL Head Ablation}
\label{app:jepa_ssl_head_ablation}

\begin{table}[htbp]
    \centering
    \caption{Comparison of aggregated macro-AUROC for JEPA pretrained with different S4 SSL head configurations (\textbf{1 Layer:} 1 S4 layers, no linear layer; \textbf{3 Layers:} 3 S4 layers, model dimension = 256; \textbf{6 Layers:} 6 S4 layers, model dimension = 256). Models were pretrained with HEEDB subset (106K samples) for 20 epochs with an S4 backbone (6 layers). The best-performing result is highlighted in boldface and underlined, while models that do not perform statistically significantly worse are also highlighted in boldface.\\}
    \label{tab:jepa_ssl_head_result}
    \begin{tabular}{lccc}
        \toprule
        \multicolumn{4}{c}{\textbf{JEPA (Finetuned)}} \\
        & \textbf{1 Layer} & \textbf{3 Layers} & \textbf{6 Layers} \\
        \midrule
        
        \multicolumn{4}{c}{\textbf{Adult ECG interpretation }} \\
        Ningbo & \underline{\textbf{0.956}} & 0.950 & 0.952 \\
        CPSC2018 & \textbf{0.955} & \underline{\textbf{0.956}} & 0.949 \\
        CPSC-Extra & 0.839 & \underline{\textbf{0.854}} & \textbf{0.847} \\
        Georgia & \textbf{0.875} & \underline{\textbf{0.879}} & 0.866 \\
        Chapman & \underline{\textbf{0.929}} & \textbf{0.928} & \textbf{0.927} \\
        SPH & \underline{\textbf{0.965}} & \textbf{0.958} & \textbf{0.953} \\
        % CODE-15\% & \underline{\textbf{0.990}} & \textbf{0.989} & \textbf{0.988} \\
        PTB-XL (all) & \textbf{0.908} & \underline{\textbf{0.910}} & \textbf{0.904} \\
        PTB-XL (sub) & \underline{\textbf{0.891}} & \textbf{0.890} & \textbf{0.880} \\
        PTB-XL (super) & \underline{\textbf{0.910}} & 0.905 & 0.907 \\
        \midrule
        
        \multicolumn{4}{c}{\textbf{Pediatric ECG interpretation }} \\
        ZZU pECG & 0.827 & \underline{\textbf{0.847}} & 0.837 \\
        \midrule
        
        \multicolumn{4}{c}{\textbf{Cardiac structure \& function }} \\
        EchoNext & \textbf{0.817} & \underline{\textbf{0.821}} & \textbf{0.817} \\
        
        \bottomrule
    \end{tabular}
    \bigskip
    
    \textit{Note:} The 1-layer SSL head performs best overall.
\end{table}

\newpage
\section{Finetuning Results (MIMIC-IV-ECG Pretrained)}
\label{app:mimic_iv_ecg_finetuning}

\begin{table}[htbp]
    \centering
    \standardfootnotesize
    \caption{Comparison of aggregated macro-AUROC under finetuning evaluation. Models were pretrained on MIMIC-IV-ECG dataset (759K samples) at 500 Hz for 20 epochs with learning rate 3e-3 using an S4 backbone (6 layers). MERL~\citep{liu2024zero} and ECGFM-KED~\citep{tian2024foundation} are external FMs pretrained on MIMIC-IV-ECG for 50 and 4 epochs with learning rates 2e-4 and 5e-5, respectively. The best-performing result is highlighted in boldface and underlined, while models that do not perform statistically significantly worse are also highlighted in boldface.\\}
    \label{tab:mimic_finetuning_500}
    \begin{tabular}{lcccccccc}
        \toprule
        & \multicolumn{2}{c}{\textbf{FMs (Finetuned)}} & \multicolumn{5}{c}{\textbf{Pretrained by Us (Finetuned)}} & \textbf{Supervised} \\
        \cmidrule(l){2-3}\cmidrule(l){4-8}\cmidrule(l){9-9}
        & \textbf{MERL} & \textbf{KED} & \textbf{Data2Vec} & \textbf{DinoSR} & \textbf{JEPA} & \textbf{CPC} & \textbf{HuBERT++} & \textbf{S4} \\
        \midrule
        
        \multicolumn{9}{c}{\textbf{Adult ECG interpretation }} \\
        Ningbo & 0.955 & 0.940 & 0.957 & \textbf{0.966} & 0.962 & \textbf{0.970} & \textbf{0.972} & \underline{\textbf{0.972}}  \\
        CPSC2018 & 0.936 & 0.930 & 0.957 & 0.965 & \textbf{0.966} & \underline{\textbf{0.971}} & \textbf{0.971} & 0.962 \\
        CPSC-Extra & 0.873 & 0.824 & 0.835 & 0.864 & 0.866 & \underline{\textbf{0.908}} & \textbf{0.895} & 0.852 \\
        Georgia & \textbf{0.912} & 0.877 & 0.873 & \textbf{0.913} & 0.889 & \textbf{0.911} & \underline{\textbf{0.917}} & \textbf{0.903} \\
        Chapman & 0.946 & 0.917 & 0.932 & 0.950 & 0.937 & 0.956 & \textbf{0.957} & \underline{\textbf{0.963}} \\
        SPH & 0.944 & 0.932 & 0.967 & 0.969 & 0.968 & \underline{\textbf{0.982}} & \textbf{0.980} & \textbf{0.981} \\
        % CODE-15\% & 0.982 & 0.964 & \textbf{0.990} & \textbf{0.990} & \textbf{0.990} & \textbf{0.990} & \underline{\textbf{0.991}} & \textbf{0.990} \\
        PTB-XL (all) & 0.925 & 0.889 & 0.915 & 0.934 & 0.924 & \textbf{0.940} & \textbf{0.937} & \underline{\textbf{0.941}} \\
        PTB-XL (sub) & \textbf{0.937} & 0.908 & 0.896 & \textbf{0.934} & 0.905 & 0.931 & 0.932 & \underline{\textbf{0.938}} \\
        PTB-XL (super) & \textbf{0.930} & 0.905 & 0.909 & 0.920 & 0.910 & 0.920 & 0.918 & \underline{\textbf{0.932}} \\
        \midrule
        
        \multicolumn{9}{c}{\textbf{Pediatric ECG interpretation }} \\
        ZZU pECG & \textbf{0.886} & 0.861 & 0.846 & 0.878 & 0.870 & \textbf{0.886} & \textbf{0.887} & \underline{\textbf{0.897}} \\
        \midrule
        
        \multicolumn{9}{c}{\textbf{Cardiac structure \& function }} \\
        EchoNext (Echo) & \textbf{0.822} & 0.806 & 0.817 & \textbf{0.825} & 0.817 & \underline{\textbf{0.828}} & \textbf{0.826} & 0.819 \\
        \bottomrule
    \end{tabular}

    \bigskip    
    \textit{Note:} CPC performs better overall. Both external FMs perform poorly.
\end{table}

\newpage
\section{Finetuning Results (HEEDB Subset Pretrained)}
\label{app:heedb_subset_finetuning}

\subsection{Small-Scale Pretraining (106K Samples)}
\label{app:heedb_106k}

\begin{table}[htbp]
    \centering
    \caption{Comparison of aggregated macro-AUROC under finetuning evaluation. Models were pretrained on HEEDB subset (106K samples) for 10 epochs using an S4 backbone (4 layers).  The best-performing result is highlighted in boldface and underlined, while models that do not perform statistically significantly worse are also highlighted in boldface.\\}
    \label{tab:heedb_106k_finetuning}
    \begin{tabular}{lcccccc}
        \toprule
        & \multicolumn{5}{c}{\textbf{Pretrained Models (Finetuned)}} & \textbf{Supervised} \\
        \cmidrule(l){2-6}\cmidrule(l){7-7}
        & \textbf{Data2Vec} & \textbf{DinoSR} & \textbf{JEPA} & \textbf{CPC} & \textbf{HuBERT++} & \textbf{S4} \\
        \midrule
        
        \multicolumn{7}{c}{\textbf{Adult ECG interpretation }} \\
        Ningbo & 0.959 & 0.961 & 0.961 & 0.955 & 0.965 & \underline{\textbf{0.972}} \\
        CPSC2018 & 0.959 & \textbf{0.964} & \textbf{0.966} & \textbf{0.964} & \underline{\textbf{0.966}} & \textbf{0.962} \\
        CPSC-Extra & 0.823 & 0.846 & 0.868 & \underline{\textbf{0.884}} & \textbf{0.870} & 0.852 \\
        Georgia & 0.890 & \textbf{0.903} & \textbf{0.908} & 0.900 & \underline{\textbf{0.909}} & \textbf{0.903} \\
        Chapman & 0.944 & 0.944 & 0.948 & 0.944 & 0.950 & \underline{\textbf{0.963}} \\
        SPH & 0.968 & \textbf{0.974} & 0.968 & 0.972 & \textbf{0.978} & \underline{\textbf{0.981}} \\
        PTB-XL (all) & 0.906 & 0.926 & 0.933 & 0.921 & 0.923 & \underline{\textbf{0.941}} \\
        PTB-XL (sub) & 0.926 & 0.925 & 0.927 & 0.925 & \textbf{0.933} & \underline{\textbf{0.938}} \\
        PTB-XL (super) & 0.923 & 0.927 & 0.924 & 0.924 & 0.928 & \underline{\textbf{0.932}} \\
        \midrule
        
        \multicolumn{7}{c}{\textbf{Pediatric ECG interpretation }} \\
        ZZU pECG & 0.865 & 0.872 & 0.883 & \textbf{0.886} & 0.881 & \underline{\textbf{0.897}} \\
        \midrule
        
        \multicolumn{7}{c}{\textbf{Cardiac structure \& function }} \\
        EchoNext & \textbf{0.823} & 0.820 & \textbf{0.823} & \underline{\textbf{0.826}} & \textbf{0.822} & \textbf{0.819} \\
        \bottomrule
    \end{tabular}
    
    \bigskip
\end{table}

\newpage
\subsection{Medium-Scale Pretraining (753K Samples)}
\label{app:heedb_753k}

\begin{table}[htbp]
    \centering
    \caption{Comparison of aggregated macro-AUROC under finetuning evaluation. Models were pretrained on HEEDB subset (753K samples) for 10 epochs using an S4 backbone (4 layers). The best-performing result is highlighted in boldface and underlined, while models that do not perform statistically significantly worse are also highlighted in boldface.\\}
    \label{tab:heedb_753k_finetuning}
    \begin{tabular}{lcccccc}
        \toprule
        & \multicolumn{5}{c}{\textbf{Pretrained Models (Finetuned)}} & \textbf{Supervised} \\
        \cmidrule(l){2-6}\cmidrule(l){7-7}
        & \textbf{Data2Vec} & \textbf{DinoSR} & \textbf{JEPA} & \textbf{CPC} & \textbf{HuBERT++} & \textbf{S4} \\
        \midrule
        
        \multicolumn{7}{c}{\textbf{Adult ECG interpretation }} \\
        Ningbo & 0.965 & 0.964 & 0.971 & \underline{\textbf{0.974}} & 0.970 & \textbf{0.972} \\
        CPSC2018 & 0.960 & \textbf{0.968} & \textbf{0.970} & \underline{\textbf{0.972}} & \textbf{0.970} & 0.962 \\
        CPSC-Extra & 0.839 & 0.883 & 0.888 & \underline{\textbf{0.906}} & \textbf{0.891} & 0.852 \\
        Georgia & 0.894 & \textbf{0.906} & \textbf{0.913} & \textbf{0.912} & \underline{\textbf{0.915}} & \textbf{0.903} \\
        Chapman & 0.946 & \textbf{0.960} & 0.957 & \textbf{0.959} & \textbf{0.958} & \underline{\textbf{0.963}} \\
        SPH & \textbf{0.976} & 0.975 & \textbf{0.976} & \underline{\textbf{0.982}} & \textbf{0.982} & \textbf{0.981} \\
        PTB-XL (all) & 0.924 & 0.935 & \underline{\textbf{0.943}} & \textbf{0.943} & \textbf{0.939} & \textbf{0.941} \\
        PTB-XL (sub) & 0.934 & \textbf{0.938} & \textbf{0.924} & \textbf{0.937} & \underline{\textbf{0.940}} & \textbf{0.938} \\
        PTB-XL (super) & 0.926 & \textbf{0.932} & \textbf{0.932} & \textbf{0.932} & \textbf{0.930} & \underline{\textbf{0.932}} \\
        \midrule
        
        \multicolumn{7}{c}{\textbf{Pediatric ECG interpretation }} \\
        ZZU pECG & 0.868 & 0.888 & 0.891 & \underline{\textbf{0.899}} & \textbf{0.896} & \textbf{0.897} \\
        \midrule
        
        \multicolumn{7}{c}{\textbf{Cardiac structure \& function }} \\
        EchoNext (Echo) & 0.816 & \textbf{0.829} & 0.826 & \underline{\textbf{0.832}} & \textbf{0.830} & 0.819 \\
        \bottomrule
    \end{tabular}
    
\end{table}

\newpage
\section{Finetuning Strategy Comparison}
\label{app:finetuning_strategy_comparison}

\begin{table}[htbp]
    \centering
    \caption{Comparison of aggregated macro-AUROC for CPC pretrained with different finetuning strategies (\textbf{Linear:} finetuning with a linear head; \textbf{Non-linear:} finetuning with a learnable query attention pooling head; \textbf{Domain Pretrain + Linear:} continued pretraining on downstream domain datasets following initial pretraining, then finetuned with a linear head). Models were pretrained with HEEDB, Emory and CODE-15\% dataset (11M samples) for 20 epochs with an S4  backbone (6 layers, deviating from the 4-layer model used in the main text). The best-performing result is highlighted in boldface and underlined, while models that do not perform statistically significantly worse are also highlighted in boldface.\\}
    \label{tab:cpc_finetuning_strategy_result}
    \begin{tabular}{lccc}
        \toprule
        \multicolumn{4}{c}{\textbf{CPC (Finetuned)}} \\
        & \textbf{Linear} & \textbf{Non-linear} & \textbf{Domain Pretrain + Linear} \\
        \midrule
        
        \multicolumn{4}{c}{\textbf{Adult ECG interpretation }} \\
        Ningbo & \underline{\textbf{0.971}} & 0.963 & \textbf{0.967} \\
        CPSC2018 & 0.968 & 0.961 & \underline{\textbf{0.974}} \\
        CPSC-Extra & \textbf{0.892} & 0.867 & \underline{\textbf{0.903}} \\
        Georgia & 0.899 & 0.889 & \underline{\textbf{0.909}} \\
        Chapman & \textbf{0.959} & 0.946 & \underline{\textbf{0.963}} \\
        SPH & \underline{\textbf{0.981}} & 0.973 & \textbf{0.977} \\
        PTB-XL (all) & \textbf{0.934} & 0.921 & \underline{\textbf{0.937}} \\
        PTB-XL (sub) & 0.921 & 0.919 & \underline{\textbf{0.934}} \\
        PTB-XL (super) & 0.916 & 0.914 & \underline{\textbf{0.921}} \\       
        
        \midrule
        
        \multicolumn{4}{c}{\textbf{Pediatric ECG interpretation }} \\
        ZZU pECG & 0.880 & \textbf{0.885} & \underline{\textbf{0.894}} \\
        \midrule
        
        \multicolumn{4}{c}{\textbf{Cardiac structure \& function }} \\
        EchoNext & 0.820 & 0.813 & \underline{\textbf{0.829}} \\
        
        \bottomrule
    \end{tabular}
    \bigskip
    
    \textit{Note:} Domain Pretrain + Linear achieves the best performance across most datasets, highlighting the benefit of domain-adaptive pretraining. The non-linear head offers no consistent advantage over the simpler linear head.
\end{table}

\newpage
\begin{table}[htbp]
    \centering
    \caption{Comparison of aggregated macro-AUROC for HuBERT++ pretrained with different finetuning strategies (\textbf{Linear:} finetuning with a linear head; \textbf{Non-linear:} finetuning with a learnable query attention pooling head; \textbf{Domain Pretrain + Linear:} continued pretraining on downstream domain datasets for 20 epochs following initial pretraining, then finetuned with a linear head). Models were pretrained with HEEDB, Emory and CODE-15\% dataset (11M samples) for 20 epochs with an S4 backbone (6 layers). The best-performing result is highlighted in boldface and underlined, while models that do not perform statistically significantly worse are also highlighted in boldface.\\}
    \label{tab:hubert_finetuning_strategy_result}
    \begin{tabular}{lccc}
        \toprule
        \multicolumn{4}{c}{\textbf{HuBERT++ (Finetuned)}} \\
        & \textbf{Linear} & \textbf{Non-linear} & \textbf{Domain Pretrain + Linear} \\
        \midrule
        
        \multicolumn{4}{c}{\textbf{Adult ECG interpretation }} \\
        Ningbo & \textbf{0.973} & 0.964 & \underline{\textbf{0.975}} \\
        CPSC2018 & 0.968 & 0.968 & \underline{\textbf{0.974}} \\
        CPSC-Extra & \textbf{0.896} & 0.863 & \underline{\textbf{0.899}} \\
        Georgia & \textbf{0.900} & 0.884 & \underline{\textbf{0.901}} \\
        Chapman & 0.959 & 0.946 & \underline{\textbf{0.964}} \\
        SPH & \textbf{0.976} & 0.975 & \underline{\textbf{0.982}} \\
        PTB-XL (all) & \underline{\textbf{0.937}} & 0.929 & \textbf{0.936} \\
        PTB-XL (sub) & \textbf{0.931} & \underline{\textbf{0.935}} & \textbf{0.931} \\
        PTB-XL (super) & 0.918 & 0.915 & \underline{\textbf{0.922}} \\
        \midrule
        
        \multicolumn{4}{c}{\textbf{Pediatric ECG interpretation }} \\
        ZZU pECG & \textbf{0.887} & \textbf{0.888} & \underline{\textbf{0.888}} \\
        \midrule
        
        \multicolumn{4}{c}{\textbf{Cardiac structure \& function }} \\
        EchoNext & \textbf{0.822} & 0.813 & \underline{\textbf{0.825}} \\        
        
        \bottomrule
    \end{tabular}
    \bigskip
    
    \textit{Note:} Domain Pretrain + Linear achieves the best performance across most datasets, highlighting the benefit of domain-adaptive pretraining. The non-linear head offers no consistent advantage over the simpler linear head.
\end{table}

\newpage
\section{Effect of Input Size}
\label{app:effect_of_input_size}

\begin{table}[htbp]
    \centering
    \caption{Comparison of aggregated macro-AUROC for data2vec pretrained for 2.5s and 5s under finetuning evaluation. Models were pretrained with HEEDB subset (106K samples) with an S4 backbone (4 layers) for 20 epochs. The best-performing result is highlighted in boldface and underlined, while models that do not perform statistically significantly worse are also highlighted in boldface.\\}
    \label{tab:data2vec_input_size_result}
    \begin{tabular}{lcc}
        \toprule
        \multicolumn{3}{c}{\textbf{data2vec (Finetuned)}} \\
        & \textbf{2.5s} & \textbf{5s} \\
        \midrule
        
        \multicolumn{3}{c}{\textbf{Adult ECG interpretation }} \\
        Ningbo & \textbf{0.957} & \underline{\textbf{0.959}} \\
        CPSC2018 & 0.953 & \underline{\textbf{0.959}} \\
        CPSC-Extra & \underline{\textbf{0.839}} & 0.823 \\
        Georgia & \underline{\textbf{0.894}} & \textbf{0.890} \\
        Chapman & \underline{\textbf{0.947}} & \textbf{0.944} \\
        SPH & \textbf{0.962} & \underline{\textbf{0.968}} \\
        PTB-XL (all) & \underline{\textbf{0.929}} & 0.906 \\
        PTB-XL (sub) & \underline{\textbf{0.931}} & \textbf{0.926} \\
        PTB-XL (super) & \underline{\textbf{0.926}} & 0.923 \\
        \midrule
        
        \multicolumn{3}{c}{\textbf{Pediatric ECG interpretation }} \\
        ZZU pECG & 0.847 & \underline{\textbf{0.865}} \\
        \midrule
        
        \multicolumn{3}{c}{\textbf{Cardiac structure \& function }} \\
        EchoNext & 0.817 & \underline{\textbf{0.823}} \\
        
        \bottomrule
    \end{tabular}
    \bigskip
    
    \textit{Note:} Input size of 2.5s achieves optimal performace across most tasks.
\end{table}

\begin{table}[htbp]
    \centering
    \caption{Comparison of aggregated macro-AUROC for DinoSR pretrained for 2.5s and 5s under finetuning evaluation. Models were pretrained with HEEDB subset (106K samples) with an S4 backbone (4 layers) for 20 epochs. The best-performing result is highlighted in boldface and underlined, while models that do not perform statistically significantly worse are also highlighted in boldface.\\}
    \label{tab:dinosr_input_size_result}
    \begin{tabular}{lcc}
        \toprule
        \multicolumn{3}{c}{\textbf{DinoSR (Finetuned)}} \\
        & \textbf{2.5s} & \textbf{5s} \\
        \midrule
        
        \multicolumn{3}{c}{\textbf{Adult ECG interpretation }} \\
        Ningbo & \textbf{0.956} & \underline{\textbf{0.961}} \\
        CPSC2018 & 0.958 & \underline{\textbf{0.964}} \\
        CPSC-Extra & \underline{\textbf{0.866}} & 0.846 \\
        Georgia & \underline{\textbf{0.904}} & \textbf{0.903} \\
        Chapman & \underline{\textbf{0.952}} & 0.944 \\
        SPH & \underline{\textbf{0.976}} & \textbf{0.974} \\
        PTB-XL (all) & \textbf{0.925} & \underline{\textbf{0.926}} \\
        PTB-XL (sub) & \underline{\textbf{0.931}} & \textbf{0.925} \\
        PTB-XL (super) & \underline{\textbf{0.928}} & \textbf{0.927} \\
        \midrule
        
        \multicolumn{3}{c}{\textbf{Pediatric ECG interpretation }} \\
        ZZU pECG & \textbf{0.869} & \underline{\textbf{0.872}} \\
        \midrule
        
        \multicolumn{3}{c}{\textbf{Cardiac structure \& function }} \\
        EchoNext & \underline{\textbf{0.824}} & \textbf{0.820} \\
        
        \bottomrule
    \end{tabular}
    \bigskip
    
    \textit{Note:} Input size of 2.5s achieves optimal performace across most tasks.
\end{table}

\begin{table}[htbp]
    \centering
    \caption{Comparison of aggregated macro-AUROC for JEPA pretrained for 2.5s and 5s under finetuning evaluation. Models were pretrained with HEEDB subset (106K samples) with an S4 backbone (4 layers) for 20 epochs. The best-performing result is highlighted in boldface and underlined, while models that do not perform statistically significantly worse are also highlighted in boldface.\\}
    \label{tab:jepa_input_size_result}
    \begin{tabular}{lcc}
        \toprule
        \multicolumn{3}{c}{\textbf{JEPA (Finetuned)}} \\
        & \textbf{2.5s} & \textbf{5s} \\
        \midrule
        
        \multicolumn{3}{c}{\textbf{Adult ECG interpretation }} \\
        Ningbo & \textbf{0.960} & \underline{\textbf{0.961}} \\
        CPSC2018 & \textbf{0.959} & \underline{\textbf{0.966}} \\
        CPSC-Extra & \underline{\textbf{0.868}} & \textbf{0.868} \\
        Georgia & \textbf{0.902} & \underline{\textbf{0.908}} \\
        Chapman & \underline{\textbf{0.952}} & \textbf{0.948} \\
        SPH & \underline{\textbf{0.969}} & \textbf{0.968} \\
        PTB-XL (all) & 0.923 & \underline{\textbf{0.933}} \\
        PTB-XL (sub) & \underline{\textbf{0.938}} & 0.927 \\
        PTB-XL (super) & \underline{\textbf{0.930}} & 0.924 \\
        \midrule
        
        \multicolumn{3}{c}{\textbf{Pediatric ECG interpretation }} \\
        ZZU pECG & 0.864 & \underline{\textbf{0.883}} \\
        \midrule
        
        \multicolumn{3}{c}{\textbf{Cardiac structure \& function }} \\
        EchoNext & \underline{\textbf{0.825}} & \textbf{0.823} \\
        
        \bottomrule
    \end{tabular}
    \bigskip
    
    \textit{Note:} Input size of 2.5s achieves optimal performace across most tasks.
\end{table}

\begin{table}[htbp]
    \centering
    \caption{Comparison of aggregated macro-AUROC for CPC pretrained for 2.5s and 5s under finetuning evaluation. Models were pretrained with HEEDB subset (106K samples) with an S4 backbone (4 layers) for 20 epochs. The best-performing result is highlighted in boldface and underlined, while models that do not perform statistically significantly worse are also highlighted in boldface.\\}
    \label{tab:cpc_input_size_result}
    \begin{tabular}{lcc}
        \toprule
        \multicolumn{3}{c}{\textbf{CPC (Finetuned)}} \\
        & \textbf{2.5s} & \textbf{5s} \\
        \midrule
        
        \multicolumn{3}{c}{\textbf{Adult ECG interpretation }} \\
        Ningbo & \underline{\textbf{0.961}} & 0.955 \\
        CPSC2018 & 0.958 & \underline{\textbf{0.964}} \\
        CPSC-Extra & \textbf{0.877} & \underline{\textbf{0.884}} \\
        Georgia & \underline{\textbf{0.901}} & \textbf{0.900} \\
        Chapman & \underline{\textbf{0.948}} & \textbf{0.944} \\
        SPH & \textbf{0.959} & \underline{\textbf{0.972}} \\
        PTB-XL (all) & \underline{\textbf{0.929}} & 0.921 \\
        PTB-XL (sub) & \underline{\textbf{0.932}} & 0.925 \\
        PTB-XL (super) & \underline{\textbf{0.929}} & 0.924 \\
        \midrule
        
        \multicolumn{3}{c}{\textbf{Pediatric ECG interpretation }} \\
        ZZU pECG & \underline{\textbf{0.888}} & \textbf{0.886} \\
        \midrule
        
        \multicolumn{3}{c}{\textbf{Cardiac structure \& function }} \\
        EchoNext & \underline{\textbf{0.826}} & \textbf{0.826} \\
        
        \bottomrule
    \end{tabular}
    \bigskip
    
    \textit{Note:} Input size of 2.5s achieves optimal performace across most tasks.
\end{table}

\begin{table}[htbp]
    \centering
    \caption{Comparison of aggregated macro-AUROC for HuBERT++ pretrained for 2.5s and 5s under finetuning evaluation. Models were pretrained with HEEDB subset (106K samples) with and S4 backbone (4 layers) for 20 epochs. The best-performing result is highlighted in boldface and underlined, while models that do not perform statistically significantly worse are also highlighted in boldface.\\}
    \label{tab:hubert_input_size_result}
    \begin{tabular}{lcc}
        \toprule
        \multicolumn{3}{c}{\textbf{HuBERT++ (Finetuned)}} \\
        & \textbf{2.5s} & \textbf{5s} \\
        \midrule
        
        \multicolumn{3}{c}{\textbf{Adult ECG interpretation }} \\
        Ningbo & \textbf{0.964} & \underline{\textbf{0.965}} \\
        CPSC2018 & \textbf{0.960} & \underline{\textbf{0.966}} \\
        CPSC-Extra & \textbf{0.866} & \underline{\textbf{0.870}} \\
        Georgia & 0.897 & \underline{\textbf{0.909}} \\
        Chapman & \textbf{0.949} & \underline{\textbf{0.950}} \\
        SPH & \textbf{0.974} & \underline{\textbf{0.978}} \\
        PTB-XL (all) & \underline{\textbf{0.925}} & \textbf{0.923} \\
        PTB-XL (sub) & \underline{\textbf{0.937}} & 0.933 \\
        PTB-XL (super) & \underline{\textbf{0.931}} & 0.928 \\
        \midrule
        
        \multicolumn{3}{c}{\textbf{Pediatric ECG interpretation }} \\
        ZZU pECG & \textbf{0.876} & \underline{\textbf{0.881}} \\
        \midrule
        
        \multicolumn{3}{c}{\textbf{Cardiac structure \& function }} \\
        EchoNext & \underline{\textbf{0.826}} & \textbf{0.822} \\
        
        \bottomrule
    \end{tabular}
    \bigskip
    
    \textit{Note:} Input size of 2.5s achieves optimal performace across most tasks.
\end{table}

\newpage
\section{Performance (HEEDB + Emory + CODE-15\%)}
\label{app:performance_full_dataset}

\subsection{Finetuning}
\label{app:finetuning_full_dataset}

\begin{table}[ht]
    \centering
    \tiny
    \caption{Comparison of aggregated macro-AUROC for classification and MAE for regression under finetuning with a linear prediction head. We highlight with $\uparrow$ tasks where higher AUROC is better and $\downarrow$ tasks where lower standardized MAE values are better. The best-performing result is highlighted in boldface and underlined, while models that do not perform statistically significantly worse are also highlighted in boldface. ${}^\dagger$ signifies evaluation of a model trained on the parent dataset (listed above).\\}
    \label{tab:finetuning_result}
        \begin{tabular}{lcccccccc}
            \toprule
            & \multicolumn{2}{c}{\textbf{FMs (Finetuned)}} & \multicolumn{5}{c}{\textbf{Pretrained by Us (Finetuned)}} & \textbf{Supervised} \\
            \cmidrule(l){2-3}\cmidrule(l){4-8}\cmidrule(l){9-9}
             & \textbf{ECGFounder} & \textbf{ECG-JEPA} & \textbf{data2vec} & \textbf{DinoSR} & \textbf{JEPA} & \textbf{CPC} & \textbf{HuBERT++} & \textbf{S4} \\
            \midrule

            \multicolumn{9}{c}{\textbf{Adult ECG interpretation }} \\
            Ningbo $\uparrow$ & \textbf{0.974} & \textbf{0.973} & 0.965 & 0.965 & 0.970 & \underline{\textbf{0.974}} & 0.969 & \textbf{0.972} \\
            CPSC2018 $\uparrow$ & 0.966 & \underline{\textbf{0.974}} & 0.953 & 0.953 & \textbf{0.971} & \textbf{0.967} & 0.962 & 0.962 \\
            CPSC-Extra $\uparrow$ & \underline{\textbf{0.906}} & \textbf{0.897} & 0.873 & 0.873 & \textbf{0.888} & \textbf{0.897} & \textbf{0.892} & 0.852 \\
            Georgia $\uparrow$ & \textbf{0.920} & \textbf{0.918} & 0.894 & 0.894 & \underline{\textbf{0.921}} & \textbf{0.920} & 0.911 & 0.903 \\
            Chapman $\uparrow$ & \textbf{0.968} & \underline{\textbf{0.972}} & 0.945 & 0.945 & \textbf{0.964} & \textbf{0.963} & 0.962 & \textbf{0.963} \\
            -Chapman (rhythm)${}^\dagger$ $\uparrow$ & \underline{\textbf{0.991}} & \textbf{0.989} & 0.977 & 0.977 & \textbf{0.989} & \textbf{0.990} & 0.987 & 0.986 \\
            SPH $\uparrow$ & \textbf{0.983} & 0.980 & 0.974 & 0.974 & 0.974 & \textbf{0.982} & \underline{\textbf{0.984}} & \textbf{0.981} \\
            PTB-XL (all) $\uparrow$ & 0.934 & \textbf{0.940} & 0.929 & 0.929 & 0.938 & \underline{\textbf{0.945}} & 0.939 & \textbf{0.941} \\
            -PTB-XL (diag) ${}^\dagger$ $\uparrow$ & \underline{\textbf{0.950}} & \textbf{0.946} & 0.941 & 0.941 & 0.935 & \textbf{0.946} & \textbf{0.946} & \textbf{0.943} \\
            -PTB-XL (form)${}^\dagger$ $\uparrow$ & 0.875 & 0.912 & 0.904 & 0.904 & \textbf{0.925} & \underline{\textbf{0.926}} & 0.906 & \textbf{0.919} \\
            -PTB-XL (rhythm)${}^\dagger$ $\uparrow$ & \underline{\textbf{0.965}} & \textbf{0.956} & 0.921 & 0.921 & \textbf{0.961} & \textbf{0.963} & 0.952 & \textbf{0.956} \\
            PTB-XL (sub) $\uparrow$ & \underline{\textbf{0.943}} & \textbf{0.935} & \textbf{0.936} & \textbf{0.936} & \textbf{0.939} & \textbf{0.941} & \textbf{0.942} & \textbf{0.938} \\
            PTB-XL (super) $\uparrow$ & \textbf{0.935} & 0.921 & 0.929 & 0.929 & \textbf{0.934} & \textbf{0.934} & \underline{\textbf{0.936}} & 0.932 \\
            \midrule

            \multicolumn{9}{c}{\textbf{Pediatric ECG interpretation }} \\
            ZZU pECG $\uparrow$ & 0.898 & \underline{\textbf{0.911}} & 0.875 & 0.875 & 0.895 & 0.896 & 0.892 & 0.897 \\
            \midrule

            \multicolumn{9}{c}{\textbf{Cardiac structure \& function }} \\
            EchoNext (Echo) $\uparrow$ & 0.817 & 0.817 & 0.824 & 0.824 & \textbf{0.830} & \underline{\textbf{0.832}} & \textbf{0.831} & 0.819 \\
            \midrule

            \multicolumn{9}{c}{\textbf{Cardiac outcomes}} \\
            MIMIC (Cardiac) $\uparrow$ & 0.768 & 0.772 & 0.778 & \textbf{0.784} & 0.780 & \underline{\textbf{0.786}} & 0.782 & 0.780 \\
            \midrule
            
            \multicolumn{9}{c}{\textbf{Non-cardiac outcomes}} \\
            MIMIC (Non-cardiac) $\uparrow$ & 0.701 & 0.711 & 0.708 & 0.719 & 0.713 & \underline{\textbf{0.722}} & 0.716 & 0.714 \\
            \midrule

            \multicolumn{9}{c}{\textbf{Acute care predictions}} \\
            MIMIC (Deterioration) $\uparrow$ & 0.717 & \textbf{0.747} & \textbf{0.754} & \textbf{0.762} & \underline{\textbf{0.764}} & \textbf{0.763} & \textbf{0.757} & \textbf{0.756} \\
            MIMIC (Mortality) $\uparrow$ & \underline{\textbf{0.810}} & \textbf{0.792} & \textbf{0.799} & \textbf{0.794} & \textbf{0.801} & \textbf{0.808} & \textbf{0.800} & \textbf{0.793} \\
            MIMIC (ICU) $\uparrow$ & 0.748 & 0.742 & 0.749 & 0.752 & 0.752 & \underline{\textbf{0.758}} & \textbf{0.753} & 0.745 \\
            \midrule

            \multicolumn{9}{c}{\textbf{Patient characteristics}} \\
            MIMIC (Sex) $\uparrow$ & 0.913 & 0.904 & 0.922 & 0.931 & 0.932 & \underline{\textbf{0.939}} & 0.932 & 0.919 \\
            MIMIC (Age) $\downarrow$ & 0.461 & 0.463 & 0.462 & 0.450 & \underline{\textbf{0.435}} & 0.445 & 0.445 & 0.455 \\
            MIMIC (Biometrics) $\downarrow$ & 0.637 & 0.640 & 0.623 & 0.610 & 0.606 & \underline{\textbf{0.596}} & 0.609 & 0.626 \\
            MIMIC (ECG Features) $\downarrow$ & 0.458 & 0.460 & 0.454 & 0.454 & \underline{\textbf{0.448}} & 0.455 & 0.450 & 0.452 \\
            MIMIC (Lab Values) $\downarrow$ & 0.679 & 0.677 & 0.678 & \textbf{0.670} & \textbf{0.670} & \underline{\textbf{0.669}} & \textbf{0.670} & 0.675 \\
            MIMIC (Vital Signs) $\downarrow$ & 0.704 & 0.703 & 0.702 & 0.701 & \underline{\textbf{0.698}} & \textbf{0.699} & \textbf{0.699} & 0.701 \\

            \bottomrule
        \end{tabular}
\end{table}

\newpage
\subsection{Frozen}
\label{app:frozen_full_dataset}

\begin{table}[ht]
    \centering
    \tiny
    \caption{Comparison of aggregated macro-AUROC for classification and MAE for regression under frozen evaluation. We highlight with $\uparrow$ tasks where higher AUROC is better and $\downarrow$ tasks where lower standardized MAE values are better. The best-performing result is highlighted in boldface and underlined, while models that do not perform statistically significantly worse are also highlighted in boldface. ${}^\dagger$ signifies evaluation of a model trained on the parent dataset (listed above).\\}
    \label{tab:frozen_result}
        \begin{tabular}{lcccccccc}
            \toprule
            & \multicolumn{2}{c}{\textbf{FMs (Frozen)}} & \multicolumn{5}{c}{\textbf{Pretrained by Us (Frozen)}} & \textbf{Supervised} \\
            \cmidrule(l){2-3}\cmidrule(l){4-8}\cmidrule(l){9-9}
             & \textbf{ECGFounder} & \textbf{ECG-JEPA} & \textbf{data2vec} & \textbf{DinoSR} & \textbf{JEPA} & \textbf{CPC} & \textbf{HuBERT++} & \textbf{S4} \\
            \midrule

            \multicolumn{9}{c}{\textbf{Adult ECG interpretation }} \\
            Ningbo $\uparrow$ & 0.961 & \textbf{0.971} & 0.895 & 0.895 & 0.941 & 0.955 & 0.941 & \underline{\textbf{0.972}} \\
            CPSC2018 $\uparrow$ & 0.966 & \underline{\textbf{0.975}} & 0.915 & 0.915 & 0.939 & 0.961 & 0.937 & 0.962 \\
            CPSC-Extra $\uparrow$ & \underline{\textbf{0.907}} & \textbf{0.902} & 0.823 & 0.823 & 0.856 & 0.872 & 0.863 & 0.852 \\
            Georgia $\uparrow$ & \underline{\textbf{0.924}} & \textbf{0.910} & 0.846 & 0.846 & 0.888 & 0.904 & 0.878 & 0.903 \\
            Chapman $\uparrow$ & \underline{\textbf{0.967}} & \textbf{0.964} & 0.914 & 0.914 & 0.941 & 0.941 & 0.940 & \textbf{0.963} \\
            -Chapman (rhythm)${}^\dagger$ $\uparrow$ & \textbf{0.983} & \underline{\textbf{0.988}} & 0.941 & 0.941 & 0.966 & \textbf{0.984} & 0.959 & \textbf{0.986} \\
            SPH $\uparrow$ & 0.966 & \textbf{0.980} & 0.923 & 0.923 & 0.964 & 0.960 & 0.964 & \underline{\textbf{0.981}} \\
            PTB-XL (all) $\uparrow$ & 0.927 & 0.934 & 0.876 & 0.876 & 0.913 & 0.928 & 0.918 & \underline{\textbf{0.941}} \\
            -PTB-XL (diag) ${}^\dagger$ $\uparrow$ & \textbf{0.940} & \textbf{0.943} & 0.888 & 0.888 & 0.922 & 0.932 & 0.933 & \underline{\textbf{0.943}} \\
            -PTB-XL (form) ${}^\dagger$ $\uparrow$ &  0.876 & 0.889 & 0.853 & 0.853 & 0.888 & \textbf{0.903} & 0.886 & \underline{\textbf{0.919}} \\
            -PTB-XL (rhythm) ${}^\dagger$ $\uparrow$ & \textbf{0.958} & \underline{\textbf{0.966}} & 0.862 & 0.862 & 0.915 & 0.946 & 0.908 & \textbf{0.956} \\
            PTB-XL (sub) $\uparrow$ & \underline{\textbf{0.939}} & \textbf{0.934} & 0.905 & 0.905 & 0.926 & \textbf{0.932} & 0.925 & \textbf{0.938} \\
            PTB-XL (super) $\uparrow$ & 0.928 & 0.917 & 0.913 & 0.913 & 0.923 & 0.907 & 0.918 & \underline{\textbf{0.932}} \\
            \midrule

            \multicolumn{9}{c}{\textbf{Pediatric ECG interpretation }} \\
            ZZU pECG $\uparrow$ & 0.891 & \underline{\textbf{0.905}} & 0.815 & 0.815 & 0.875 & 0.872 & 0.869 & \textbf{0.897} \\
            \midrule

            \multicolumn{9}{c}{\textbf{Cardiac structure \& function }} \\
            EchoNext (Echo) $\uparrow$ & 0.803 & 0.811 & 0.799 & 0.799 & \textbf{0.815} & \underline{\textbf{0.820}} & \textbf{0.819} & \textbf{0.819} \\
            \midrule

            \multicolumn{9}{c}{\textbf{Cardiac outcomes}} \\
            MIMIC (Cardiac) $\uparrow$ & 0.745 & 0.757 & 0.752 & 0.758 & 0.764 & 0.774 & 0.766 & \textbf{\underline{0.780}} \\
            \midrule

            \multicolumn{9}{c}{\textbf{Non-cardiac outcomes}} \\
            MIMIC (Non-cardiac) $\uparrow$ & 0.671 & 0.688 & 0.676 & 0.688 & 0.691 & 0.706 & 0.695 & \textbf{\underline{0.714}} \\
            \midrule

            \multicolumn{9}{c}{\textbf{Acute care predictions}} \\
            MIMIC (Deterioration) $\uparrow$ & 0.697 & 0.702 & 0.705 & \textbf{0.727} & \textbf{0.745} & \textbf{0.741} & \textbf{0.726} & \textbf{\underline{0.756}} \\
            MIMIC (Mortality) $\uparrow$ & \textbf{0.769} & \textbf{0.788} & 0.740 & 0.761 & \textbf{0.773} & \textbf{0.783} & \textbf{0.785} & \textbf{\underline{0.793}} \\
            MIMIC (ICU) $\uparrow$ & 0.731 & 0.734 & 0.720 & 0.727 & 0.736 & \underline{\textbf{0.750}} & 0.737 & \textbf{0.745} \\
            \midrule

            \multicolumn{9}{c}{\textbf{Patient characteristics}} \\
            MIMIC (Sex) $\uparrow$ & 0.872 & 0.894 & 0.893 & 0.885 & 0.905 & \underline{\textbf{0.920}} & 0.914 & \textbf{0.919} \\
            MIMIC (Age) $\downarrow$ & 0.515 & 0.484 & 0.504 & 0.511 & 0.486 & 0.465 & 0.473 & \textbf{\underline{0.455}} \\
            MIMIC (Biometrics) $\downarrow$ & 0.702 & 0.700 & 0.666 & 0.671 & 0.655 & 0.637 & 0.646 & \textbf{\underline{0.626}} \\
            MIMIC (ECG Features) $\downarrow$ & 0.489 & 0.477 & 0.487 & 0.487 & 0.468 & 0.468 & 0.463 & \textbf{\underline{0.452}} \\
            MIMIC (Lab Values) $\downarrow$ & 0.703 & 0.694 & 0.693 & 0.698 & 0.684 & \textbf{0.676} & 0.680 & \textbf{\underline{0.675}} \\
            MIMIC (Vital Signs) $\downarrow$ & 0.719 & 0.716 & 0.712 & 0.714 & 0.709 & \textbf{0.702} & 0.704 & \textbf{\underline{0.701}} \\

            \bottomrule
        \end{tabular}
\end{table}

\newpage
\subsection{Linear}
\label{app:linear_full_dataset}

\begin{table}[ht]
    \centering
    \tiny
    \caption{Comparison of aggregated macro-AUROC for classification and MAE for regression under linear evaluation. We highlight with $\uparrow$ tasks where higher AUROC is better and $\downarrow$ tasks where lower standardized MAE values are better. The best-performing result is highlighted in boldface and underlined, while models that do not perform statistically significantly worse are also highlighted in boldface. ${}^\dagger$ signifies evaluation of a model trained on the parent dataset (listed above).\\}
    \label{tab:linear_result}
        \begin{tabular}{lcccccccc}
            \toprule
            & \multicolumn{2}{c}{\textbf{FMs (Linear)}} & \multicolumn{5}{c}{\textbf{Pretrained by Us (Linear)}} & \textbf{Supervised} \\
            \cmidrule(l){2-3}\cmidrule(l){4-8}\cmidrule(l){9-9}
             & \textbf{ECGFounder} & \textbf{ECG-JEPA} & \textbf{data2vec} & \textbf{DinoSR} & \textbf{JEPA} & \textbf{CPC} & \textbf{HuBERT++} & \textbf{S4} \\
            \midrule

            \multicolumn{9}{c}{\textbf{Adult ECG interpretation }} \\
            Ningbo $\uparrow$ & \textbf{0.970} & \textbf{0.970} & 0.814 & 0.848 & 0.911 & 0.899 & 0.793 & \underline{\textbf{0.972}} \\
            CPSC2018 $\uparrow$ & 0.964 & \underline{\textbf{0.975}} & 0.814 & 0.809 & 0.895 & 0.896 & 0.781 & 0.962 \\
            CPSC-Extra $\uparrow$ & \underline{\textbf{0.910}} & \textbf{0.902} & 0.780 & 0.774 & 0.857 & 0.779 & 0.711 & 0.852 \\
            Georgia $\uparrow$ & \underline{\textbf{0.923}} & \textbf{0.920} & 0.771 & 0.818 & 0.855 & 0.845 & 0.735 & 0.903 \\
            Chapman $\uparrow$ & \underline{\textbf{0.968}} & \textbf{0.962} & 0.823 & 0.850 & 0.894 & 0.854 & 0.775 & \textbf{0.963} \\
            -Chapman (rhythm)${}^\dagger$ & \textbf{0.987} & \underline{\textbf{0.989}} & 0.830 & 0.892 & 0.954 & 0.940 & 0.814 & \textbf{0.986} \\
            SPH $\uparrow$ & 0.975 & 0.967 & 0.869 & 0.867 & 0.940 & 0.922 & 0.840 & \underline{\textbf{0.981}} \\
            PTB-XL (all) $\uparrow$ & 0.931 & 0.928 & 0.799 & 0.805 & 0.878 & 0.866 & 0.753 & \underline{\textbf{0.941}} \\
            -PTB-XL (diag)${}^\dagger$ $\uparrow$ & \underline{\textbf{0.947}} & 0.925 & 0.807 & 0.800 & 0.886 & 0.882 & 0.760 & \textbf{0.943} \\
            -PTB-XL (form)${}^\dagger$ $\uparrow$ & 0.874 & \textbf{0.908} & 0.761 & 0.767 & 0.842 & 0.796 & 0.701 & \underline{\textbf{0.919}} \\
            -PTB-XL (rhythm)${}^\dagger$ $\uparrow$ & \textbf{0.961} & \underline{\textbf{0.969}} & 0.811 & 0.863 & 0.903 & 0.916 & 0.799 & 0.956 \\
            PTB-XL (sub) $\uparrow$ & \underline{\textbf{0.945}} & 0.916 & 0.823 & 0.838 & 0.896 & 0.881 & 0.813 & \textbf{0.938} \\
            PTB-XL (super) $\uparrow$ & 0.924 & 0.911 & 0.826 & 0.822 & 0.880 & 0.860 & 0.809 & \underline{\textbf{0.932}} \\
            \midrule

            \multicolumn{9}{c}{\textbf{Pediatric ECG interpretation }} \\
            ZZU pECG $\uparrow$ & \underline{\textbf{0.900}} & \textbf{0.891} & 0.763 & 0.757 & 0.844 & 0.846 & 0.726 & \textbf{0.897} \\
            \midrule

            \multicolumn{9}{c}{\textbf{Cardiac structure \& function }} \\
            EchoNext $\uparrow$ & 0.795 & 0.806 & 0.790 & 0.795 & 0.805 & 0.799 & 0.785 & \underline{\textbf{0.819}} \\
            \midrule

            \multicolumn{9}{c}{\textbf{Cardiac outcomes}} \\
            MIMIC (Cardiac) $\uparrow$ & 0.751 & 0.751 & 0.704 & 0.722 & 0.746 & 0.752 & 0.701 & \textbf{\underline{0.780}} \\
            \midrule

            \multicolumn{9}{c}{\textbf{Non-cardiac outcomes}} \\
            MIMIC (Non-cardiac) $\uparrow$ & 0.671 & 0.675 & 0.638 & 0.652 & 0.673 & 0.682 & 0.633 & \textbf{\underline{0.714}} \\
            \midrule

            \multicolumn{9}{c}{\textbf{Acute care predictions}} \\
            MIMIC (Deterioration) $\uparrow$ & 0.713 & 0.720 & 0.670 & 0.681 & \textbf{0.731} & \textbf{0.728} & 0.684 & \textbf{\underline{0.756}} \\
            MIMIC (Mortality) $\uparrow$ & \textbf{0.774} & \textbf{0.782} & 0.698 & 0.742 & \textbf{0.762} & \textbf{0.761} & 0.720 & \textbf{\underline{0.793}} \\
            MIMIC (ICU) $\uparrow$ & 0.730 & \textbf{0.736} & 0.686 & 0.697 & 0.725 & 0.732 & 0.701 & \textbf{\underline{0.745}} \\
            \midrule

            \multicolumn{9}{c}{\textbf{Patient characteristics}} \\
            MIMIC (Sex) $\uparrow$ & 0.872 & 0.883 & 0.817 & 0.826 & 0.870 & 0.882 & 0.809 & \textbf{\underline{0.919}} \\
            MIMIC (Age) $\downarrow$ & 0.511 & 0.489 & 0.632 & 0.580 & 0.552 & 0.534 & 0.608 & \textbf{\underline{0.455}} \\
            MIMIC (Biometrics) $\downarrow$ & 0.700 & 0.685 & 0.733 & 0.713 & 0.689 & 0.680 & 0.732 & \textbf{\underline{0.626}} \\
            MIMIC (ECG Features) $\downarrow$ & 0.488 & 0.490 & 0.572 & 0.536 & 0.492 & 0.501 & 0.549 & \textbf{\underline{0.452}} \\
            MIMIC (Lab Values) $\downarrow$ & 0.693 & 0.695 & 0.709 & 0.711 & 0.692 & 0.684 & 0.704 & \textbf{\underline{0.675}} \\
            MIMIC (Vital Signs) $\downarrow$ & 0.716 & 0.714 & 0.724 & 0.722 & 0.710 & 0.707 & 0.719 & \textbf{\underline{0.701}} \\

            \bottomrule
        \end{tabular}
\end{table}

\newpage
\section{Rankings}\label{app:rankings}

\begin{table}[ht!]
    \centering
    \tiny
    \caption{Statistical ranking of FMs across evaluation modes and datasets. Rankings (Finetuned / Frozen / Linear) are assigned based on statistical equivalence groups determined by bootstrap testing, where models not performing significantly worse than the best model share the same rank. Lower ranks indicate better performance. ${}^\dagger$ signifies evaluation of a model trained on the parent dataset (listed above).\\}
    \label{tab:detailed_ranking_table}
        \begin{tabular}{lcccccccc}
            \toprule
            & \multicolumn{2}{c}{\textbf{FMs}} & \multicolumn{5}{c}{\textbf{Pretrained by Us}} & \textbf{Supervised} \\
            \cmidrule(l){2-3}\cmidrule(l){4-8}\cmidrule(l){9-9}
             & \textbf{ECGFounder} & \textbf{ECG-JEPA} & \textbf{data2vec} & \textbf{DinoSR} & \textbf{JEPA} & \textbf{CPC} & \textbf{HuBERT++} & \textbf{S4} \\
            \midrule

            \multicolumn{9}{c}{\textbf{Adult ECG interpretation }} \\
            Ningbo & 1/3/1 & 1/1/1 & 7/7/7 & 7/7/6 & 5/5/4 & 1/3/4 & 5/5/7 & 1/1/1 \\
            CPSC2018 & 4/2/2 & 1/1/1 & 7/7/6 & 7/7/6 & 1/5/4 & 1/2/4 & 4/5/8 & 4/2/2 \\
            CPSC-Extra & 1/1/1 & 1/1/1 & 6/7/5 & 6/7/5 & 1/3/3 & 1/3/5 & 1/3/8 & 6/3/3 \\
            Georgia & 1/1/1 & 1/1/1 & 7/7/7 & 7/7/6 & 1/3/4 & 1/3/4 & 5/6/8 & 5/3/3 \\
            Chapman & 1/1/1 & 1/1/1 & 7/7/7 & 7/7/5 & 1/4/4 & 1/4/5 & 6/4/8 & 1/1/1 \\
            -Chapman (rhythm)${}^\dagger$ & 1/1/1 & 1/1/1 & 7/7/7 & 7/7/6 & 1/5/4 & 1/1/5 & 5/5/7 & 5/1/1 \\
            SPH & 1/3/2 & 5/1/2 & 6/7/6 & 6/7/6 & 6/3/4 & 1/3/5 & 1/3/8 & 1/1/1 \\
            PTB-XL (all) & 4/2/2 & 1/2/2 & 7/7/6 & 7/7/6 & 4/5/4 & 1/2/4 & 4/5/8 & 1/1/1 \\
            -PTB-XL (diag)${}^\dagger$ & 1/1/1 & 1/1/3 & 6/7/6 & 6/7/6 & 6/4/4 & 1/4/4 & 1/4/8 & 1/1/1 \\
            -PTB-XL (form)${}^\dagger$ & 8/3/3 & 4/3/1 & 4/7/5 & 4/7/5 & 1/3/3 & 1/1/5 & 4/3/8 & 1/1/1 \\
            -PTB-XL (rhythm)${}^\dagger$ & 1/1/1 & 1/1/1 & 7/7/7 & 7/7/6 & 1/5/4 & 1/4/4 & 6/5/7 & 1/1/3 \\
            PTB-XL (sub) & 1/1/1 & 1/1/3 & 1/7/6 & 1/7/6 & 1/5/4 & 1/1/5 & 1/5/8 & 1/1/1 \\
            PTB-XL (super) & 1/2/2 & 8/4/3 & 5/6/6 & 5/6/6 & 1/3/4 & 1/8/5 & 1/4/8 & 5/1/1 \\
            \midrule
            \multicolumn{9}{c}{\textbf{Pediatric ECG interpretation }} \\
            ZZU pECG & 2/3/1 & 1/1/1 & 7/7/6 & 7/7/6 & 2/4/4 & 2/4/4 & 6/4/8 & 2/1/1 \\
            \midrule
            \multicolumn{9}{c}{\textbf{Cardiac structure \& function }} \\
            EchoNext (Echo) & 4/5/5 & 4/5/2 & 4/7/5 & 4/7/5 & 1/1/2 & 1/1/2 & 1/1/5 & 4/1/1 \\

            \midrule
            \multicolumn{9}{c}{\textbf{Cardiac outcomes }} \\
            MIMIC (Cardiac) & 7/7/2 & 7/5/2 & 6/7/7 & 1/5/6 & 3/3/5 & 1/2/2 & 3/3/7 & 3/1/1 \\
            \midrule

            \multicolumn{9}{c}{\textbf{Non-cardiac outcomes }} \\
            MIMIC (Non-cardiac) & 8/8/5 & 5/4/3 & 7/7/7 & 2/4/6 & 5/4/3 & 1/2/2 & 3/3/8 & 3/1/1 \\
            \midrule

            \multicolumn{9}{c}{\textbf{Acute care predictions }} \\
            MIMIC (Deterioration) & 8/6/5 & 1/6/1 & 1/6/5 & 1/1/5 & 1/1/1 & 1/1/1 & 1/1/5 & 1/1/1 \\
            MIMIC (Mortality) & 1/1/1 & 1/1/1 & 1/8/6 & 1/7/6 & 1/1/1 & 1/1/1 & 1/1/6 & 1/1/1 \\
            MIMIC (ICU) & 3/3/3 & 7/3/1 & 3/7/8 & 3/7/6 & 3/3/3 & 1/1/3 & 1/3/6 & 7/1/1 \\
            \midrule
            
            \multicolumn{9}{c}{\textbf{Patient characteristics }} \\
            MIMIC (Sex) & 7/8/4 & 8/5/2 & 5/5/7 & 2/7/6 & 2/4/4 & 1/1/2 & 2/3/8 & 6/1/1 \\
            MIMIC (Age) & 6/7/3 & 6/4/2 & 6/6/8 & 4/7/6 & 1/4/5 & 2/2/4 & 2/3/7 & 5/1/1 \\
            MIMIC (Biometrics) & 7/7/5 & 7/7/2 & 5/5/7 & 2/6/6 & 2/4/4 & 1/2/2 & 2/3/7 & 5/1/1 \\
            MIMIC (ECG Features) & 7/8/2 & 8/5/3 & 4/6/8 & 4/6/6 & 1/3/4 & 4/3/5 & 2/2/7 & 3/1/1 \\
            MIMIC (Lab Values) & 7/8/3 & 5/5/3 & 7/5/7 & 1/7/7 & 1/4/3 & 1/1/2 & 1/3/6 & 5/1/1 \\
            MIMIC (Vital Signs) & 7/8/4 & 7/6/4 & 4/5/8 & 4/6/7 & 1/4/3 & 1/1/2 & 1/3/6 & 4/1/1 \\

            \bottomrule
        \end{tabular}
\end{table}

\begin{table}[ht!]
    \centering
    \tiny
    \caption{Median statistical rankings of FMs across evaluation modes by categories. Rankings (Finetuned/Frozen/Linear) represent the median performance position across all datasets within each category. Lower values indicate better overall performance.\\}
    \label{tab:median_ranking_table}
    \resizebox{\textwidth}{!}{
        \begin{tabular}{lcccccccc}
            \toprule
            & \multicolumn{2}{c}{\textbf{FMs}} & \multicolumn{5}{c}{\textbf{Pretrained by Us}} & \textbf{Supervised} \\
            \cmidrule(l){2-3}\cmidrule(l){4-8}\cmidrule(l){9-9}
             & \textbf{ECGFounder} & \textbf{ECG-JEPA} & \textbf{data2vec} & \textbf{DinoSR} & \textbf{JEPA} & \textbf{CPC} & \textbf{HuBERT++} & \textbf{S4} \\
            \midrule

            Adult ECG interpretation & 1/1/1 & 1/1/1 & 7/7/6 & 7/7/6 & 1/4/4 & 1/3/5 & 4/5/8 & 1/1/1 \\
            Pediatric ECG interpretation & 2/3/1 & 1/1/1 & 7/7/6 & 7/7/6 & 2/4/4 & 2/4/4 & 6/4/8 & 2/1/1 \\
            Cardiac structure \& function & 4/5/5 & 4/5/2 & 4/7/5 & 4/7/5 & 1/1/2 & 1/1/2 & 1/1/5 & 4/1/1 \\
            Cardiac outcomes & 7/7/2 & 7/5/2 & 6/7/7 & 1/5/6 & 3/3/5 & 1/2/2 & 3/3/7 & 3/1/1 \\
            Non-cardiac outcomes & 8/8/5 & 5/4/3 & 7/7/7 & 2/4/6 & 5/4/3 & 1/2/2 & 3/3/8 & 3/1/1 \\
            Acute care predictions & 3/3/3 & 1/3/1 & 1/7/6 & 1/7/6 & 1/1/1 & 1/1/1 & 1/1/6 & 1/1/1 \\
            Patient characteristics & 7/8/3.5 & 7/5/2.5 & 5/5/7.5 & 3/6.5/6 & 1/4/4 & 1/1.5/2 & 2/3/7 & 5/1/1 \\

            \bottomrule
        \end{tabular}
    }
\end{table}

\newpage

\section{Comparison between CPC and external CPC}
\label{app:cpc_vs_external_cpc}
Table~\ref{tab:finetuning_result_cpc_old} provides a direct comparison to the ECG-CPC model from \citep{al-masud2026benchmarking}. The model show a relatively similar level of performance with slight advantages for the CPC model trained as part of this work, most likely due to a slightly larger training dataset and longer pretraining.
\begin{table}[htbp]
    \centering
    % \tiny
    \caption{Comparison of aggregated macro-AUROC between CPC and external CPC~\citep{al-masud2026benchmarking} for classification and MAE for regression under finetuning with a linear prediction head. We highlight with $\uparrow$ tasks where higher AUROC is better and $\downarrow$ tasks where lower standardized MAE values are better. The best-performing result is highlighted in boldface and underlined, while models that do not perform statistically significantly worse are also 
highlighted in boldface. ${}^\dagger$ signifies evaluation of a model trained on the parent dataset (listed above). Both CPC and CPC (external) use 4 layers. CPC is pretrained on HEEDB, HEEDB-Emory, and CODE-15\% for 10 epochs with a learning rate of 3e-3, while CPC (external) is pretrained on HEEDB for 2 epochs with a learning rate of 1e-3.\\}
    \label{tab:finetuning_result_cpc_old}
        \begin{tabular}{lcc}
            \toprule
             & \textbf{CPC} & \textbf{CPC (external)} \\
            \midrule

            \multicolumn{3}{c}{\textbf{Adult ECG interpretation}} \\
            Ningbo $\uparrow$ & \underline{\textbf{0.974}} & \textbf{0.973} \\
            CPSC2018 $\uparrow$ & \textbf{0.967} & \underline{\textbf{0.969}} \\
            CPSC-Extra $\uparrow$ & \textbf{0.897} & \underline{\textbf{0.898}} \\
            Georgia $\uparrow$ & \underline{\textbf{0.920}} & \textbf{0.913} \\
            Chapman $\uparrow$ & \underline{\textbf{0.963}} & \textbf{0.962} \\
            -Chapman (rhythm)${}^\dagger$ $\uparrow$ & \underline{\textbf{0.990}} & 0.987 \\
            SPH $\uparrow$ & \underline{\textbf{0.982}} & \textbf{0.981} \\
            PTB-XL (all) $\uparrow$ & 0.945 & \underline{\textbf{0.949}} \\
            -PTB-XL (diag)${}^\dagger$ $\uparrow$ & 0.946 & \underline{\textbf{0.951}} \\
            -PTB-XL (form)${}^\dagger$ $\uparrow$ & \textbf{0.926} & \underline{\textbf{0.934}} \\
            -PTB-XL (rhythm)${}^\dagger$ $\uparrow$ & \underline{\textbf{0.963}} & \textbf{0.959} \\
            PTB-XL (sub) $\uparrow$ & \underline{\textbf{0.941}} & \textbf{0.940} \\
            PTB-XL (super) $\uparrow$ & \underline{\textbf{0.934}} & \textbf{0.934} \\
            \midrule

            \multicolumn{3}{c}{\textbf{Pediatric ECG interpretation}} \\
            ZZU pECG $\uparrow$ & \underline{\textbf{0.896}} & \textbf{0.892} \\
            \midrule

            \multicolumn{3}{c}{\textbf{Cardiac structure \& function}} \\
            EchoNext (Echo) $\uparrow$ & \underline{\textbf{0.832}} & \textbf{0.831} \\
            \midrule

            \multicolumn{3}{c}{\textbf{Cardiac outcomes}} \\
            MIMIC (Cardiac) $\uparrow$ & \underline{\textbf{0.786}} & 0.781 \\
            \midrule

            \multicolumn{3}{c}{\textbf{Non-cardiac outcomes}} \\
            MIMIC (Non-cardiac) $\uparrow$ & \underline{\textbf{0.722}} & 0.719 \\
            \midrule

            \multicolumn{3}{c}{\textbf{Acute care predictions}} \\
            MIMIC (Deterioration) $\uparrow$ & \textbf{0.763} & \underline{\textbf{0.764}} \\
            MIMIC (Mortality) $\uparrow$ & \underline{\textbf{0.808}} & \textbf{0.803}  \\
            MIMIC (ICU) $\uparrow$ & \underline{\textbf{0.758}} & 0.753 \\
            \midrule

            \multicolumn{3}{c}{\textbf{Patient characteristics}} \\
            MIMIC (Sex) $\uparrow$ & \underline{\textbf{0.939}} & 0.933 \\
            MIMIC (Age) $\downarrow$ & 0.445 & \underline{\textbf{0.437}} \\
            MIMIC (Biometrics) $\downarrow$ & \underline{\textbf{0.596}} & 0.604 \\
            MIMIC (ECG Features) $\downarrow$ & 0.455 & \underline{\textbf{0.451}} \\
            MIMIC (Lab Values) $\downarrow$ & \underline{\textbf{0.669}} & 0.673 \\
            MIMIC (Vital Signs) $\downarrow$ & \underline{\textbf{0.699}} & \textbf{0.700}  \\

            \bottomrule
        \end{tabular}
\end{table}

\newpage
\section{Intra- and Inter-Model Representation Similarity}
\label{app:intra_and_inter_model_rep_sim}

\subsection{Intra-model CKA Analysis}

\begin{figure}[htbp]
\centering
% First row
\begin{subfigure}{0.32\textwidth}
    \centering
    \includegraphics[width=\textwidth]{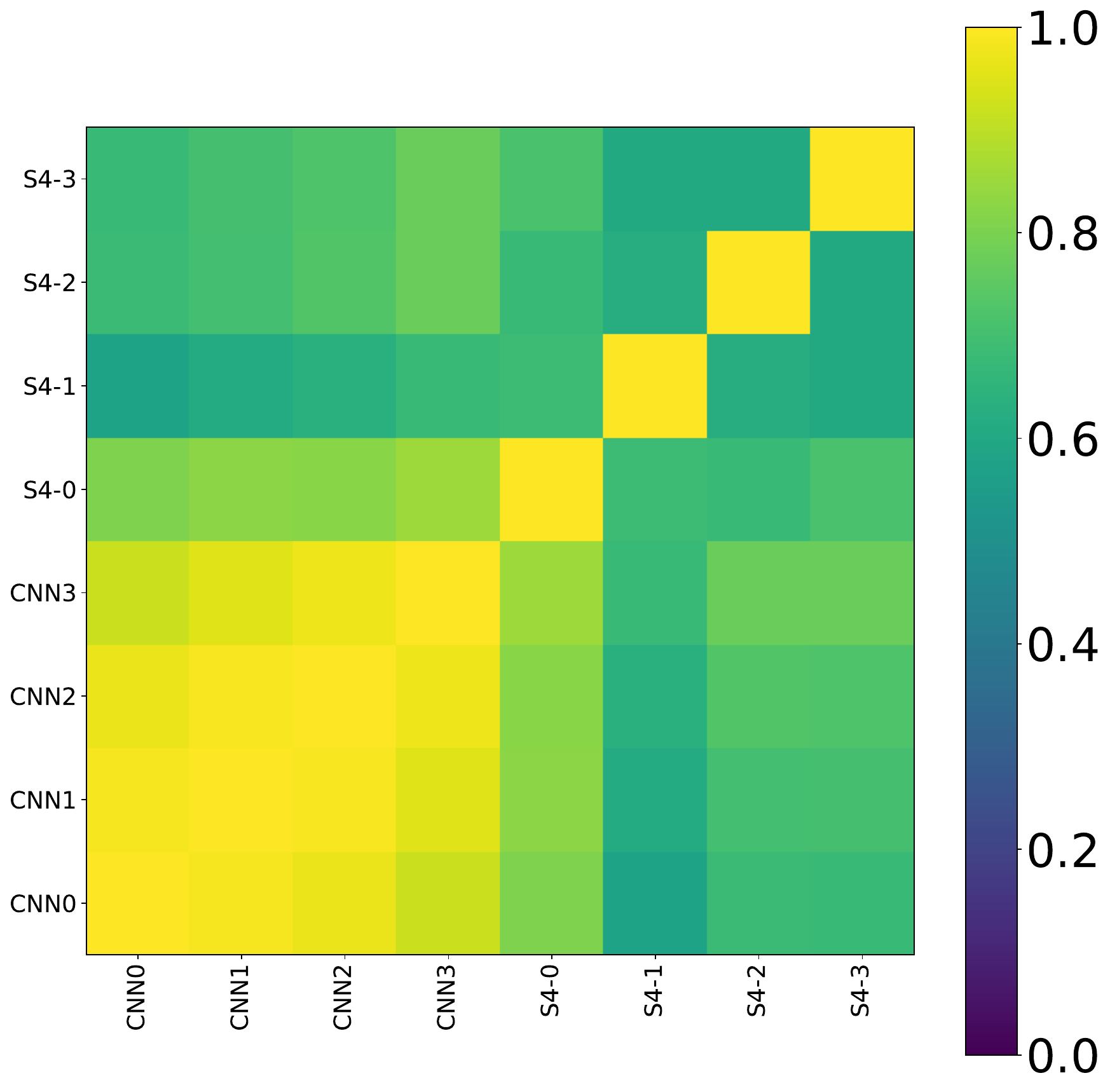}
    \caption{Data2Vec}
    \label{fig:heedb_data2vec_cka}
\end{subfigure}
\hfill
\begin{subfigure}{0.32\textwidth}
    \centering
    \includegraphics[width=\textwidth]{resources/dinosr_full_dataset.pdf}
    \caption{DinoSR}
    \label{fig:heedb_dinosr_cka}
\end{subfigure}
\hfill
\begin{subfigure}{0.32\textwidth}
    \centering
    \includegraphics[width=\textwidth]{resources/jepa_full_dataset.pdf}
    \caption{JEPA}
    \label{fig:heedb_jepa_cka}
\end{subfigure}

\vspace{0.5em} % Vertical spacing between rows

% Second row
\begin{subfigure}{0.32\textwidth}
    \centering
    \includegraphics[width=\textwidth]{resources/cpc_full_dataset.pdf}
    \caption{CPC}
    \label{fig:heedb_cpc_cka}
\end{subfigure}
% \hfill
\begin{subfigure}{0.32\textwidth}
    \centering
    \includegraphics[width=\textwidth]{resources/hubert++_full_dataset.pdf}
    \caption{HuBERT++}
    \label{fig:heedb_hubert_cka}
\end{subfigure}

\caption{Layer-wise representation similarity within each pretraining objective, measured by CKA with a Gaussian RBF kernel ($\sigma=1.0$) on 2,500 PTB-XL samples. Warmer colors indicate greater similarity between layer pairs.}
\label{fig:heedb_intra-model_cka_analysis}
\end{figure}

\subsection{Inter-model CKA Analysis}
\label{app:inter_model_cka_analysis}

\begin{figure}[htbp]
\centering
\begin{subfigure}{0.32\textwidth}
    \centering
    \includegraphics[width=\textwidth]{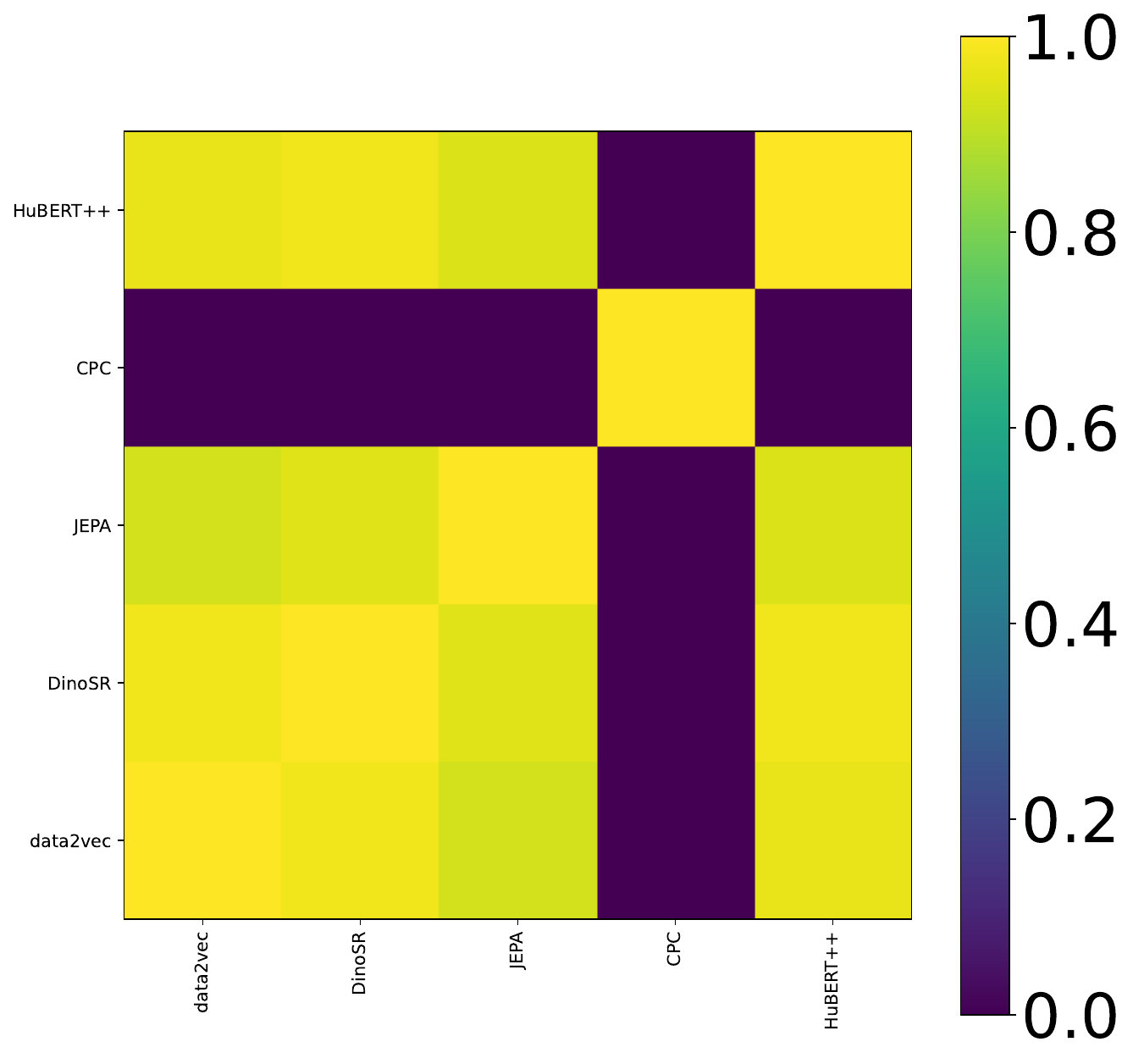}
    \caption{Early stage}
    \label{fig:cka_early_stage}
\end{subfigure}
\hfill
\begin{subfigure}{0.32\textwidth}
    \centering
    \includegraphics[width=\textwidth]{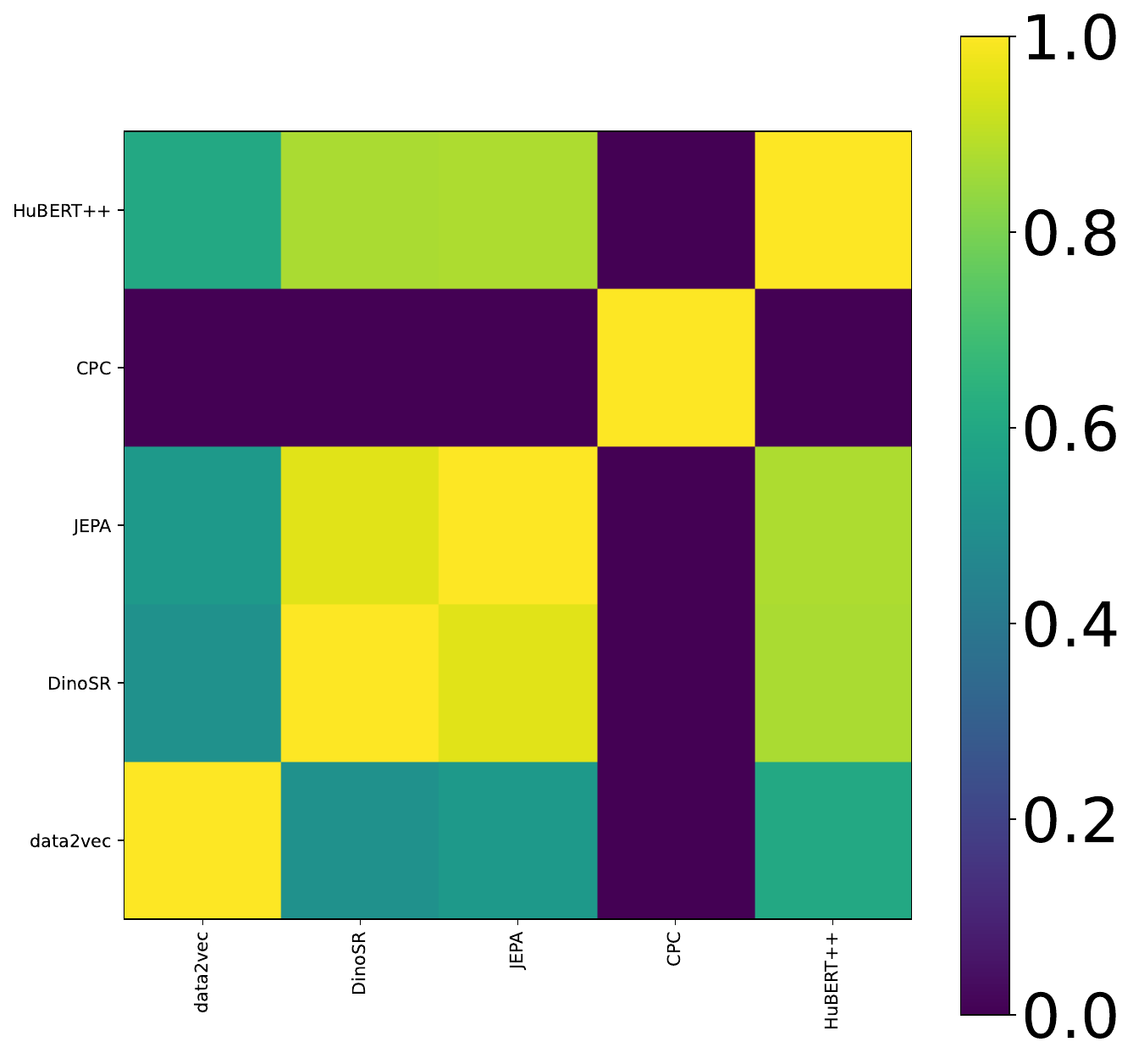}
    \caption{Mid stage}
    \label{fig:cka_mid_stage}
\end{subfigure}
\hfill
\begin{subfigure}{0.32\textwidth}
    \centering
    \includegraphics[width=\textwidth]{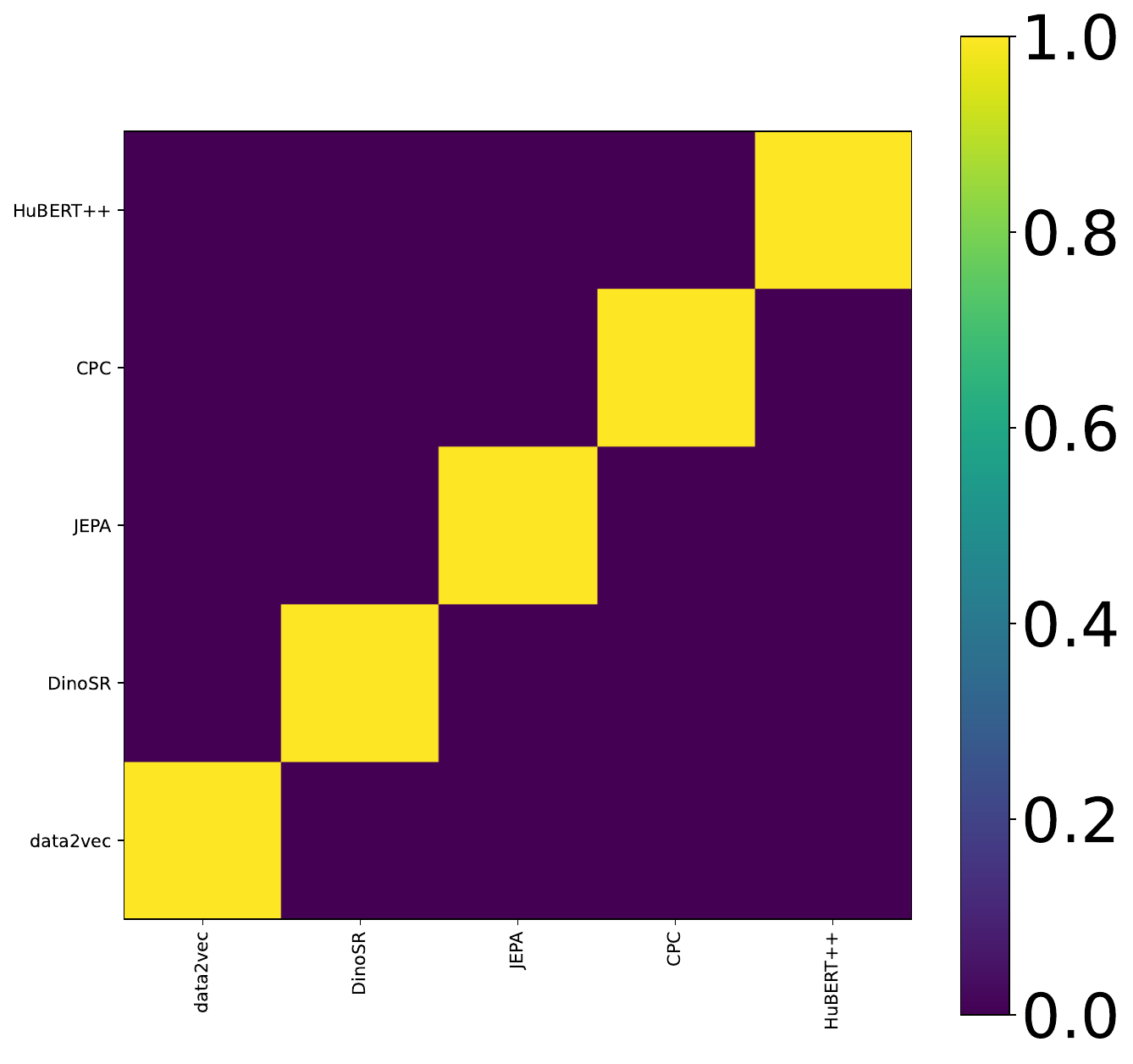}
    \caption{Late stage}
    \label{fig:cka_late_stage}
\end{subfigure}
\caption{Inter-model representational similarity across network depths. CKA heatmaps comparing corresponding stages across five FMs (Data2Vec, DinoSR, JEPA, HuBERT++). Higher values (yellow) indicate similar representations between layers. CKA computed using Gaussian RBF kernel ($\sigma=1.0$) on 256 samples of PTB-XL (all) dataset per model.}
\label{fig:inter-model_cka_analysis}
\end{figure}

\newpage
\section{Scaling Analysis}
\label{app:scaling_analysis}

\subsection{Validation Loss Scaling with Pretraining Data}
\label{app:val_loss_scaling}
\begin{figure}[htbp]
    \centering
    % First row
    \begin{subfigure}{0.32\textwidth}
        \centering
        \includegraphics[width=\textwidth]{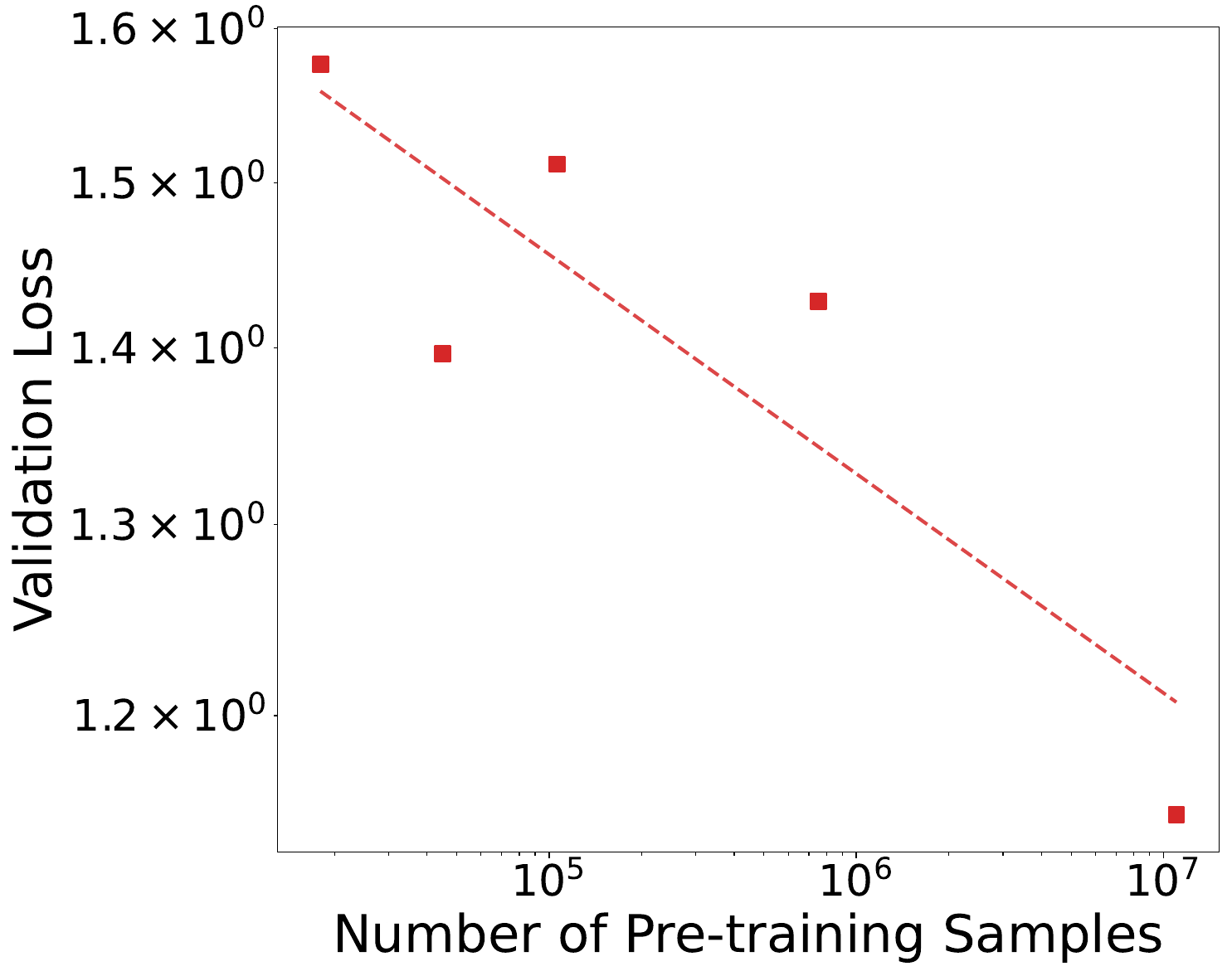}
        \caption{data2vec}
        \label{fig:app_scaling_data2vec}
    \end{subfigure}
    \hfill
    \begin{subfigure}{0.32\textwidth}
        \centering
        \includegraphics[width=\textwidth]{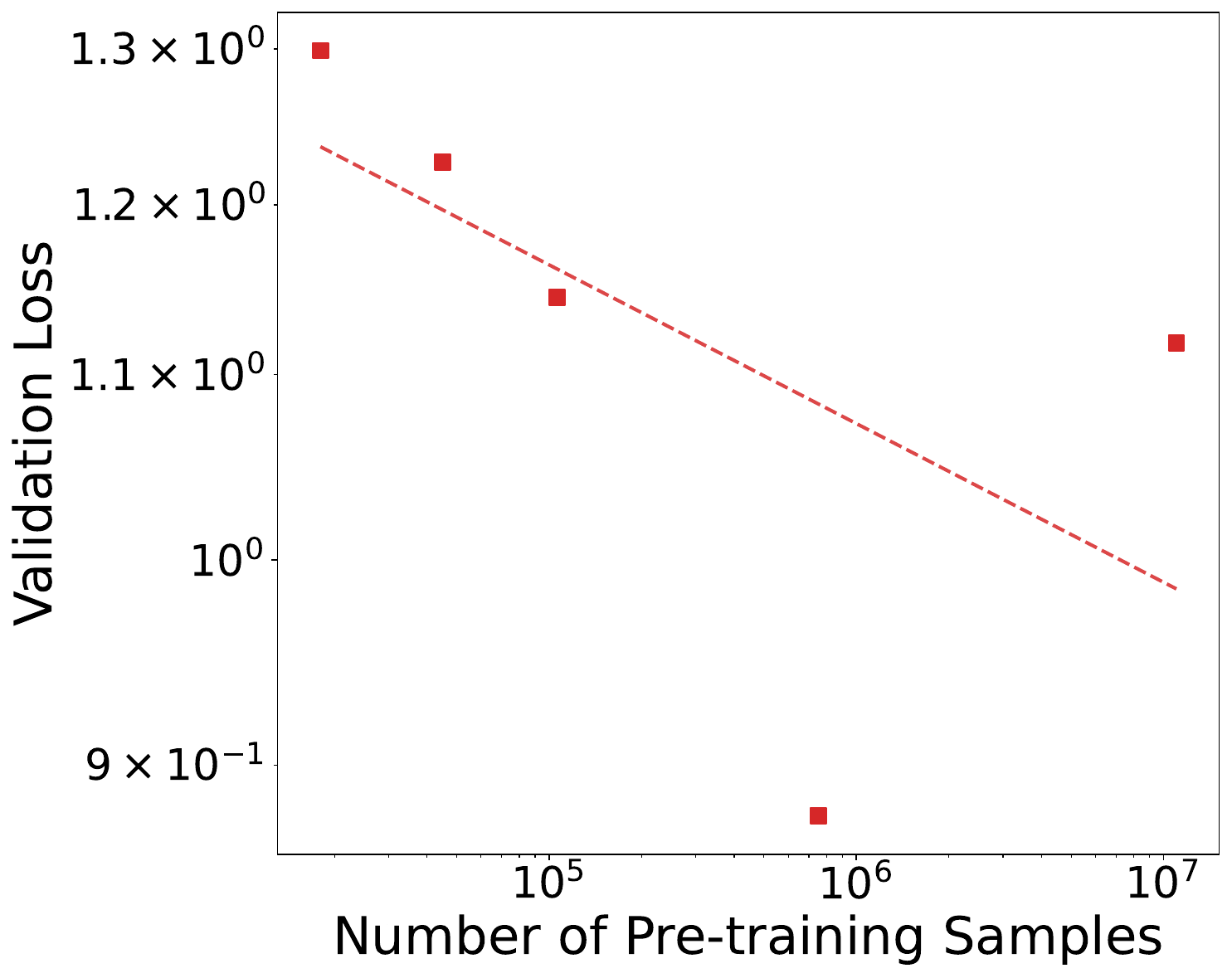}
        \caption{DinoSR}
        \label{fig:app_scaling_dinosr}
    \end{subfigure}
    \hfill
    \begin{subfigure}{0.32\textwidth}
        \centering
        \includegraphics[width=\textwidth]{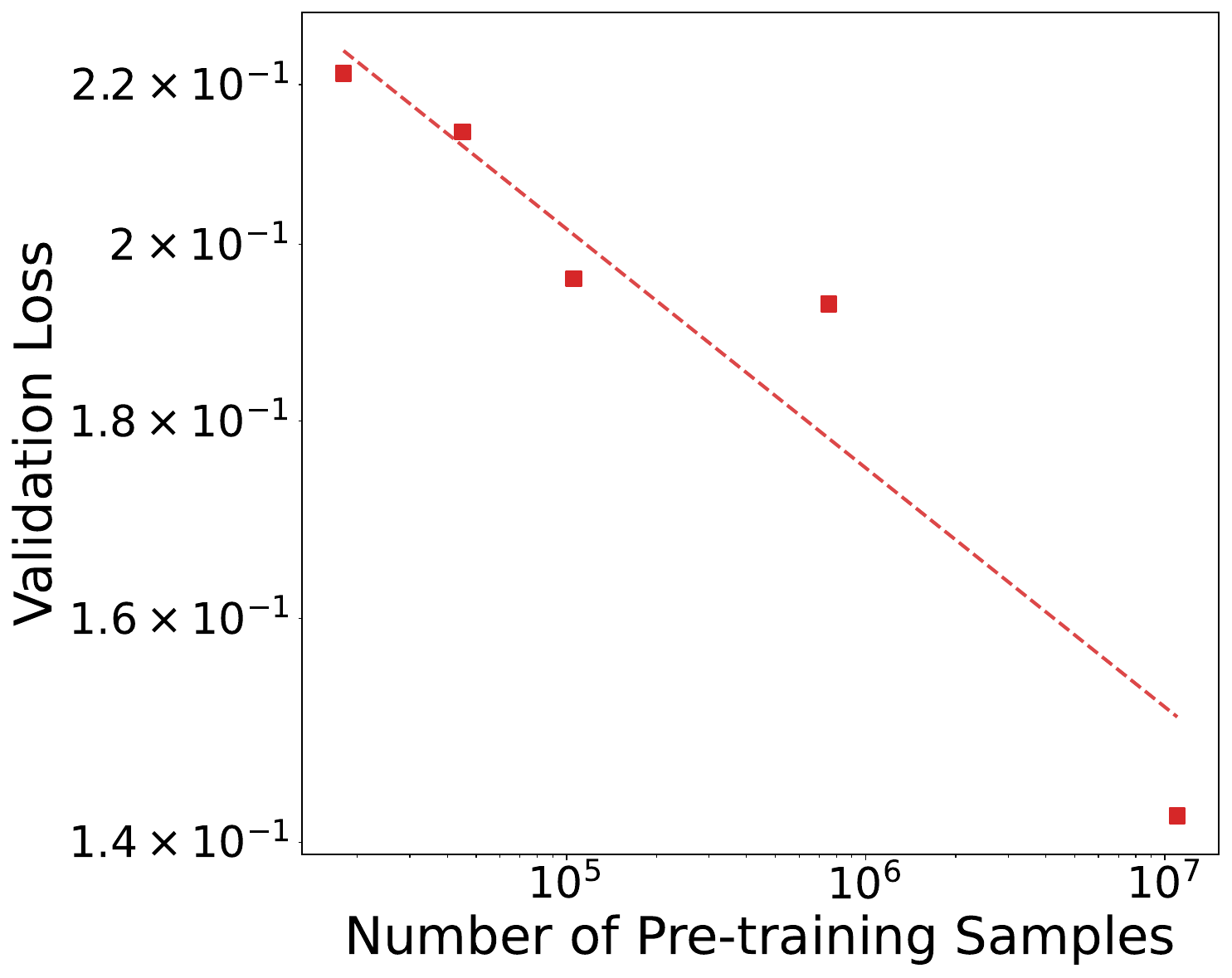}
        \caption{JEPA}
        \label{fig:app_scaling_jepa}
    \end{subfigure}
    
    \vspace{0.5em}

    % Second row
    \begin{subfigure}{0.32\textwidth}
        \centering
        \includegraphics[width=\textwidth]{resources/scaling_cpc_val_loss.pdf}
        \caption{CPC}
        \label{fig:app_scaling_cpc}
    \end{subfigure}
    \quad\quad\quad\quad
    \begin{subfigure}{0.32\textwidth}
        \centering
        \includegraphics[width=\textwidth]{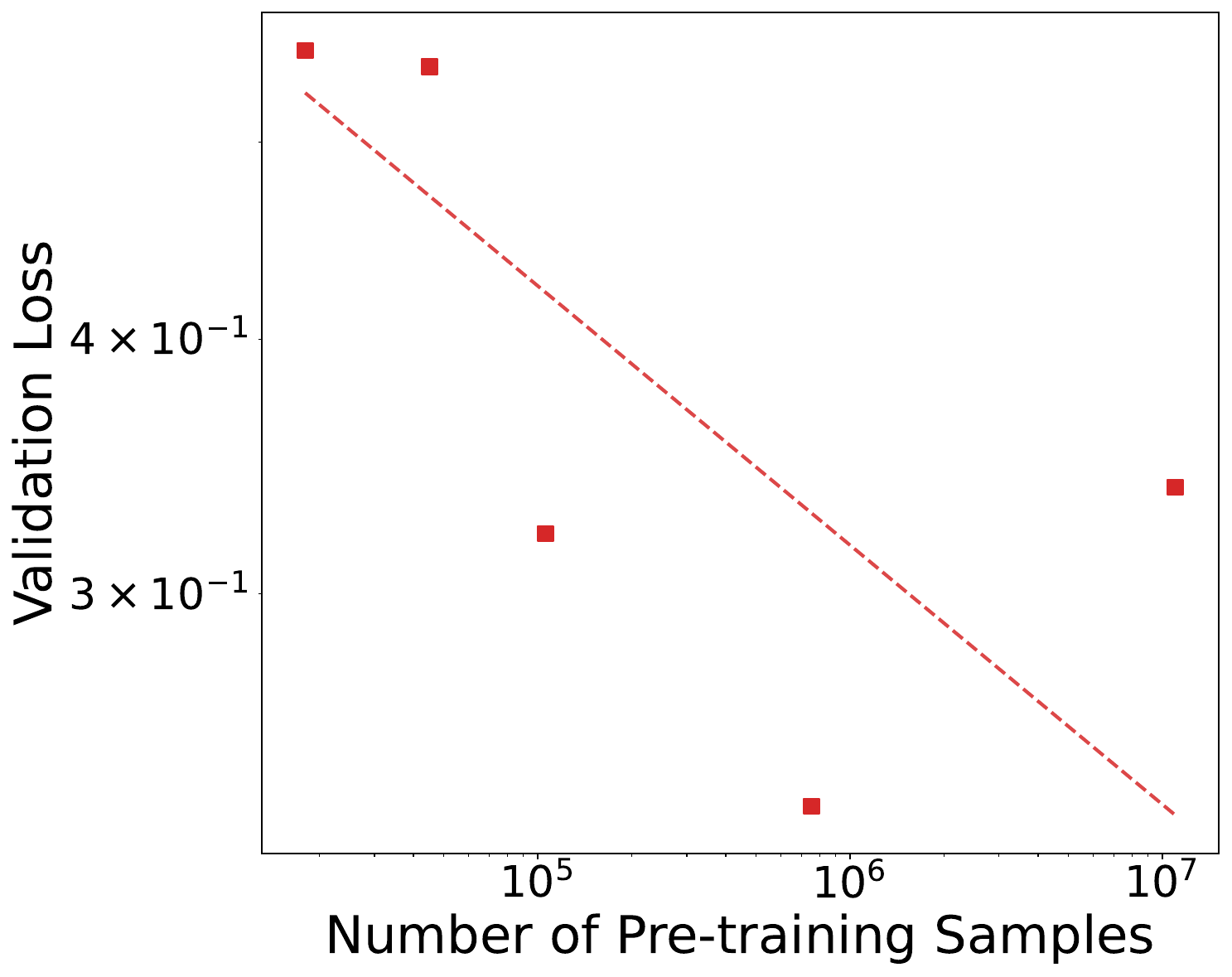}
        \caption{HuBERT++}
        \label{fig:app_scaling_hubert++}
    \end{subfigure}
    \caption{Validation loss as a function of pretraining dataset size for each model. Dashed lines show power-law fits of the form $f(n) = C \cdot N^{-\alpha}$.}
\end{figure}

\begin{table}[ht]
    \centering
    \caption{Power-law fit parameters $f(n) = C \cdot N^{-\alpha}$ for validation loss across models.\newline}
    \label{tab:power_law_fits}
    \begin{tabular}{lrrrr}
        \toprule
        \textbf{Model} & \textbf{$C$} & \textbf{$\alpha$} & \textbf{$R^2$} \\
        \midrule
        data2vec  & $2.304$               & $0.0399$ & $0.7592$ \\
        DinoSR    & $1.749$               & $0.03538$ & $0.3616$ \\
        JEPA      & $0.4119$              & $0.06194$ & $0.9127$ \\
        CPC       & $1.601 \times 10^{4}$ & $0.1889$ & $0.9702$ \\
        HuBERT++  & $1.838$               & $0.1272$ & $0.5606$ \\
        \bottomrule
    \end{tabular}
\end{table}

\newpage
\subsection{Pretraining Dataset Scaling with AUROC}
\label{app:scaling_auroc}
We fit scaling curves of the parametric form $C\cdot N^{-\alpha} + L_0$, where $N$ is the training set size, $\alpha$ governs the rate of improvement, and $L_0$ captures the residual error floor. Not all method/dataset combination show a consistent scaling. Based on fit quality, the clearest scaling behavior overall is seen on the PTB-XL (super) dataset/task, on a comparably coarse prediction task with only 5 classes.

\begin{figure}[htbp]
    \centering
    % First row
    \begin{subfigure}{0.32\textwidth}
        \centering
        \includegraphics[width=\textwidth]{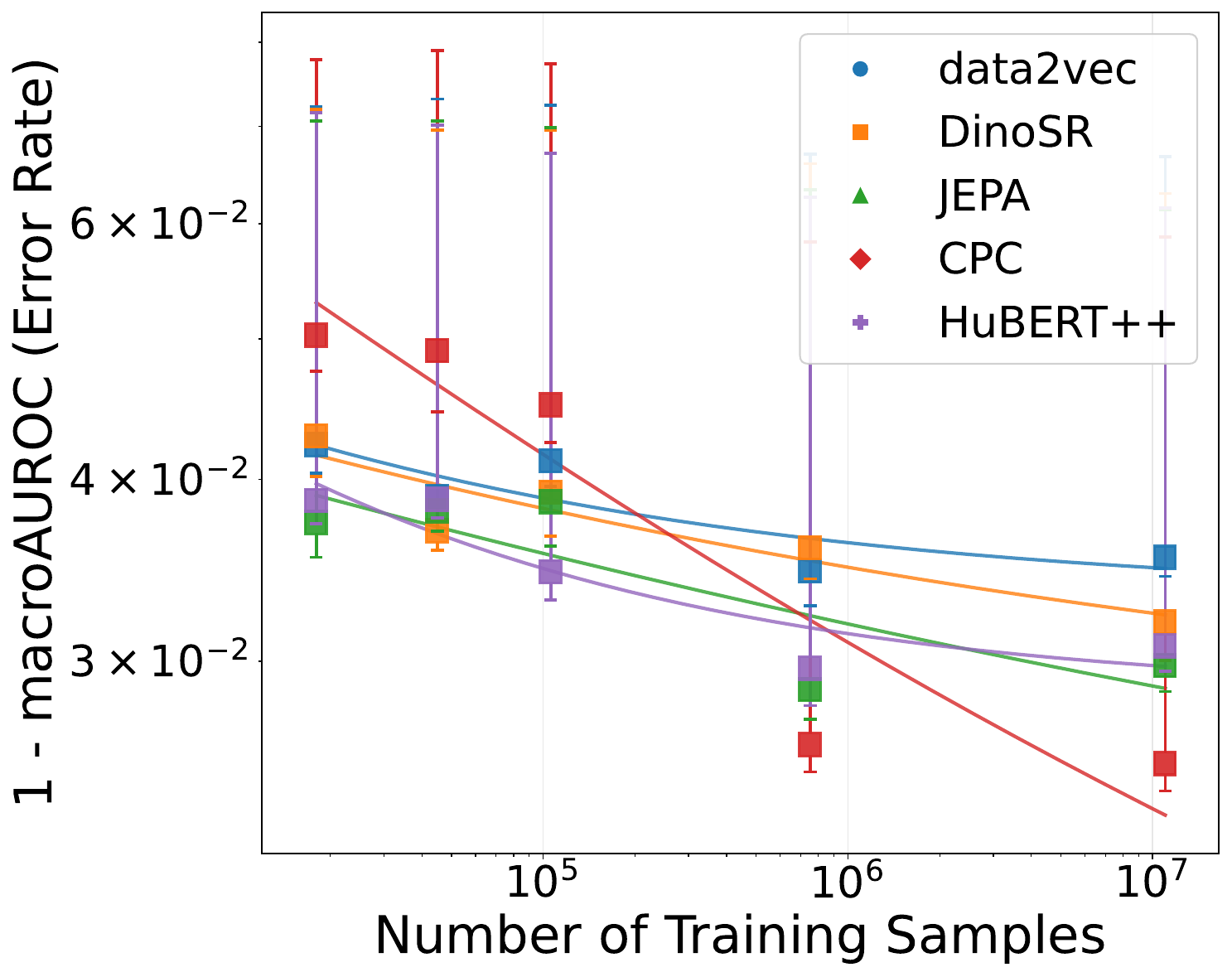}
        \caption{Ningbo}
        \label{fig:app_scaling_ningbo}
    \end{subfigure}
    \hfill
    \begin{subfigure}{0.32\textwidth}
        \centering
        \includegraphics[width=\textwidth]{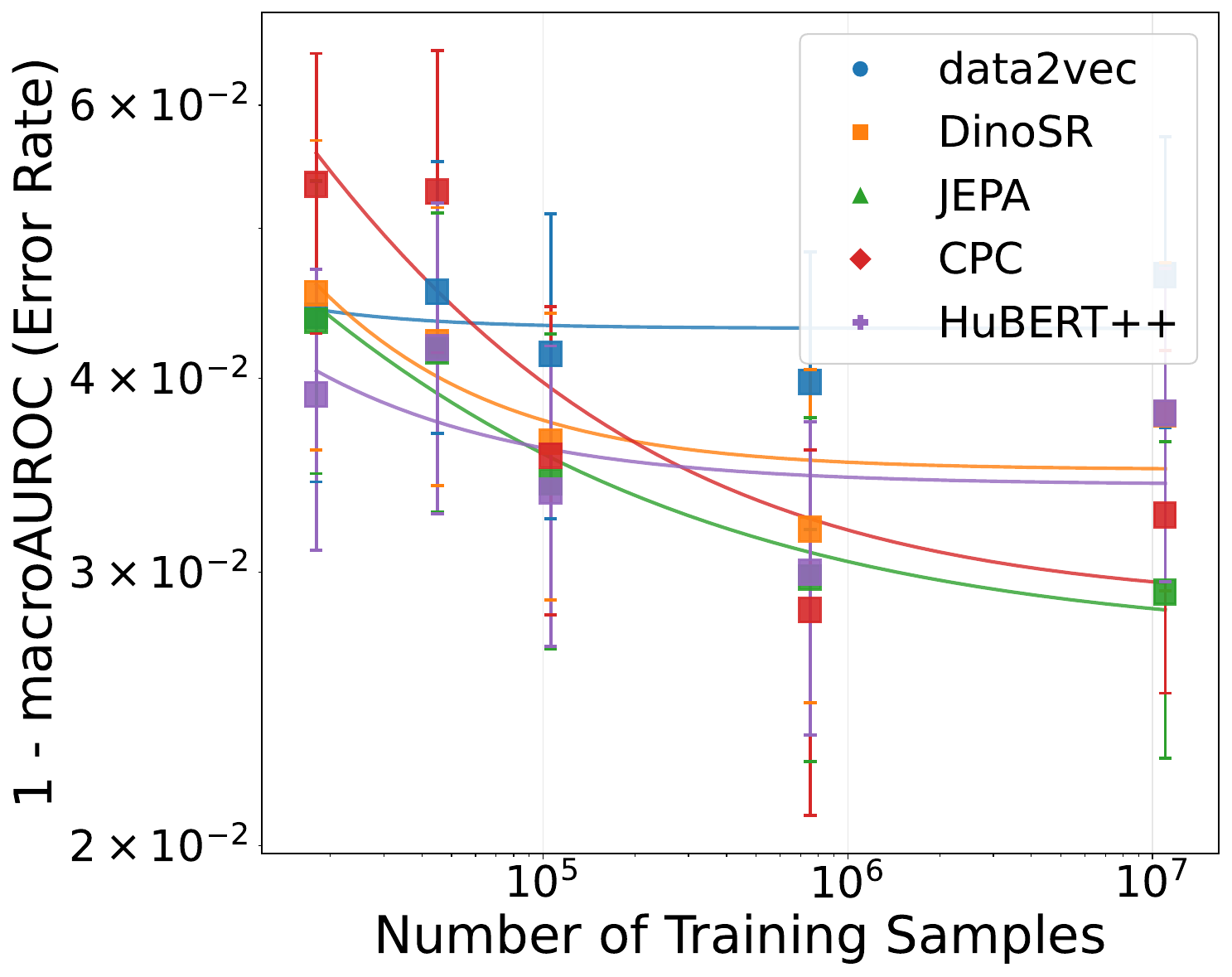}
        \caption{CPSC2018}
        \label{fig:app_scaling_cpsc2018}
    \end{subfigure}
    \hfill
    \begin{subfigure}{0.32\textwidth}
        \centering
        \includegraphics[width=\textwidth]{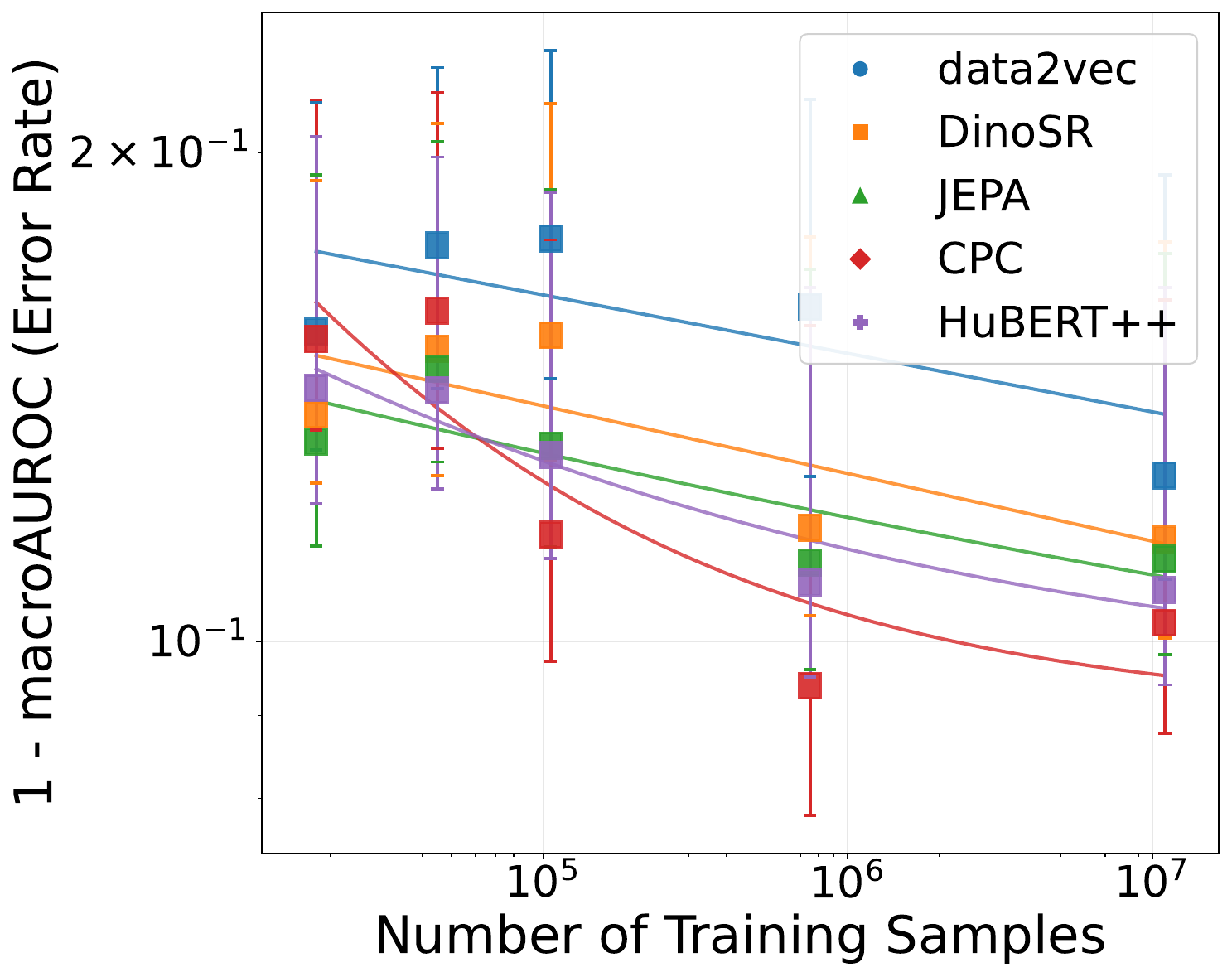}
        \caption{CPSC-Extra}
        \label{fig:app_scaling_cpsc_extra}
    \end{subfigure}
    
    \vspace{0.5em}

    % Second row
    \begin{subfigure}{0.32\textwidth}
        \centering
        \includegraphics[width=\textwidth]{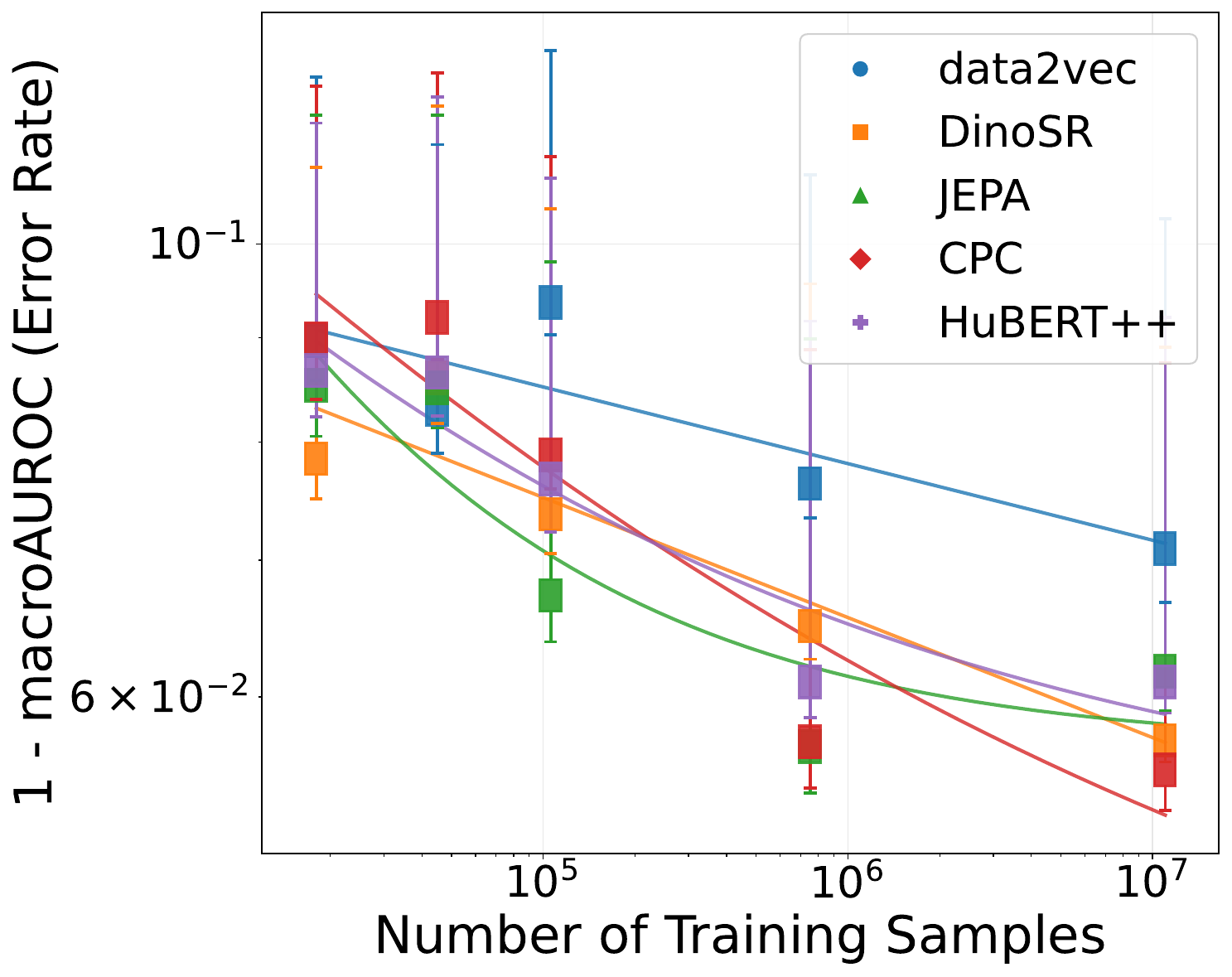}
        \caption{PTB-XL (all)}
        \label{fig:app_scaling_ptbxl_all}
    \end{subfigure}
    \hfill
    \begin{subfigure}{0.32\textwidth}
        \centering
        \includegraphics[width=\textwidth]{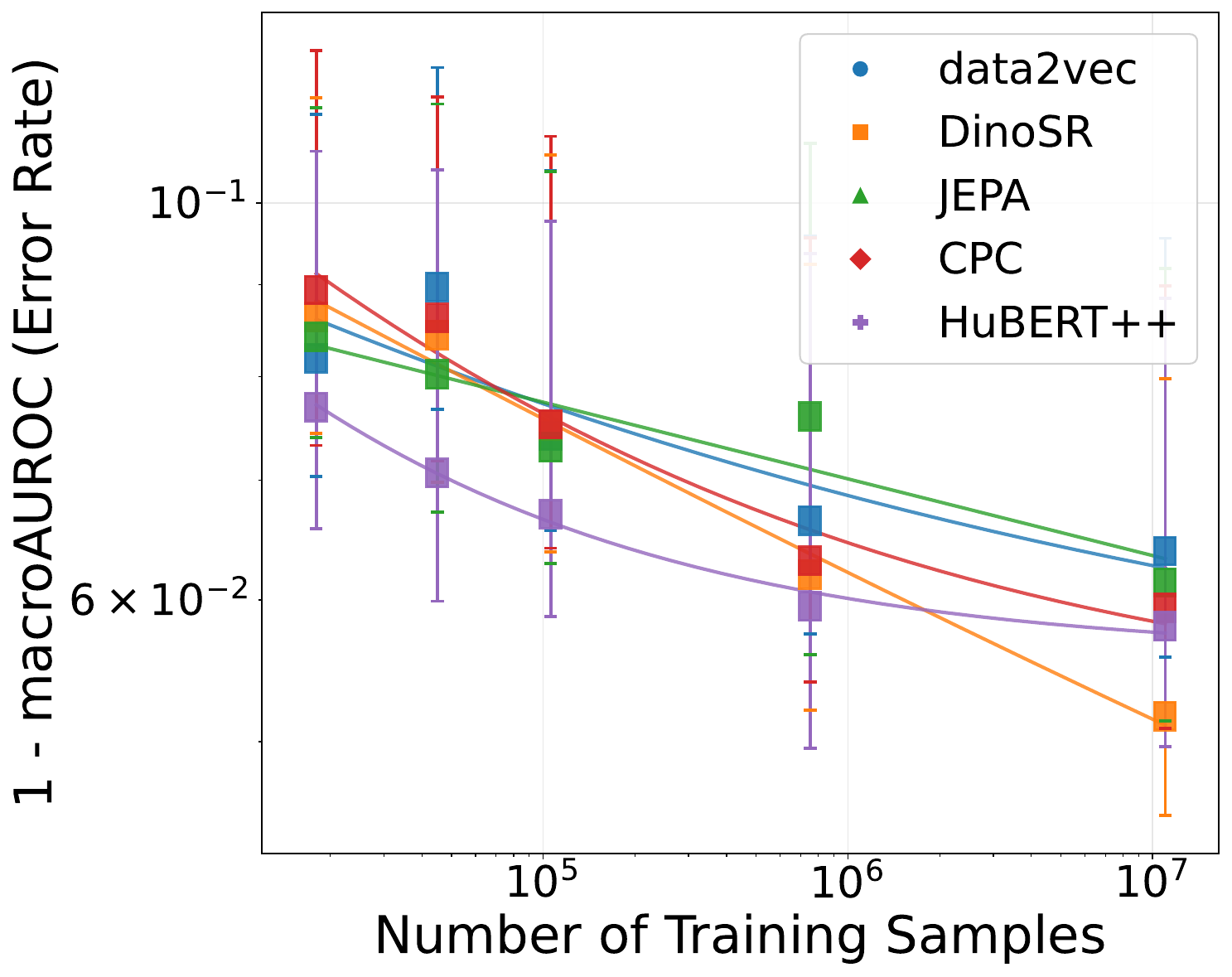}
        \caption{PTB-XL (sub)}
        \label{fig:app_scaling_ptbxl_sub}
    \end{subfigure}
    \hfill
    \begin{subfigure}{0.32\textwidth}
        \centering
        \includegraphics[width=\textwidth]{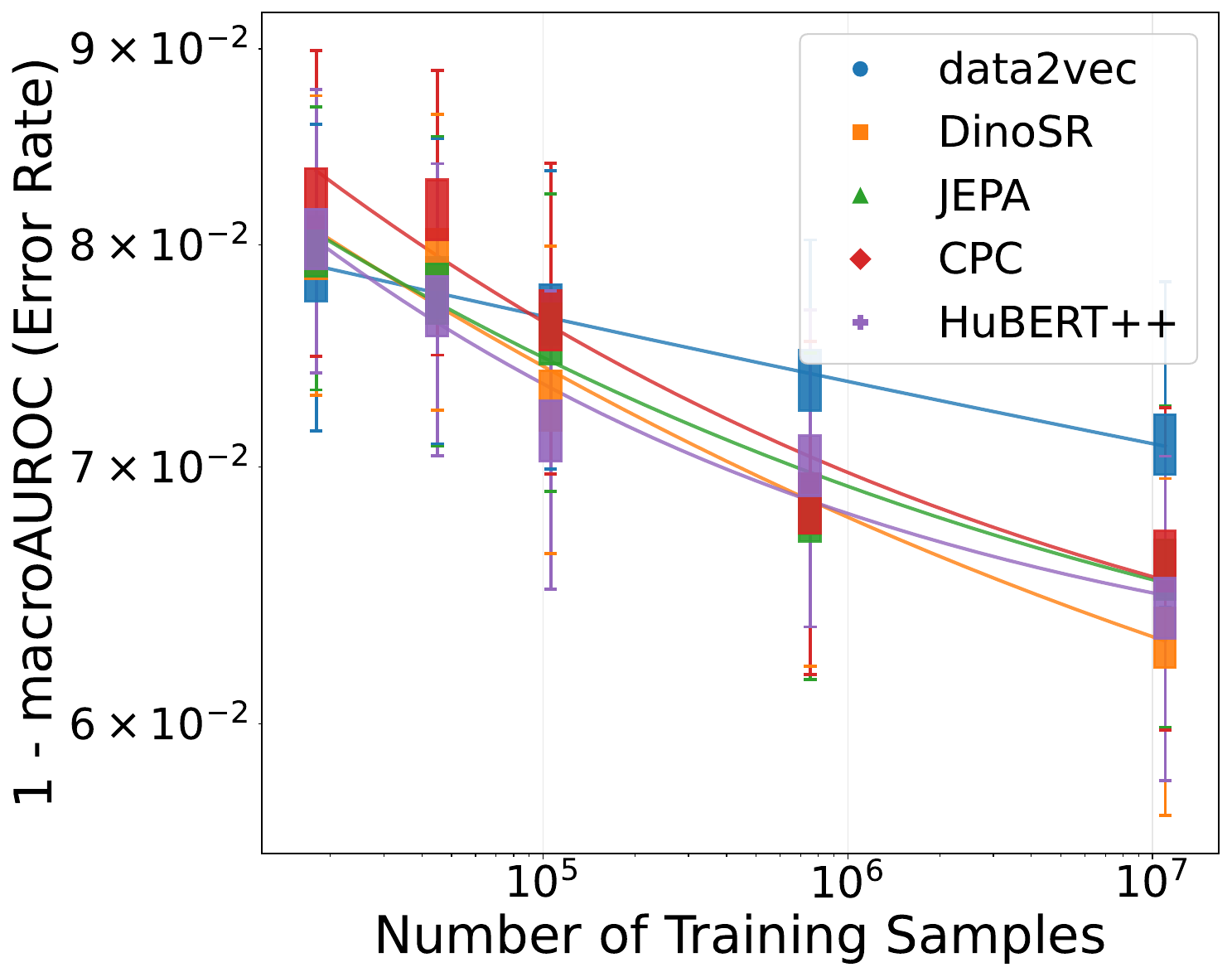}
        \caption{PTB-XL (super)}
        \label{fig:app_scaling_ptbxl_super}
    \end{subfigure}

    \vspace{0.5em}

    % Third row
    \begin{subfigure}{0.32\textwidth}
        \centering
        \includegraphics[width=\textwidth]{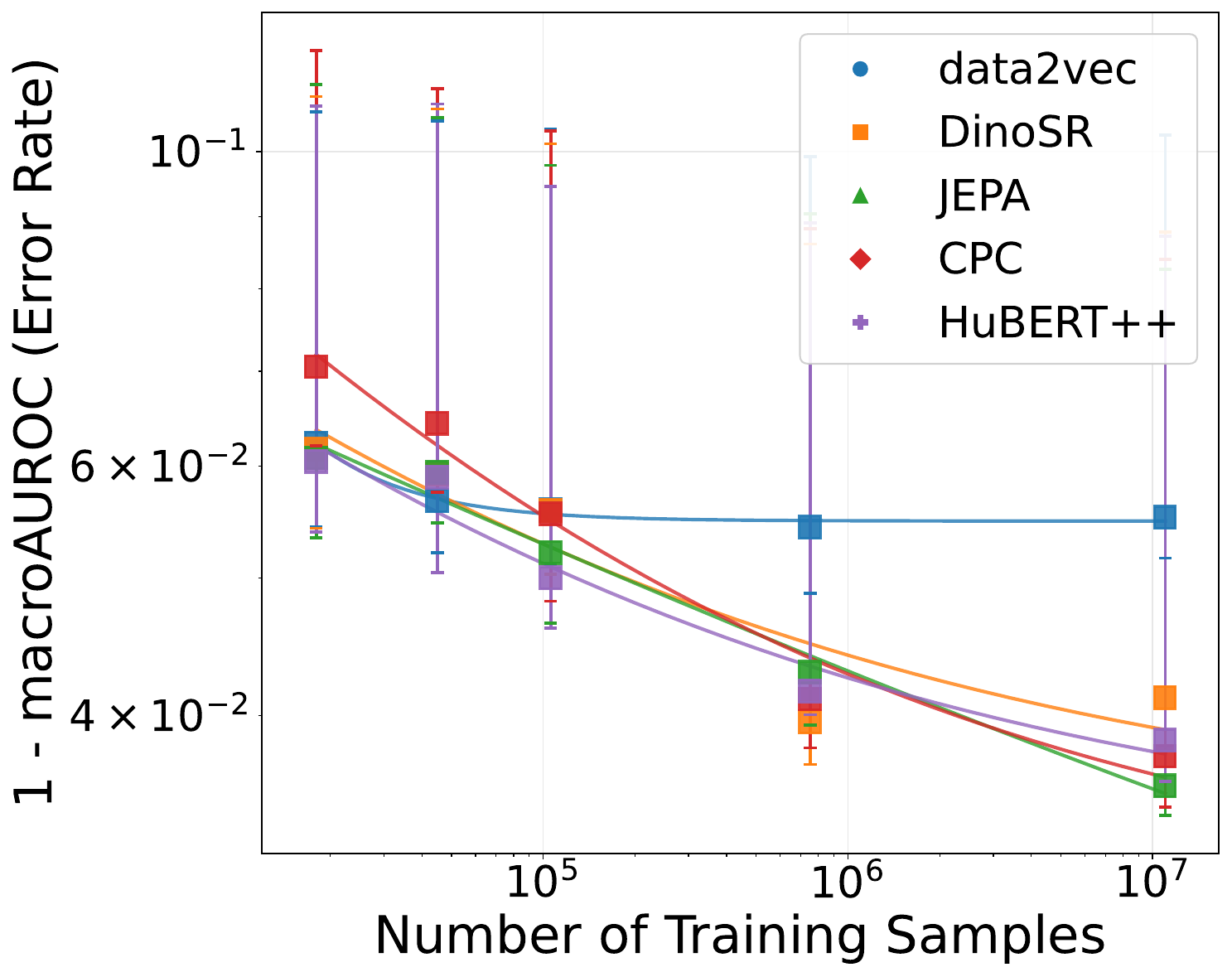}
        \caption{Chapman}
        \label{fig:app_scaling_chapman}
    \end{subfigure}
    \hfill
    \begin{subfigure}{0.32\textwidth}
        \centering
        \includegraphics[width=\textwidth]{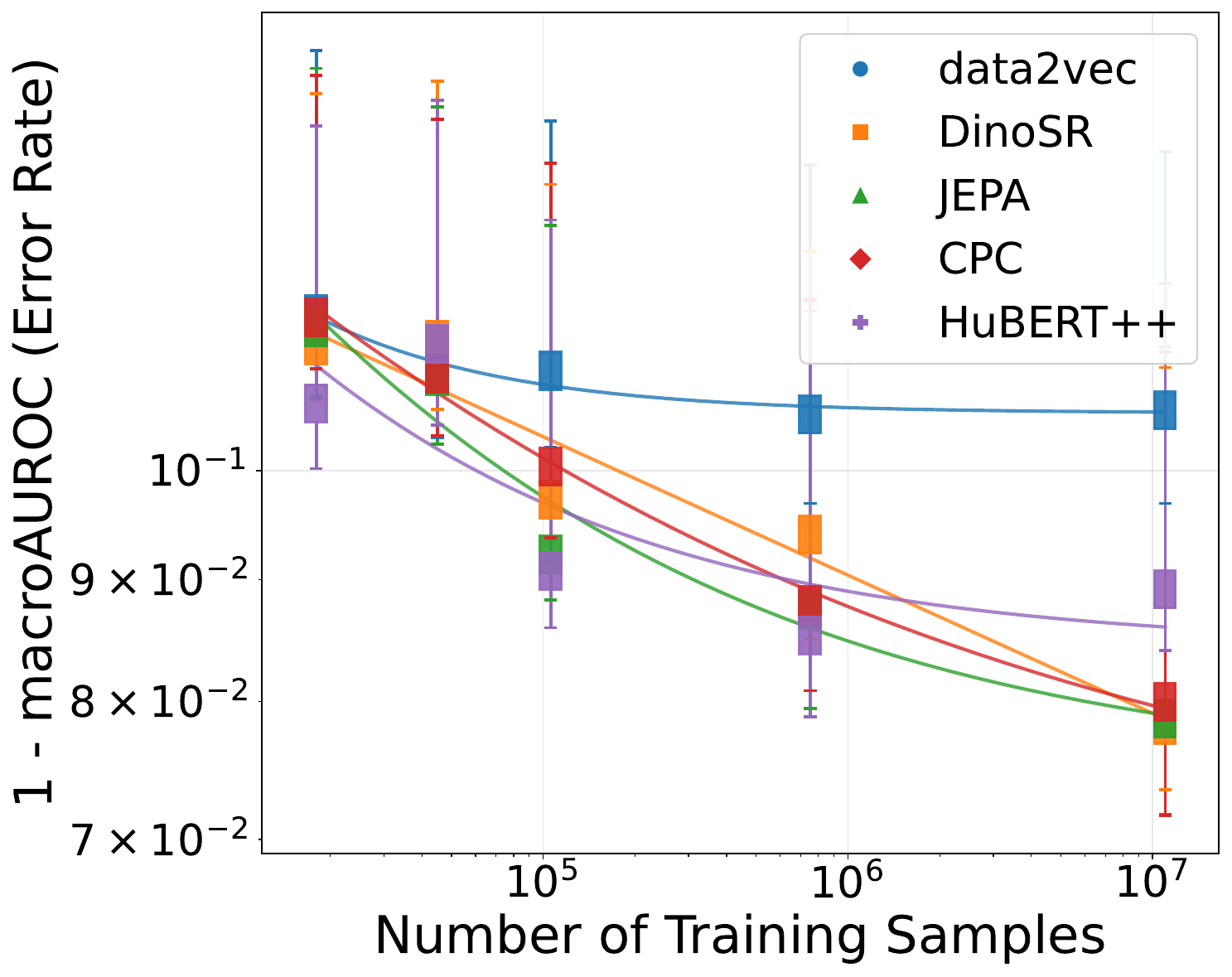}
        \caption{Georgia}
        \label{fig:app_scaling_georgia}
    \end{subfigure}
    \hfill
    \begin{subfigure}{0.32\textwidth}
        \centering
        \includegraphics[width=\textwidth]{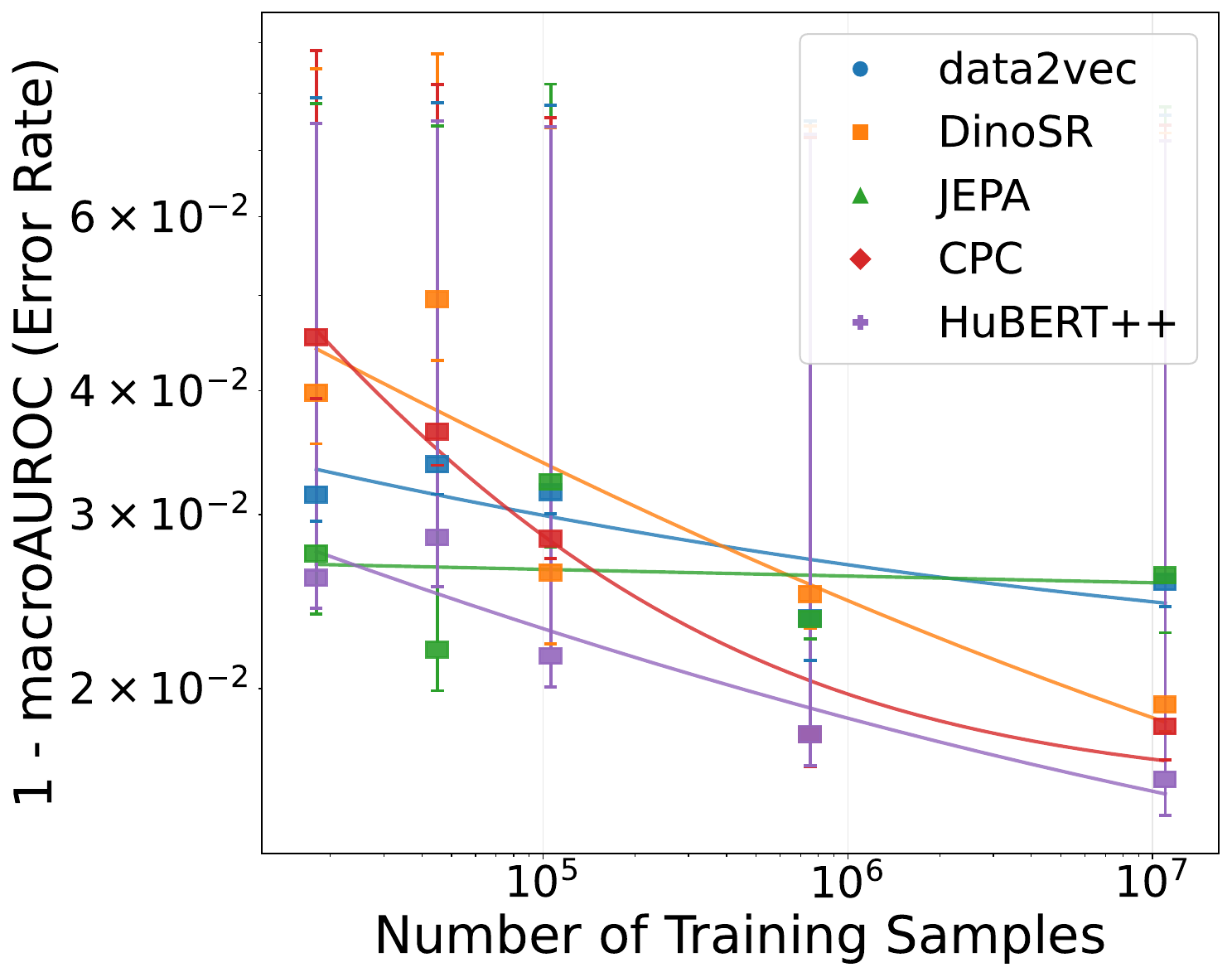}
        \caption{SPH}
        \label{fig:app_scaling_sph}
    \end{subfigure}

    \vspace{0.5em}

    % Fourth row
    \begin{subfigure}{0.32\textwidth}
        \centering
        \includegraphics[width=\textwidth]{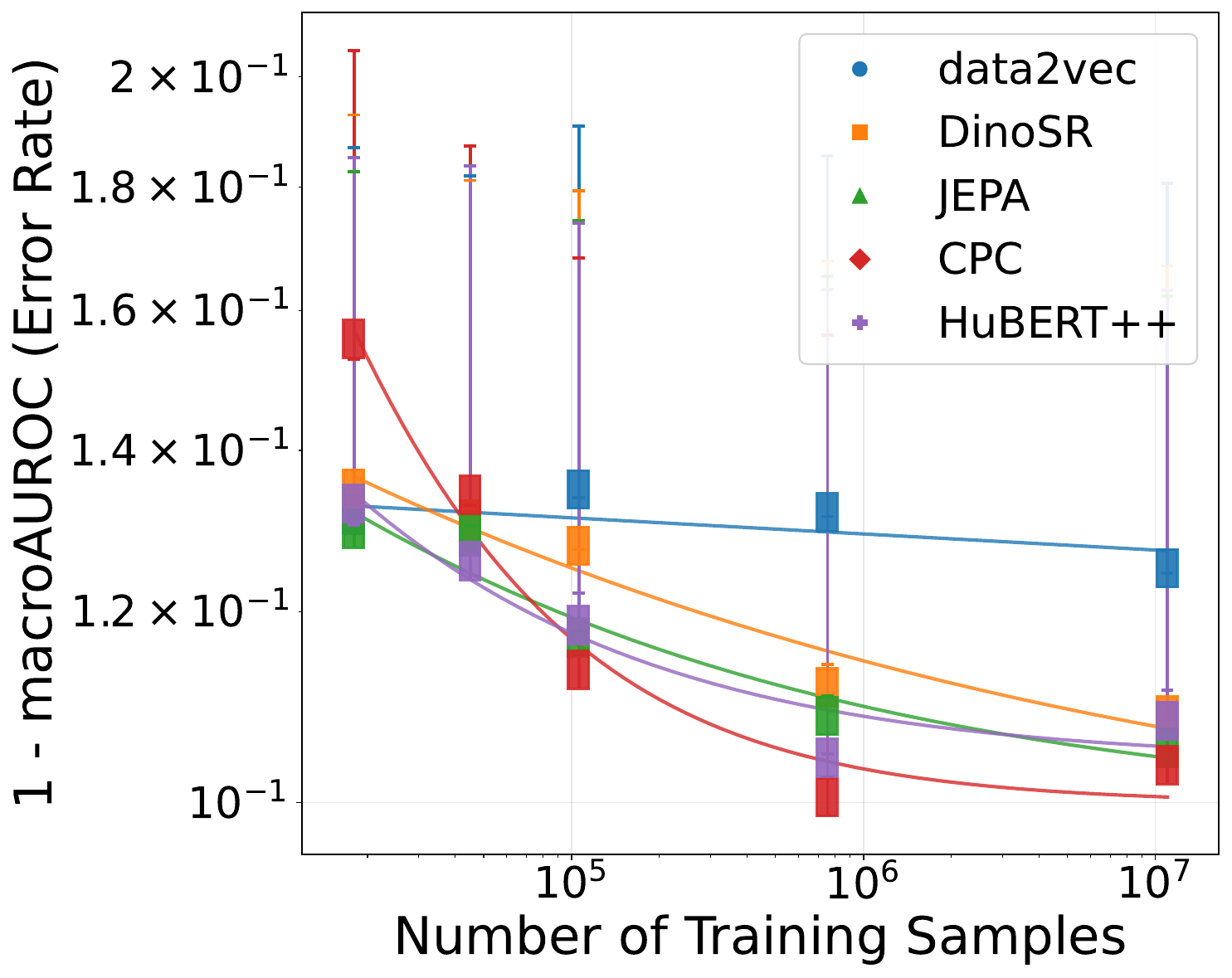}
        \caption{ZZU pECG}
        \label{fig:app_scaling_zzu_pecg}
    \end{subfigure}
    \quad\quad\quad\quad
    \begin{subfigure}{0.32\textwidth}
        \centering
        \includegraphics[width=\textwidth]{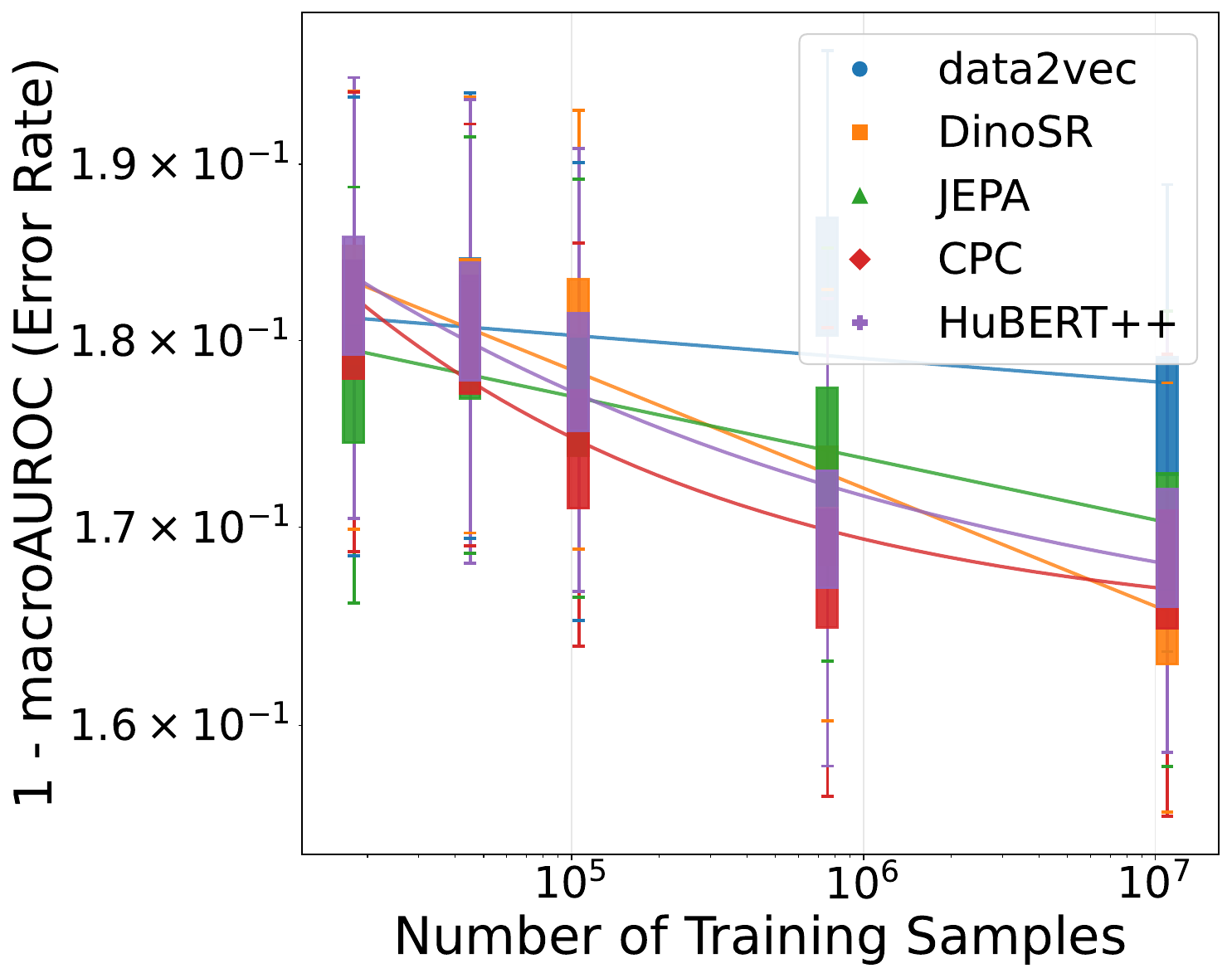}
        \caption{EchoNext}
        \label{fig:app_scaling_echonext}
    \end{subfigure}

    \caption{Scaling analysis for adult ECG interpretation task category datasets.}
\label{fig:app_scaling_sample_size}
\end{figure}

% \newpage

\begin{table}[htbp!]
    \centering
    \caption{Fit parameters for the scaling analysis of Ningbo dataset.\newline}
    \begin{tabular}{lcccc}
        \toprule
        \textbf{Model} & $C$ & $\alpha$ & $L_0$ & $R^2$ \\
        \midrule
        data2vec & 0.1352 & 0.2761 & 0.0332 & 0.757 \\
        DinoSR   & 0.0631 & 0.1436 & 0.0261 & 0.827 \\
        JEPA     & 0.0581 & 0.1014 & 0.0175 & 0.708 \\
        CPC      & 0.2428 & 0.1752 & 0.0093 & 0.894 \\
        HuBERT++ & 0.3262 & 0.3438 & 0.0285 & 0.861 \\
        \bottomrule
    \end{tabular}
    \label{tab:scaling_params_ningbo}
\end{table}

\begin{table}[htbp!]
    \centering
    \caption{Fit parameters for the scaling analysis of CPSC2018 dataset.\newline}
    \begin{tabular}{lcccc}
        \toprule
        \textbf{Model} & $C$ & $\alpha$ & $L_0$ & $R^2$ \\
        \midrule
        data2vec & 48.6179 & 1.0801 & 0.0431 & 0.033 \\
        DinoSR   & 38.3061 & 0.8317 & 0.0349 & 0.753 \\
        JEPA     &  0.9634 & 0.4093 & 0.0271 & 0.939 \\
        CPC      &  4.4522 & 0.5198 & 0.0286 & 0.818 \\
        HuBERT++ &  5.6429 & 0.6937 & 0.0342 & 0.294 \\
        \bottomrule
    \end{tabular}
    \label{tab:scaling_params_cpsc2018}
\end{table}

\begin{table}[htbp!]
    \centering
    \caption{Fit parameters for the scaling analysis of CPSC-Extra dataset.\newline}
    \begin{tabular}{lcccc}
        \toprule
        \textbf{Model} & $C$ & $\alpha$ & $L_0$ & $R^2$ \\
        \midrule
        data2vec & 0.2475 & 0.0360 & 0.0000 & 0.518 \\
        DinoSR   & 0.2257 & 0.0417 & 0.0000 & 0.595 \\
        JEPA     & 0.1709 & 0.0869 & 0.0678 & 0.676 \\
        CPC      & 4.0753 & 0.4125 & 0.0902 & 0.789 \\
        HuBERT++ & 0.5334 & 0.2324 & 0.0924 & 0.912 \\
        \bottomrule
    \end{tabular}
    \label{tab:scaling_params_cpsc_extra}
\end{table}

\begin{table}[htbp!]
    \centering
    \caption{Fit parameters for the scaling analysis of PTB-XL (all) dataset.\newline}
    \begin{tabular}{lcccc}
        \toprule
        \textbf{Model} & $C$ & $\alpha$ & $L_0$ & $R^2$ \\
        \midrule
        data2vec & 0.1311 & 0.0376 & 0.0000 & 0.695 \\
        DinoSR   & 0.1479 & 0.0588 & 0.0000 & 0.847 \\
        JEPA     & 3.2329 & 0.4721 & 0.0566 & 0.827 \\
        CPC      & 0.4017 & 0.1964 & 0.0359 & 0.893 \\
        HuBERT++ & 0.4598 & 0.2535 & 0.0513 & 0.903 \\
        \bottomrule
    \end{tabular}
    \label{tab:scaling_params_ptbxl_all}
\end{table}

\begin{table}[htbp!]
    \centering
    \caption{Fit parameters for the scaling analysis of PTB-XL (sub) dataset.\newline}
    \begin{tabular}{lcccc}
        \toprule
        \textbf{Model} & $C$ & $\alpha$ & $L_0$ & $R^2$ \\
        \midrule
        data2vec & 0.1791 & 0.1625 & 0.0497 & 0.756 \\
        DinoSR   & 0.2114 & 0.1041 & 0.0120 & 0.982 \\
        JEPA     & 0.1268 & 0.0429 & 0.0000 & 0.845 \\
        CPC      & 0.5924 & 0.2752 & 0.0513 & 0.963 \\
        HuBERT++ & 1.0934 & 0.4022 & 0.0559 & 0.992 \\
        \bottomrule
    \end{tabular}
    \label{tab:scaling_params_ptbxl_sub}
\end{table}

\begin{table}[htbp!]
    \centering
    \caption{Fit parameters for the scaling analysis of PTB-XL (super) dataset.\newline}
    \begin{tabular}{lcccc}
        \toprule
        \textbf{Model} & $C$ & $\alpha$ & $L_0$ & $R^2$ \\
        \midrule
        data2vec & 0.0555 & 0.0366 & 0.0402 & 0.994 \\
        DinoSR   & 0.1156 & 0.1372 & 0.0506 & 0.964 \\
        JEPA     & 0.1207 & 0.1687 & 0.0575 & 0.967 \\
        CPC      & 0.1717 & 0.1942 & 0.0580 & 0.948 \\
        HuBERT++ & 0.2071 & 0.2407 & 0.0606 & 0.958 \\
        \bottomrule
    \end{tabular}
    \label{tab:scaling_params_ptbxl_super}
\end{table}

\begin{table}[htbp!]
    \centering
    \caption{Fit parameters for the scaling analysis of Chapman dataset.\newline}
    \begin{tabular}{lcccc}
        \toprule
        \textbf{Model} & $C$ & $\alpha$ & $L_0$ & $R^2$ \\
        \midrule
        data2vec & 7209.1044 & 1.4070 & 0.0549 & 0.984 \\
        DinoSR   &    0.3651 & 0.2530 & 0.0330 & 0.881 \\
        JEPA     &    0.1618 & 0.1190 & 0.0117 & 0.979 \\
        CPC      &    0.7315 & 0.2909 & 0.0296 & 0.979 \\
        HuBERT++ &    0.3795 & 0.2583 & 0.0318 & 0.955 \\
        \bottomrule
    \end{tabular}
    \label{tab:scaling_params_chapman}
\end{table}

\begin{table}[htbp!]
    \centering
    \caption{Fit parameters for the scaling analysis of Georgia dataset.\newline}
    \begin{tabular}{lcccc}
        \toprule
        \textbf{Model} & $C$ & $\alpha$ & $L_0$ & $R^2$ \\
        \midrule
        data2vec &  13.1679 & 0.7297 & 0.1057 & 0.935 \\
        DinoSR   &   0.2007 & 0.0593 & 0.0018 & 0.923 \\
        JEPA     &   1.2382 & 0.3460 & 0.0744 & 0.946 \\
        CPC      &   0.4540 & 0.2273 & 0.0680 & 0.992 \\
        HuBERT++ &   1.7855 & 0.4301 & 0.0843 & 0.663 \\
        \bottomrule
    \end{tabular}
    \label{tab:scaling_params_georgia}
\end{table}

\begin{table}[htbp!]
    \centering
    \caption{Fit parameters for the scaling analysis of SPH dataset.\newline}
    \begin{tabular}{lcccc}
        \toprule
        \textbf{Model} & $C$ & $\alpha$ & $L_0$ & $R^2$ \\
        \midrule
        data2vec & 0.0707 & 0.1692 & 0.0199 & 0.661 \\
        DinoSR   & 0.2519 & 0.2001 & 0.0087 & 0.672 \\
        JEPA     & 0.0222 & 0.0089 & 0.0064 & 0.013 \\
        CPC      & 4.1350 & 0.5014 & 0.0157 & 0.981 \\
        HuBERT++ & 0.0954 & 0.1712 & 0.0097 & 0.829 \\
        \bottomrule
    \end{tabular}
    \label{tab:scaling_params_sph}
\end{table}

\begin{table}[htbp!]
    \centering
    \caption{Fit parameters for the scaling analysis of ZZU pECG dataset.\newline}
    \begin{tabular}{lcccc}
        \toprule
        \textbf{Model} & $C$ & $\alpha$ & $L_0$ & $R^2$ \\
        \midrule
        data2vec &  0.1417 & 0.0067 & 0.0000 & 0.369 \\
        DinoSR   &  0.2564 & 0.1844 & 0.0944 & 0.947 \\
        JEPA     &  0.5077 & 0.2785 & 0.0988 & 0.931 \\
        CPC      & 57.4340 & 0.7056 & 0.0999 & 0.977 \\
        HuBERT++ &  3.0097 & 0.4690 & 0.1040 & 0.933 \\
        \bottomrule
    \end{tabular}
    \label{tab:scaling_params_zzu_pecg}
\end{table}

\begin{table}[htbp!]
    \centering
    \caption{Fit parameters for the scaling analysis of EchoNext dataset.\newline}
    \begin{tabular}{lcccc}
        \toprule
        \textbf{Model} & $C$ & $\alpha$ & $L_0$ & $R^2$ \\
        \midrule
        data2vec & 0.1796 & 0.0032 & 0.0072 & 0.196 \\
        DinoSR   & 0.2111 & 0.0160 & 0.0028 & 0.958 \\
        JEPA     & 0.1944 & 0.0082 & 0.0001 & 0.880 \\
        CPC      & 0.4532 & 0.3308 & 0.1647 & 0.921 \\
        HuBERT++ & 0.1656 & 0.2115 & 0.1627 & 0.938 \\
        \bottomrule
    \end{tabular}
    \label{tab:scaling_params_echonext}
\end{table}

\begin{table}[htbp!]
    \centering
    \caption{Weighted average of the scaling exponent $\bar{\alpha}$ across datasets, 
    using $R^2$-weighted and $R^4$-weighted ($(R^2)^2$) aggregation.\newline}
    \begin{tabular}{lcc}
        \toprule
        \textbf{Model} & $\bar{\alpha}\ (R^2\ \text{weighted})$ & $\bar{\alpha}\ ((R^2)^2\ \text{weighted})$ \\
        \midrule
        data2vec  & 0.3813 & 0.4479 \\
        DinoSR    & 0.1772 & 0.1692 \\
        JEPA      & 0.2084 & 0.2131 \\
        CPC       & 0.3474 & 0.3471 \\
        HuBERT++  & 0.3118 & 0.3023 \\
        \bottomrule
    \end{tabular}
    \label{tab:weighted_alpha}
\end{table}

\newpage
\subsection{Correlation Between Pre-training Validation Loss and Downstream Performance}
\label{app:correlation_plot}

\begin{figure}[htbp]
    \centering
    % First row
    \begin{subfigure}{0.32\textwidth}
        \centering
        \includegraphics[width=\textwidth]{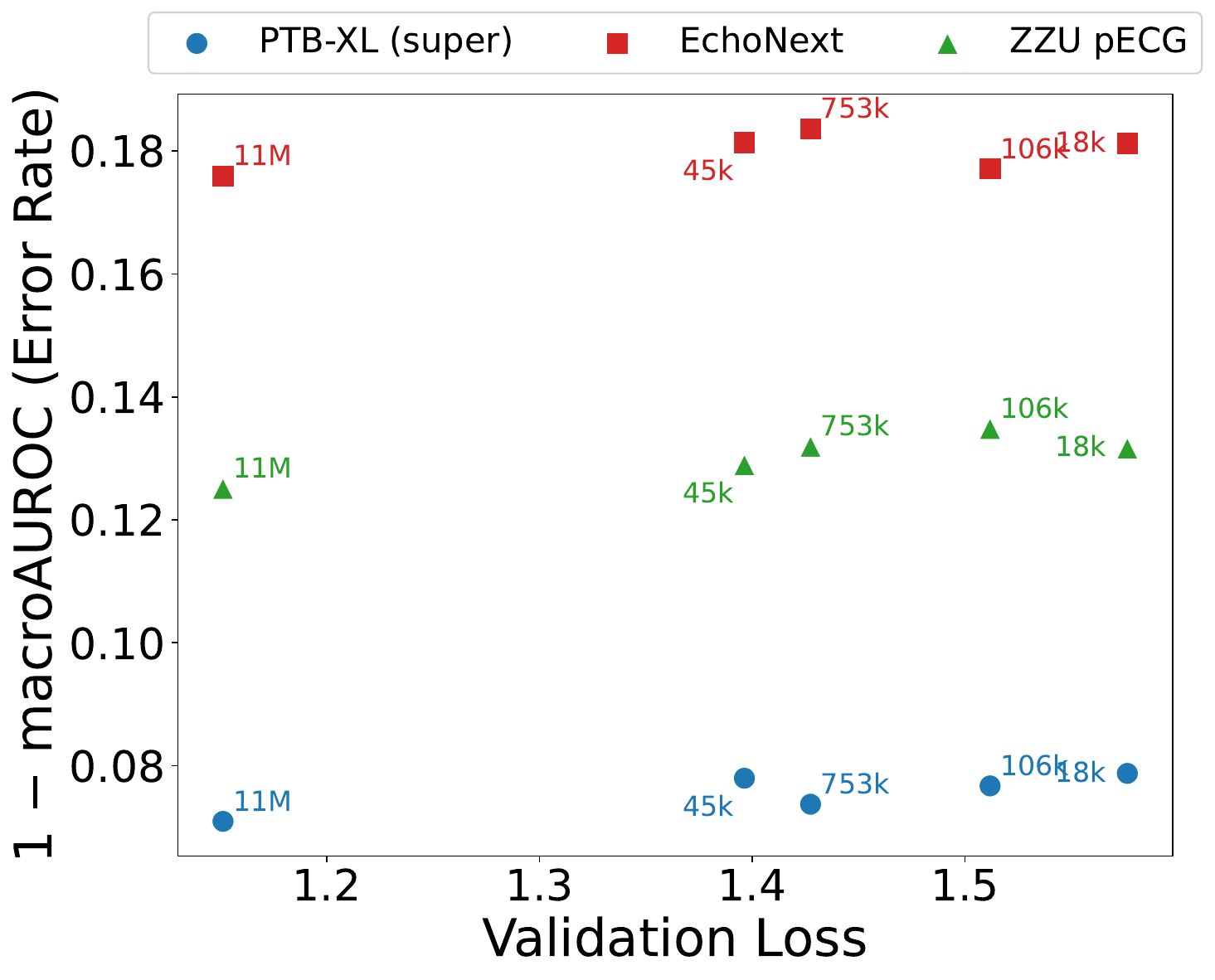}
        \caption{data2vec}
        \label{fig:app_correlation_data2vec}
    \end{subfigure}
    \hfill
    \begin{subfigure}{0.32\textwidth}
        \centering
        \includegraphics[width=\textwidth]{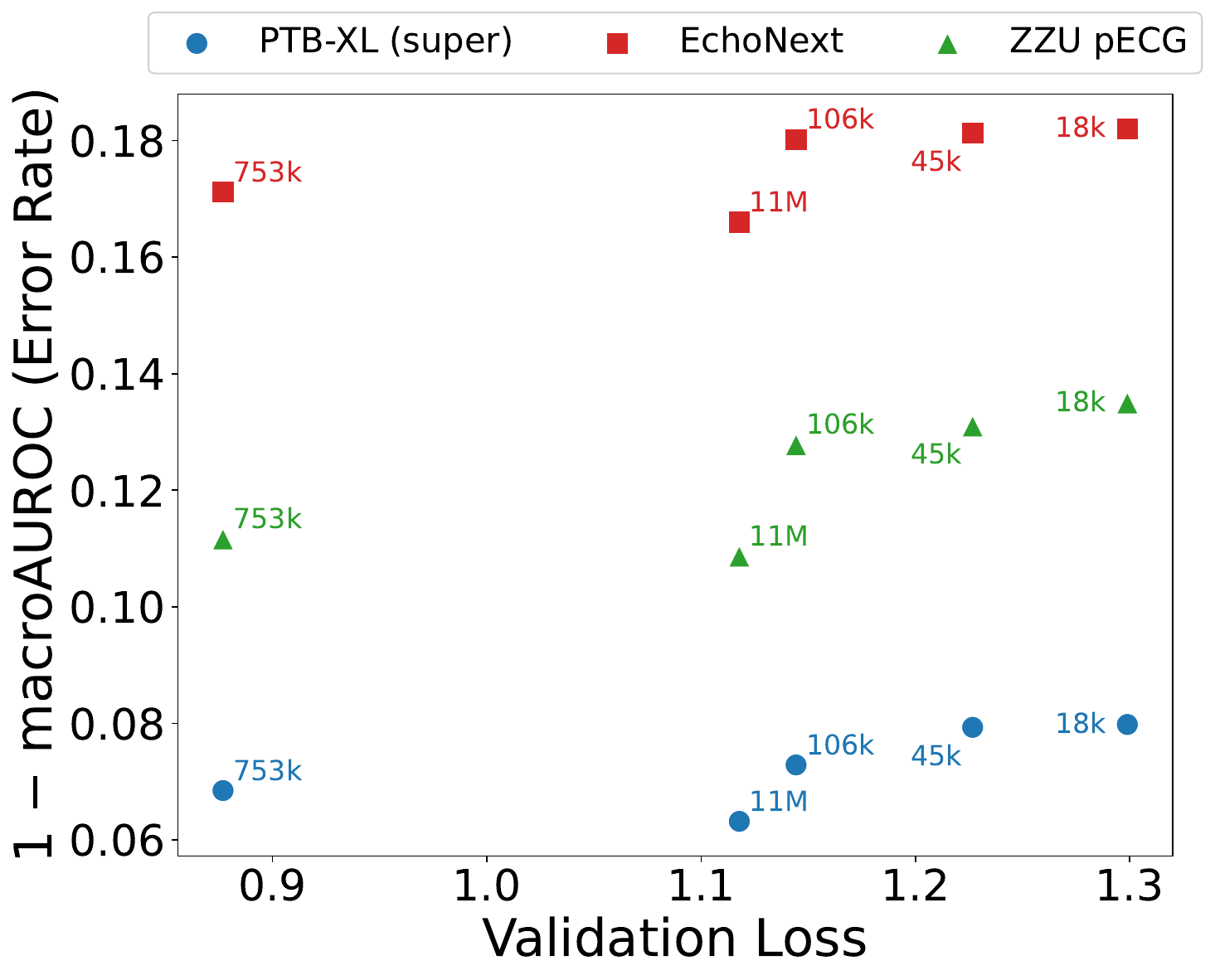}
        \caption{DinoSR}
        \label{fig:app_correlation_dinosr}
    \end{subfigure}
    \hfill
    \begin{subfigure}{0.32\textwidth}
        \centering
        \includegraphics[width=\textwidth]{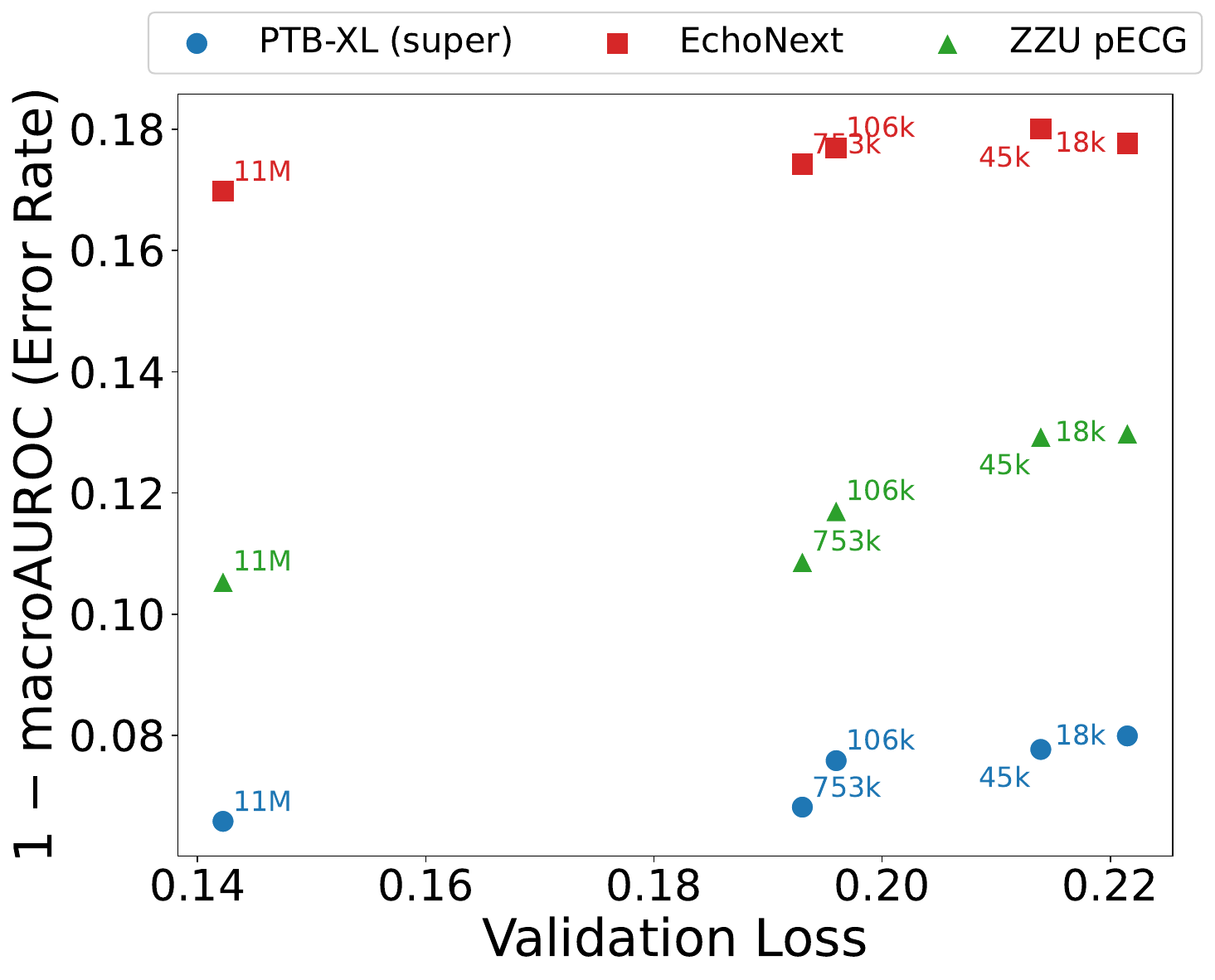}
        \caption{JEPA}
        \label{fig:app_correlation_jepa}
    \end{subfigure}
    
    \vspace{0.5em}

    % Second row
    \begin{subfigure}{0.32\textwidth}
        \centering
        \includegraphics[width=\textwidth]{resources/val_loss_auroc_cpc.pdf}
        \caption{CPC}
        \label{fig:app_correlatoin_cpc}
    \end{subfigure}
    \quad\quad\quad\quad
    \begin{subfigure}{0.32\textwidth}
        \centering
        \includegraphics[width=\textwidth]{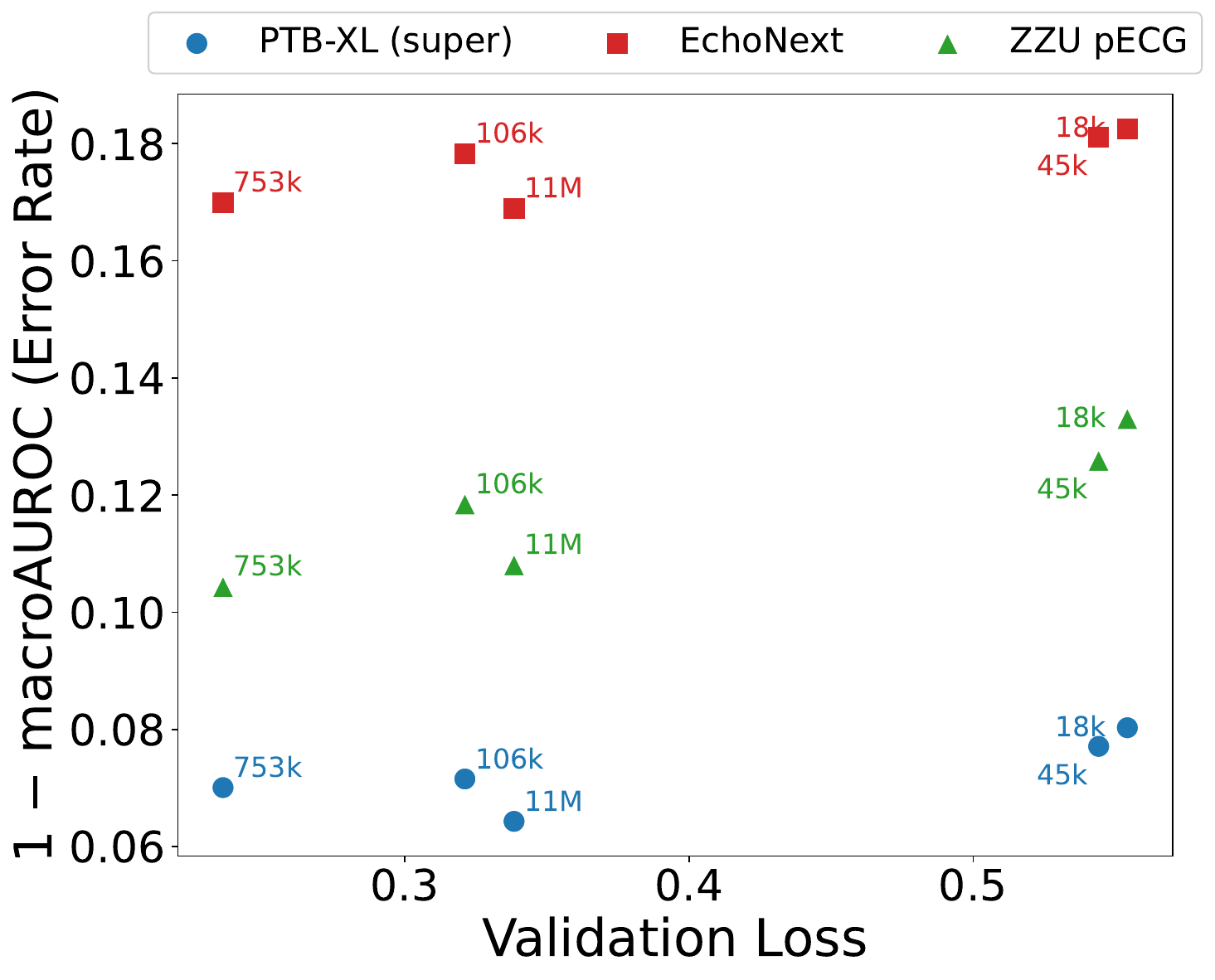}
        \caption{HuBERT++}
        \label{fig:app_correlation_hubert++}
    \end{subfigure}
    \caption{Correlation between pre-training validation loss and downstream classification error (1\,--\,macro-AUROC) on PTB-XL (super) dataset.}
\end{figure}

\begin{table}[ht]
    \centering
    \caption{Spearman's rank correlation ($r$) between pre-training validation loss and downstream error rate (1\,--\,macro-AUROC) for five self-supervised models across three datasets.\newline}
    \begin{tabular}{lcccccc}
        \toprule
        & \multicolumn{2}{c}{\textbf{PTB-XL (super)}} 
        & \multicolumn{2}{c}{\textbf{EchoNext}} 
        & \multicolumn{2}{c}{\textbf{ZZU pECG}} \\
        \cmidrule(lr){2-3}\cmidrule(lr){4-5}\cmidrule(lr){6-7}
        \textbf{Model} & $r$ & $p$ & $r$ & $p$ & $r$ & $p$ \\
        \midrule
        data2vec  & 0.700 & 0.188 & 0.200 & 0.747 & 0.700 & 0.188 \\
        DinoSR    & 0.900 & 0.037 & 0.900 & 0.037 & 0.900 & 0.037 \\
        JEPA      & 1.000 & 0.000 & 0.900 & 0.037 & 1.000 & 0.000 \\
        CPC       & 1.000 & 0.000 & 1.000 & 0.000 & 0.900 & 0.037 \\
        HuBERT++  & 0.700 & 0.188 & 0.700 & 0.188 & 0.900 & 0.037 \\
        \bottomrule
    \end{tabular}
    \label{tab:spearman_correlation}
\end{table}

\newpage
\section{EchoNext Label Efficiency}
\label{app:echonext_label_efficiency}
To analyze label efficiency, we use the same parametric Ansatz as used in Appendix~\ref{app:scaling_auroc}.
\begin{table}[htbp!]
    \centering
    \caption{Fit parameters for the EchoNext label efficiency.\newline}
    \begin{tabular}{lcccc}
        \toprule
        \textbf{Model} & $C$ & $\alpha$ & $L_0$ & $R^2$ \\
        \midrule
        data2vec & 0.5207 & 0.1107 & 0.0000 & 0.984 \\
        DinoSR   & 0.4843 & 0.1274 & 0.0231 & 0.997 \\
        JEPA     & 0.4448 & 0.0986 & 0.0000 & 0.959 \\
        CPC      & 0.4565 & 0.1059 & 0.0000 & 0.988 \\
        HuBERT++ & 0.4684 & 0.1033 & 0.0000 & 0.980 \\
        \bottomrule
    \end{tabular}
    \label{tab:echonext_label_efficiency_params}
\end{table}

% \newpage
\section{Computational Efficiency Analysis}
\label{app:computation_efficiency_analysis}

\begin{table}[htbp]
    \centering
    \scriptsize
    \caption{Unified comparison of computational cost, memory usage, and inference efficiency for all ECG representation learning models. Reported metrics include: (i) GFLOPs for forward (F) and backward (B) passes (lower is better), measured with batch size 1 on an NVIDIA L40; (ii) peak GPU memory during inference (lower is better), measured on PTB-XL (all) with batch size 64; and (iii) throughput (samples/s; higher is better) and latency (ms/sample; lower is better) under the same hardware and batch size. Timesteps correspond to the input sequence length for each model, and parameter counts include all trainable weights. The best model is highlighted in bold face and underlined.\\}
    \begin{tabular}{lcccccc}
        \toprule
        \textbf{Model} & \textbf{Timesteps} & \textbf{Parameters} $\downarrow$ & \textbf{GFLOP (F/B)} $\downarrow$ & \textbf{GPU Mem (MB)} $\downarrow$ & \textbf{Thr}$\uparrow$/ \textbf{Lat}$\downarrow$ \\
        \midrule        
        ECGFounder (CNN) & 1250 & 33.8M & \underline{\textbf{0.602}} / \underline{\textbf{5.066}} & \underline{\textbf{187.37}} & \underline{\textbf{2220.67}} / \underline{\textbf{0.450}} \\
        ECG-JEPA (Transformer) & 2500 & 87.2M & 73.877 / 221.6 & 2136.79 & 98.71 / 10.131 \\
        \midrule
        data2vec (SSM) & 600 & \underline{\textbf{3M}} & 1.741 / 5.213 & 482.59 & 441.10 / 2.267 \\
        DinoSR (SSM) & 600 & \underline{\textbf{3M}} & 1.741 / 5.213 & 482.59 & 436.51 / 2.291 \\
        JEPA (SSM) & 600 & \underline{\textbf{3M}} & 1.741 / 5.213 & 482.59 & 409.39 / 2.443 \\
        CPC (SSM) & 600 & \underline{\textbf{3M}} & 1.741 / 5.213 & 480.53 & 401.14 / 2.493 \\
        HuBERT++ (SSM) & 600 & \underline{\textbf{3M}} & 1.741 / 5.213 & 482.59 & 418.21 / 2.391 \\
        \bottomrule
    \end{tabular}
    \label{tab:efficiency}
\end{table}

\newpage
\section{HuBERT++}
\label{app:hubertpp}
\subsection{Design decisions}
We briefly describe the design decisions underlying HuBERT++. We draw on strong non-contrastive SSL methods such as I-JEPA \citep{assran2023self}, HuBERT \citep{hsu2021hubert}, DINO \citep{caron2021emerging}, DinoSR \citep{liu2023dinosr} and CAPI \citep{darcet2025cluster} from the vision and speech domain. Its design is most closely related to DinoSR and CAPI. We follow the categorization of design choices put forward by \citet{darcet2025cluster}:
\begin{enumerate}
    \item We use exponential moving averages of a student network as prediction target (CAPI Figure 3) like I-JEPA, HuBERT, CAPI
    \item We use clustering as loss (CAPI Figure 4) like CAPI. To avoiding backpropagation through the clustering, we therefore need another non-SGD cluster update.
    \item We use Sinkhorn-Knopp like DINO and unlike DinoSR, which encourages equiparticipation and prevents collapse.
    \item We use granular, soft prediction targets (unlike DinoSR), which would also allow us to use different temperatures (as in DINO).
    \item We use cluster assignmenta using Sinkhorn-Knopp optimal transport (unlike CAPI, which uses a quite ad-hoc procedure). This has the nice side effect that prediction target computation and cluster updates happen consistently
\end{enumerate}

\subsection{Pseudo-code}
We provide pseudo-code for the three most crucial components of the algorithm.
\paragraph{Computing EMA targets}
\begin{lstlisting}[language=Python, numbers=left, basicstyle=\small\ttfamily, breaklines=true]
    @torch.no_grad()
    def get_ema_targets(self, x):
        """
        Get soft targets from EMA path using Sinkhorn-Knopp.
        
        Args:
            x: input images [B, C, T]
        
        Returns:
            targets: [B, K] soft assignment probabilities
        """
        # Extract and normalize EMA features
        z_ema = self.ema_encoder(x)
        z_ema = self.ema_projector(z_ema)
        z_ema = F.normalize(z_ema, dim=1)
        
        # Compute similarity to prototypes
        logits = z_ema @ self.ema_prototypes.T / self.temperature  # [B, K]
        
        # Apply Sinkhorn-Knopp for balanced soft assignments
        targets = sinkhorn_knopp(
            logits, 
            num_iters=self.sinkhorn_iters,
            epsilon=self.sinkhorn_epsilon
        )
        
        return targets, z_ema
\end{lstlisting}

\paragraph{Update cluster prototypes}

\begin{lstlisting}[language=Python, numbers=left, basicstyle=\small\ttfamily, breaklines=true]
   @torch.no_grad()
    def update_prototypes(self, ema_features, assignments):
        """
        Update prototypes using soft assignments from Sinkhorn-Knopp.
        
        Args:
            ema_features: [B, D] normalized features from EMA encoder
            assignments: [B, K] soft assignment matrix from Sinkhorn-Knopp
        """
        if self.step_count < self.freeze_prototypes_steps:
            return
        
        # Compute new prototype positions as weighted average of features
        new_prototypes = assignments.T @ ema_features  # [K, D]
        new_prototypes = F.normalize(new_prototypes, dim=1)
        
        # EMA update of prototypes
        self.ema_prototypes.data = (
            self.prototype_momentum * self.ema_prototypes.data + 
            (1 - self.prototype_momentum) * new_prototypes
        )
        self.ema_prototypes.data = F.normalize(self.ema_prototypes.data, dim=1)
\end{lstlisting}

\paragraph{Forward method}

\begin{lstlisting}[language=Python, numbers=left, basicstyle=\small\ttfamily, breaklines=true]
    def forward(self, x):
        """
        Forward pass with single augmented view.
        
        Args:
            x: augmented view [B, C, T]
        
        Returns:
            loss: scalar loss value
        """
        # Get targets from EMA path (no gradients)
        with torch.no_grad():
            targets, z_ema = self.get_ema_targets(x)
        
        # Get predictions from online path (with gradients)
        z_online = self.online_encoder(x)
        z_online = self.mask(z_online) # optional masking in latent space
        z_online = self.online_projector(z_online)
        z_online = self.online_predictor(z_online)
        z_online = F.normalize(z_online, dim=1)
        
        # Compute logits for online features against prototypes
        logits_online = z_online @ self.ema_prototypes.T / self.temperature
        
        # Cross-entropy loss: predict EMA targets with online features
        loss = -torch.sum(targets * F.log_softmax(logits_online, dim=1), dim=1).mean()
        
        # Update prototypes (no gradient flow)
        with torch.no_grad():
            self.update_prototypes(z_ema, targets)
        
        self.step_count += 1
        
        return loss
\end{lstlisting}

\newpage
\subsection{Comparison with HuBERT-ECG}
\label{app:hubert_ecg_vs_hubert++}

\begin{table}[htbp]
    \centering
    \caption{Comparison of aggregated macro-AUROC for external FM HuBERT-ECG~\citep{coppola2024hubert} and HuBERT++ pretrained for full pretraining dataset. The best-performing result is highlighted in boldface and underlined, while models that do not perform statistically significantly worse are also highlighted in boldface.\\}
    \label{tab:hubert_ecg_vs_hubert++}
    \begin{tabular}{lcc}
        \toprule
        \multicolumn{3}{c}{\textbf{FMs (Finetuned)}} \\
        & \textbf{HuBERT-ECG} & \textbf{HuBERT++} \\
        \midrule
        
        \multicolumn{3}{c}{\textbf{Adult ECG interpretation }} \\
        Ningbo & 0.958 & \underline{\textbf{0.969}} \\
        CPSC2018 & \textbf{0.956} & \underline{\textbf{0.962}} \\
        CPSC-Extra & 0.876 & \underline{\textbf{0.892}} \\
        Georgia & 0.883 & \underline{\textbf{0.911}} \\
        Chapman & 0.941 & \underline{\textbf{0.962}} \\
        SPH & 0.953 & \underline{\textbf{0.984}} \\
        PTB-XL (all) & 0.915 & \underline{\textbf{0.939}} \\
        PTB-XL (sub) & 0.918 & \underline{\textbf{0.942}} \\
        PTB-XL (super) & 0.908 & \underline{\textbf{0.936}} \\
        \midrule
        
        \multicolumn{3}{c}{\textbf{Pediatric ECG interpretation }} \\
        ZZU pECG & \textbf{0.883} & \underline{\textbf{0.892}} \\
        \midrule
        
        \multicolumn{3}{c}{\textbf{Cardiac structure \& function }} \\
        EchoNext & 0.792 & \underline{\textbf{0.831}} \\
        
        \bottomrule
    \end{tabular}
    \bigskip
    
    \textit{Note:} HuBERT++ performs best for all datasets.
\end{table}

\newpage
\section{Dataset Details}
\label{app:dataset_details}

\begin{table}[htbp]
    \centering
    \caption{Overview of datasets used for pre-training and downstream evaluation, along with their sample sizes and licenses.\newline}
    \begin{tabular}{lrc}
        \toprule
        \textbf{Dataset} & \textbf{Samples} & \textbf{License} \\
        \midrule
        \multicolumn{3}{c}{\textbf{Pre-training}} \\
        HEEDB~\citep{koscova2025bdsp, koscova2026harvard}          & 9,922,934 & BDSP Credentialed Health Data License 1.5.0 \\
        HEEDB-Emory~\citep{koscova2025bdsp, koscova2026harvard}    & 950,441 & BDSP Credentialed Health Data License 1.5.0 \\
        CODE-15\%~\citep{ribeiro_2021_4916206}      & 340,285    & Creative Commons Attribution 4.0 International \\
        \midrule
        \multicolumn{3}{c}{\textbf{Downstream Evaluation}} \\
        Ningbo~\citep{reyna2022physionet}         & 34,808     & Open Data Commons Open Database License v1.0 \\
        CPSC2018~\citep{reyna2021cinc,reyna2022physionet}       & 6,867      & Open Data Commons Open Database License v1.0 \\
        CPSC-Extra~\citep{reyna2021cinc,reyna2022physionet}     & 3,441      & Open Data Commons Open Database License v1.0 \\
        Georgia~\citep{reyna2021cinc,reyna2022physionet}        & 10,286     & Open Data Commons Open Database License v1.0 \\
        Chapman~\citep{Zheng2020}        & 10,646     & Creative Commons Attribution 4.0 International \\%CC-BY 4.0 is their figshare default
        SPH~\citep{liu2022large,liu_wang_chen_zhang_li_bian_shu_chen_2022}            & 25,770     & Creative Commons Zero 1.0 Universal \\ %CC0 is the standard figshare default   
        PTB-XL~\citep{wagner2022ptbxl,wagner2020ptb}         & 21,799     & Creative Commons Attribution 4.0 International \\
        ZZU pECG~\citep{tan2025pediatric,jian2025}       & 12,328     & Creative Commons Attribution 4.0 International \\
        EchoNext~\citep{elias2025echonext,Hughes2026}       & 82,543     & PhysioNet Restricted Health Data License 1.5.0 \\
        MIMIC-IV-ECG~\citep{PhysioNet-mimic-iv-ecg-1.0}   & 182,076    & Open Data Commons Open Database License v1.0 \\
        \bottomrule
    \end{tabular}
    \label{tab:datasets_licenses}
\end{table}

\newpage
\section{Experimental Setup}
\label{app:experimental_setup}

\paragraph{Input preprocessing}
All models operate on standard 12-lead ECG signals sampled at 240 Hz. Raw waveforms are segmented into non-overlapping windows of 2.5\,s (600 timesteps) and passed directly to the encoder. No data augmentation is applied during pretraining beyond the masking or context selection strategies intrinsic to each objective. Pretraining data are partitioned into 10 folds, with 9 folds used for training
and the remaining fold held out for validation; no held-out test set is used during pretraining.

\paragraph{Shared encoder architecture}
All five pretraining objectives share an identical encoder that is kept fixed across every comparison, ensuring that any observed performance differences are attributable solely to the pretraining loss and its associated components. The encoder consists of a CNN stem followed by an S4-based sequential backbone. The CNN stem comprises four convolutional layers with output feature dimensions $[512, 512, 512, 512]$, kernel sizes $[3, 1, 1, 1]$, strides $[2, 1, 1, 1]$, and dilations $[1, 1, 1, 1]$; batch normalization is applied after each convolutional layer (no layer normalization in the stem). The S4 backbone uses 4 layers with model dimension 512, state dimension 8, dropout rate 0.2, and no prenormalization or batch normalization inside the S4 blocks. The backbone operates in non-causal mode for all objectives except CPC, which requires a causal backbone consistent with its autoregressive prediction objective.

\paragraph{Shared training configuration}
All models are optimized with Adam, learning rate 3e-3, weight decay 1e-3, and batch size 64. Training is performed in full (fp32) precision on a single NVIDIA RTX PRO 6000 Blackwell Server Edition GPU. Models are trained for 10 epochs.

The pretraining corpus at each scale corresponds to specific HEEDB subsets: the 18K-sample run uses subset S0001-1987, the 45K-sample run uses subset S0001-1990, the 106K-sample run uses subset S0001-2019, and the 753K-sample run uses subset S0001-2007. The full 11M-sample corpus combines all available HEEDB cohorts, HEEDB-Emory, and CODE-15\%.

\paragraph{data2vec configurations}
The EMA teacher averages the contextualized representations produced by the top 2 S4 backbone layers (layers 3 and 4). Layer normalization is applied to the averaged teacher targets before the regression loss is computed. The SSL prediction head is a single-layer non-causal S4 module with the same model dimension as the backbone ($d{=}512$).

\paragraph{DinoSR configurations}
The EMA teacher produces discrete cluster pseudo-labels via online $k$-means quantization using two codebooks each of size 256 (codebook EMA momentum\,${=}\,0.9$). Layer normalization is applied to teacher features prior to quantization. The cross-entropy prediction loss uses a softmax temperature of 1.0. The SSL prediction head is a single-layer non-causal S4 module. A span masking strategy is applied in the latent space: positions are independently sampled as span midpoints with probability 0.065, and a contiguous span of 10 tokens centered on each midpoint is masked, yielding an effective masking rate of approximately 65\% of the input sequence.

\paragraph{JEPA configurations}
A multi-block masking strategy is employed following~\cite{assran2023self}. The context encoder receives an unmasked region uniformly sampled from $[85\%,\,100\%]$ of the input (1 context block), and the predictor targets 4 non-overlapping prediction blocks, each spanning $[15\%,\,20\%]$ of the input, with a minimum of 64 tokens retained and overlap between prediction blocks disallowed. The EMA teacher produces latent representations from the top backbone layer, to which layer normalization is applied twice: once at the individual layer level and once to the final aggregated targets. The prediction loss is the smooth-$\ell_1$ loss with $\beta{=}1.0$. The SSL prediction head is a single-layer non-causal S4 module.

\paragraph{CPC configurations}
No masking or EMA teacher is used. The model employs a causal S4 backbone and learns to predict 14 future latent steps from the causal context representation via a contrastive objective, drawing all negative samples exclusively from within the same input sequence. The SSL head is a linear projection without bias term.

\paragraph{HuBERT++ configurations}
The EMA teacher produces soft cluster assignment targets via Sinkhorn--Knopp optimal transport applied online to two codebooks of sizes 128 and 256. The prediction loss is a KL divergence between the student's log-probabilities and the teacher's soft assignments, computed on both masked and unmasked tokens with $\alpha{=}0.75$ weighting the masked token loss and $(1{-}\alpha)$ weighting the unmasked token loss. The same span masking strategy (midpoint probability 0.065, span length 10) is applied in the latent space. The SSL head is a multi-layer perceptron (MLP). Full design details and pseudo-code are provided in Appendix~\ref{app:hubertpp}.

\paragraph{Downstream evaluation}
Downstream evaluation follows the protocol of~\citep{al-masud2026benchmarking} across 26 clinically relevant tasks from 10 public datasets covering 7 task categories. All downstream runs use Adam, learning rate 1e-3, weight decay 1e-3, batch size 64, and 100 epochs. Three evaluation modes are considered. \emph{Finetuning (linear head):} the pretrained encoder is optimized end-to-end together with a linear prediction head using layer-dependent learning rates: the head is updated at the base rate (1e-3), the predictor module at 1e-4, and the CNN stem at 1e-5 (discriminative factor 0.1 applied per group). Sequence-level predictions are obtained by mean-pooling the encoder output before the linear head. \emph{Frozen:} encoder weights are kept fixed and a learnable query-attention pooling head~\citep{bardes2024vjepa} (16 attention heads, no bias) is trained at the base learning rate, operating directly on the full token sequence. \emph{Linear:} encoder weights are kept fixed and a single linear layer is trained on mean-pooled representations at the base learning rate. Binary cross-entropy is used for classification tasks and MAE for regression tasks with $z$-normalized targets. Macro-averaged AUROC serves as the primary metric for classification; standardized MAE is used for regression. Statistical significance is assessed via pairwise empirical bootstrapping on the held-out test set, with rankings determined by statistical equivalence groups (ties indicate no statistically significant difference at the chosen confidence level).

\paragraph{Computational resources}
All pretraining and downstream evaluation experiments were conducted on NVIDIA RTX PRO 6000 Blackwell Server Edition GPUs for pretraining and NVIDIA L40 GPUs for downstream evaluation, with each pretraining run using a single GPU. To accelerate data loading, all pretraining datasets were copied to \texttt{\$TMPDIR} (a per-job local NVMe-backed file system on the HPC cluster) prior to each run; without this step, I/O bottlenecks would significantly delay training. Approximate wall-clock training times per objective at each pretraining scale are summarized in Table~\ref{tab:compute}.

\begin{table}[htbp]
    \centering
    \caption{Approximate wall-clock pretraining time per run (single NVIDIA RTX PRO 6000 Blackwell Server Edition). `m' stands for minute, `h' for hour. \newline}
    \label{tab:compute}
    \begin{tabular}{lccc}
        \toprule
        \textbf{Objective} & \textbf{Small (106K, 10 ep.)} & \textbf{Medium (753K, 10 ep.)} & \textbf{Full (11M, 10 ep.)} \\
        \midrule
        data2vec   & 33.9 m & 3.9 h  & 31.2 h \\
        DinoSR     & 37.7 m & 4.3 h & 38.6 h \\
        JEPA       & 35.2 m & 4.0 h & 34.8 h \\
        CPC        & 42.4 m & 4.8 h & 45.1 h \\
        HuBERT++   & 27.6 m & 3.0 h & 26.3 h \\
        \bottomrule
    \end{tabular}
\end{table}

\end{document}